\documentclass[aps,prd,onecolumn,preprintnumbers,superscriptaddress,nofootinbib]{revtex4-2}
\usepackage[table]{xcolor}   
\usepackage{array}           
\usepackage{booktabs}       
\usepackage{graphicx}
\usepackage{epstopdf}
\usepackage{amsmath}
\usepackage{amsfonts}
\usepackage{amssymb}
\usepackage{appendix}
\usepackage{enumerate}
\usepackage{natbib}
\usepackage{comment}
\usepackage{bbold}
\usepackage[shortlabels]{enumitem}
\usepackage{slashed}
\usepackage{subfigure}
\usepackage{setspace}
\usepackage{footnote}
\usepackage{lipsum}
\usepackage{multirow}
\usepackage[colorlinks = true,
            linkcolor = blue,
            urlcolor  = blue,
            citecolor = blue,
            anchorcolor = blue]{hyperref}
\usepackage[capitalize]{cleveref}
\usepackage{braket}
\usepackage[compat=1.1.0]{tikz-feynman}
\usepackage{array,multirow}
\usepackage{physics}
\usepackage{feynmp-auto}
\usepackage[normalem]{ulem}
\usepackage{url}
\usepackage{units}
\usepackage[normalem]{ulem}
\usepackage{orcidlink}
\usepackage{lipsum}

\newcommand{\graymathbox}[1]{
  \tikz[baseline=(X.base)]{
    \node[
      fill=gray!8,
      draw=gray!80,
      line width=0.8pt,
      rounded corners=3pt,
      inner xsep=5pt,
      inner ysep=4pt
    ] (X) {$\displaystyle #1$};
  }
}

\begin{document}

\title{Four, One, and None: Quantifying the Ultra-High-Energy Neutrino Anomaly Across ANITA-IV, KM3NeT, and IceCube}

    \author{Dibya~S.~Chattopadhyay~\orcidlink{0000-0003-2323-3950}}
	\email{dibya.chattopadhyay@okstate.edu}
	\affiliation{Department of Physics, Oklahoma State University, Stillwater, OK 74078, USA}
    
	\author{Carlos~A.~Arg\"uelles~\orcidlink{0000-0003-4186-4182}}
	\email{carguelles@fas.harvard.edu}
	\affiliation{Department of Physics \& Laboratory for Particle Physics and Cosmology, Harvard University, Cambridge, MA 02138, USA}
    
    \author{Vedran~Brdar~\orcidlink{0000-0001-7027-5104}}
	\email{vedran.brdar@okstate.edu}
	\affiliation{Department of Physics, Oklahoma State University, Stillwater, OK 74078, USA}


\begin{abstract}
The four near-horizon neutrino-like events reported by ANITA-IV and the ultra-high-energy track-like event KM3-230213A observed by KM3NeT imply neutrino event rates that are in tension with the absence of corresponding events at IceCube.
In this work, we perform a joint analysis of these events, taking into account the absence of any corresponding ones at IceCube.
We construct semi-analytic, energy- and direction-dependent effective areas for the three detectors and account for the time-dependent ANITA-IV and KM3NeT exposures.
For a diffuse all-sky power-law flux varying both the normalization and the spectral index, the measured event rates across the three detectors are not reproduced.
The best-fit configuration, corresponding to a tension of $\sim7.5\sigma$, predicts approximately five IceCube events while strongly underpredicting the ANITA-IV and KM3NeT counts.
In contrast to the diffuse scenario, the tension can be substantially alleviated if the events arise from short-duration transients that occur exactly along the observed directions during the ANITA-IV and KM3NeT detection windows. Such a realization, however, is highly fine-tuned. If rare transients are instead distributed randomly across the full sky over the $\sim15$-year IceCube observation period, additional sources inevitably contribute to the IceCube exposure.
For benchmark populations with a probability of approximately $10^{-4}$ to produce four favorable transients at ANITA-IV, the best-fit configuration of sources, out of $10^5$ Monte Carlo realizations, remains in $5.9\sigma$ tension.
We conclude that, within the Standard Model, directional and temporal variations alone can not reconcile the ANITA-IV and KM3NeT observations with the IceCube null result, under both a diffuse all-sky flux and a rare-transient source hypothesis.
\end{abstract}


\maketitle

\section{Introduction}
\label{sec:introduction}

Over the last 15 years, high-energy neutrino astronomy has moved from first detection to a robust observational program, led by the IceCube Neutrino Observatory~\cite{IceCube:2006tjp,IceCube:2016zyt}, situated at the Amundsen–Scott South Pole Station~\cite{IceCube:2013cdw,IceCube:2013low,IceCube:2014stg}.
IceCube has detected astrophysical neutrinos in the TeV to PeV range, providing multiple measurements of the diffuse high-energy neutrino spectrum~\cite{IceCube:2020wum,IceCube:2024fxo}.
However, the situation at even higher energies appears less clear.
In the EeV regime, several experiments probe complementary neutrino energy ranges and directions in the sky~\cite{ANITA:2019wyx,IceCube:2018fhm,IceCubeCollaborationSS:2025jbi,KM3NeT:2025ccp}.
While IceCube benefits from its large instrumented volume and long exposure, radio-detection experiments such as ANITA (Antarctic Impulsive Transient Antenna)~\cite{ANITA:2008mzi} are sensitive to enormous effective volumes for Earth-skimming and near-horizon showers~\cite{ANITA:2008wdk,ANITA:2019wyx,ANITA:2021xxh}.
KM3NeT~\cite{KM3Net:2016zxf}, another large-volume Cherenkov neutrino telescope, is located in the Mediterranean Sea and therefore provides complementary sky coverage.
In parallel, a growing network of operating, under-construction, and proposed next-generation facilities---including IceCube-Gen2~\cite{IceCube-Gen2:2020qha}, Baikal-GVD~\cite{Baikal-GVD:2018isr}, P-ONE~\cite{P-ONE:2020ljt}, TRIDENT~\cite{TRIDENT:2022hql}, PUEO~\cite{PUEO:2020bnn} and RNO-G~\cite{RNO-G:2020rmc}, GRAND~\cite{GRAND:2018iaj}, POEMMA~\cite{POEMMA:2020ykm}, and TAMBO~\cite{TAMBO:2025jio}---will chart the ultra-high energy (UHE) neutrino sky~\cite{Arguelles:2024ncf}.

A number of UHE events have been recorded in the ANITA and KM3NeT experiments~\cite{ANITA:2016vrp,ANITA:2018sgj,ANITA:2020gmv,KM3NeT:2025ccp}. 
Over the four flights of the ANITA experiment, a total of six anomalous tau-neutrino-like events have been reported~\cite{ANITA:2016vrp,ANITA:2018sgj,ANITA:2020gmv,ANITA:2021xxh}. 
During the flights of ANITA-I and III, a total of two steeply upgoing events, compatible with a $\tau$-lepton produced by an up-going $\nu_\tau$, were observed~\cite{ANITA:2016vrp,ANITA:2018sgj,2020JCAP...01..012S}.
The inferred trajectories of these two events imply that the parent neutrinos would have traversed approximately $5450$~km and $7000$~km through Earth~\cite{ANITA:2016vrp,ANITA:2018sgj,Romero-Wolf:2018zxt,2020JCAP...01..012S}.
Such steep angles of emergence are highly anomalous, since at such UHEs, the incoming $\nu_\tau$ would be expected to undergo severe attenuation while traversing the Earth, leading to an extremely small survival probability~\cite{Romero-Wolf:2018zxt,Fox:2018syq}.
This tension is further reinforced by the absence of similar steeply upgoing events in IceCube and Pierre Auger~\cite{IceCube:2020gbx,PierreAuger:2025hvl}.

Meanwhile, ANITA-IV has reported four non-inverted, near-horizon cosmic-ray-like events that are difficult to accommodate within standard expectations~\cite{ANITA:2021xxh,ANITA:2020gmv,Bertolez-Martinez:2023scp}.
If interpreted as $\tau$ showers from Earth-skimming $\nu_\tau$, then the neutrino fluxes needed to account for these four ANITA-IV events would overshoot existing limits from IceCube~\cite{ANITA:2021xxh,Bertolez-Martinez:2023scp,Arguelles:2025ewg}.
The expected anthropogenic background in ANITA-IV is estimated to be at most $0.37_{-0.17}^{+0.27}$ events, independent of the polarity considered~\cite{ANITA:2020gmv}.
On the other hand, while none of the four ANITA-IV events occurred at sky locations that were simultaneously visible to the Pierre Auger Observatory, the fluence required for ANITA-IV to observe these events is in strong tension with Pierre Auger limits over a wide range of energies. The four ANITA-IV events are also in tension with the ANITA in-ice Askaryan neutrino channel at energies above $10$~EeV~\cite{ANITA:2021xxh}.

More recently, KM3NeT has reported the detection of KM3-230213A, with a median incoming neutrino energy of 220 PeV~\cite{KM3NeT:2025npi,KM3NeT:2025ccp}.
Therefore, the energy of this event is at least an order of magnitude above the most energetic event ever observed at IceCube~\cite{IceCube:2021rpz}.
Given IceCube’s larger effective area and longer exposure, at least several events would be expected in IceCube at comparable energies, leading to a tension at the level of roughly $(2-3.5)\sigma$ depending on whether the source is assumed to be diffuse or transient~\cite{Li:2025tqf,KM3NeT:2025npi,KM3NeT:2025ccp,IceCubeCollaborationSS:2025jbi,Palmisano:2025abd}.

The four ANITA-IV events, when considered together with the IceCube non-observation of any similar UHE neutrinos, already point to a substantial tension in the expected event rates~\cite{ANITA:2020gmv,ANITA:2021xxh,Bertolez-Martinez:2023scp,Arguelles:2025ewg}. Such an anomaly is further exacerbated by the observation of a UHE neutrino by KM3NeT, and a corresponding null observation at IceCube~\cite{KM3NeT:2025npi,KM3NeT:2025ccp,IceCubeCollaborationSS:2025jbi,Li:2025tqf,Brdar:2025azm,Dev:2025czz,Farzan:2025ydi,Palmisano:2025abd,Arguelles:2025ewg}.
In this work, we perform, for the first time, a joint statistical analysis of the four ANITA-IV events, the single KM3NeT event, and the absence of corresponding UHE events in IceCube. We do not consider the steeply upgoing ANITA-I and ANITA-III events. Instead, we focus on the four near-horizon ANITA-IV events and the KM3NeT event, for which the principal anomaly is an inconsistency in the expected event counts -- or, equivalently, in the flux normalization required to account for the ANITA-IV and KM3NeT observations without overproducing events in IceCube.

We find that for a diffuse all-sky flux, the tension reaches $7.5\sigma$, increasing to $7.9\sigma$ for a fixed $E_\nu^{-2}$ spectrum. This tension is driven primarily by the mismatch between the ANITA-IV observations and the IceCube null result. In fact, a dedicated analysis using only these two detectors yields a tension of similar magnitude.
Allowing the neutrinos to originate from a population of rare transient sources reduces the anomaly to $5.9\sigma$.
For the transient scenario, we consider rare neutrino sources distributed over the full sky and fix the probability of observing at least four transients in the optimal observation window of ANITA-IV to $10^{-4}$.
Such a statistical treatment naturally predicts some sources in the optimal IceCube angular window and avoids artificially reducing the tension by placing transients only in favorable locations and time windows.
Under both the diffuse and the transient hypotheses, the best-fit configurations require extreme upward fluctuations at ANITA-IV and KM3NeT, together with a downward fluctuation at IceCube.

\begin{figure}
    \centering
    \includegraphics[width=0.5\linewidth]{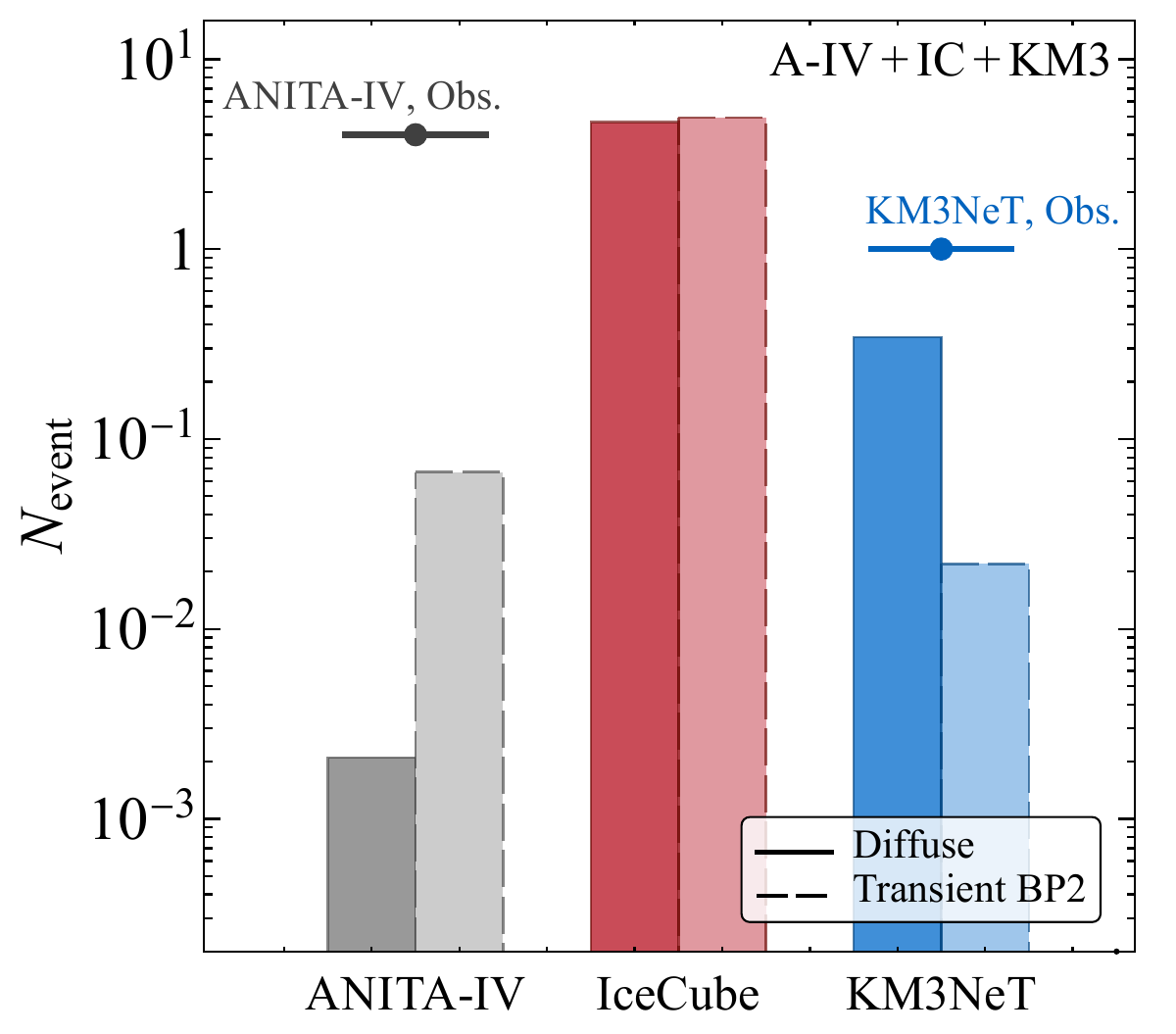}
    \caption{Best-fit expected event numbers at ANITA-IV (gray), IceCube (red), and KM3NeT (blue) for the diffuse-flux scenario and for the transient benchmark (BP2), corresponding to 185 transient sources with 10-day duration each, randomly distributed over 5400 days of IceCube runtime. Solid histogram boundaries with darker shading show the diffuse result, while dashed boundaries with lighter shading represent the result for the transient benchmark BP2. The best-fit expectation values are obtained for an $E_\nu^{-2}$ neutrino spectrum.  The solid gray and blue markers indicate the observed event counts, with four events at ANITA-IV, one at KM3NeT, and none at IceCube. In both diffuse and transient scenarios, fitting the ANITA-IV and KM3NeT observations leads to a substantial expected event yield at IceCube, while the expected ANITA-IV and KM3NeT event counts remain well below their observed values.}
    \label{fig:best_fit_event_all_diffuse_transient}
\end{figure}

These results are summarized in Fig.~\ref{fig:best_fit_event_all_diffuse_transient}, where the corresponding best-fit event yields are shown. The diffuse-flux results are shown with solid boundaries, while the transient-source results are shown with dashed boundaries and lighter colors. For a diffuse flux, we show the expected number of events for a best-fit flux, for a $\gamma=2$ spectral index. The joint ANITA-IV + KM3NeT + IceCube analysis leads to best-fit values that correspond to $\sim 5$ expected events at IceCube, with $\mathcal{O}(10^{-3})$ expected events at ANITA-IV and $\mathcal{O}(10^{-1})$ expected events at KM3NeT (ARCA-21, corresponding to 21 of the 230 strings deployed).
This is in severe conflict with the null observation at IceCube, and the observation of four and one events at ANITA-IV and KM3NeT, respectively.
For the transient scenario, in Fig.~\ref{fig:best_fit_event_all_diffuse_transient}, we present the results for a benchmark point corresponding to $185$ sources with a $10$~day duration each, distributed uniformly across the sky and over $5400$~days (see more details in~\cref{sec:transient}).
The best-fit transient scenario, chosen out of $10^5$ Monte Carlo (MC) realizations, improves the expected event rates for ANITA-IV to $\sim\mathcal{O}(10^{-1})$ by exploiting the time and direction-dependent detector exposures.
The transient scenario, however, yields a decreased expected number of events at KM3NeT.
In both the diffuse and transient scenarios, a simultaneous upward fluctuation at ANITA-IV and KM3NeT and a downward fluctuation at IceCube remains necessary.

In the remainder of the work, we will explain how we reached the results shown in~\cref{fig:best_fit_event_all_diffuse_transient}.
In section~\ref{sec:details} we summarize the relevant details of the observed events.
In~\cref{sec:directional_exposure}, we calculate the time-dependent directional effective areas and exposures of ANITA-IV, IceCube, and KM3NeT, and assess their implications for steady and transient-source interpretations of the observed events. In~\cref{sec:diffuse} we quantify the diffuse-flux interpretation using a Poisson likelihood. In~\cref{sec:transient} we perform the analysis for transient sources, first for fixed event directions and then for an all-sky transient population. We conclude in~\cref{sec:conclusion}. 

\section{The ANITA-IV and KM3NeT events}
\label{sec:details}

In this section, we discuss all the relevant details of the four ANITA-IV events and the one KM3NeT event observed, contrasting them with the lack of any such UHE observations at IceCube.

ANITA-IV has reported four non-inverted, near-horizon, cosmic-ray-like events~\cite{ANITA:2020gmv}. Their polarity and geometry make them qualitatively different from ordinary downgoing cosmic-ray air showers reflected from the Antarctic ice surface. The four upgoing events are observationally consistent with $\tau$-induced extreme air showers (EASs) from Earth-skimming $\nu_\tau$~\cite{ANITA:2021xxh}. Note that these events have also been reported to be in tension with non-observation in Pierre Auger and ANITA’s Askaryan in-ice neutrino channel above $10^{19}$~eV~\cite{ANITA:2021xxh}.

\begin{table}[t]
\centering
\renewcommand{\arraystretch}{1.16}
\resizebox{\textwidth}{!}{%
\begin{tabular}{|l|c|c|c|c|c|}
\hline
Detector
& \multicolumn{4}{c|}{ANITA-IV}
& KM3NeT (ARCA-21) \\
\hline

Event ID
& 4098827
& 19848917
& 50549772
& 72164985
& KM3-230213A \\

Event label
& \textbf{A}
& \textbf{B}
& \textbf{C}
& \textbf{D}
& \textbf{KM3-230213A} \\
\hline

UTC date
& \; 2016 December~03 \quad
& \; 2016 December~08 \quad
& \; 2016 December~16 \quad
& \; 2016 December~22 \quad
& 2023 February~13 \\

UTC time
& 10:03:27
& 11:44:54
& 15:03:19
& 06:28:14
& 01:16:47 \\

Elevation angle ($^\circ$)
& $-6.17 \pm 0.21$
& $-6.71 \pm 0.20$
& $-6.73 \pm 0.20$
& $-6.12 \pm 0.10$
& $0.60$ \\

Horizon angle ($^\circ$)
& $-5.92 \pm 0.02$
& $-6.06 \pm 0.02$
& $-5.92 \pm 0.02$
& $-5.93 \pm 0.02$
& $-$ \\

Sky location RA ($^\circ$)
& $331.52$
& $15.14$
& $232.18$
& $224.00$
& $94.3$ \\

Sky location dec. ($^\circ$)
& $15.17$
& $-6.31$
& $2.03$
& $3.83$
& $-7.8$ \\
\hline
\end{tabular}
}
\caption{Event times, local arrival directions, and equatorial coordinates of the four anomalous $\nu_\tau$-like ANITA-IV near-horizon events and the track-like KM3-230213A event~\cite{ANITA:2020gmv,KM3NeT:2025npi}. The ANITA-IV events were observed slightly below the payload horizon and originate from four distinct sky locations. KM3-230213A was observed by KM3NeT in its ARCA-21 configuration from a nearly horizontal direction, approximately $0.6^\circ$ above the local horizon. The uncertainties on the KM3NeT equatorial coordinates correspond to the quoted $68\%$ C.L. directional uncertainty.}
\label{tab:anita_iv_location_time}
\end{table}

In~\cref{tab:anita_iv_location_time}, we show the details regarding the location and timing of the four anomalous ANITA-IV neutrino-like events~\cite{ANITA:2020gmv}.
For brevity, from here on, we will denote the four ANITA-IV events as Event A, B, C, and D.
Comparing the elevation angle and the horizon angle, we note that all four events come from below the horizon.
Using the apparent source position (latitude, longitude, and altitude), as well as the exact position (latitude, longitude, and altitude) of the ANITA-IV balloon, we calculate the exact direction of the sky from which the neutrinos arrived. The four anomalous events originate from four distinct points in the sky, implying that their astrophysical interpretation would require either a diffuse all-sky source population or four separate transient sources that were active during the ANITA-IV window of observation. 
The parent $\nu_\tau$ energy for the four ANITA-IV events can be estimated, but the value depends on the assumed power-law spectrum. For an $E_\nu^{-2}$ spectrum, the $1\sigma$ energy ranges for the four events span $(2.3-42.4)$~EeV~\cite{ANITA:2021xxh}.

The KM3NeT experiment, in its ARCA-21 configuration (at approximately $10\%$ completion), observed a UHE $\nu_\mu$ track-like event on 13th February 2023, at 01:16:47 UTC~\cite{KM3NeT:2025npi}.
This event, referred to as KM3-230213A, is consistent with a muon energy of $120^{+110}_{-60}$~PeV, with a $90\%$ C.L. energy range of $(35-380)$~PeV.
Assuming an incident neutrino spectrum proportional to $E_\nu^{-2}$, the inferred incoming $\nu_\mu$ energy lies in the range $(0.072-2.6)$~EeV at $90\%$ C.L.
At such UHEs, however, a $\tau$-lepton can also produce a track-like signature in detectors such as KM3NeT and IceCube~\cite{IceCube:2012lak,Kistler:2016ask,IceCube:2020fpi}.
In the continuous-energy-loss approximation, the relative energy-loss rates of muons and tau leptons imply that the latter, with an energy approximately one order of magnitude larger than that of the former, can yield a comparable energy-deposition profile~\cite{Dutta:2000hh,Bigas:2008ff,Jeong:2017mzv}. 
The event may therefore also be interpreted as originating from a $\nu_\tau$ with an energy of approximately $(0.7$--$26)$~EeV~\cite{Celli:2026neutrino}. 
Note that a $\tau$-lepton decays to a final state involving $\mu$ approximately $17\%$ of the time~\cite{ParticleDataGroup:2024cfk}.
Therefore, we must consider both the $\nu_\mu$ and $\nu_\tau$ effective areas in our analysis.

The sky coordinates of the KM3-230213A event are: RA = $(94.3\pm 1.5)^\circ$, dec. = $(-7.8\pm 1.5)^\circ$, at $68\%$ C.L. (see~\cref{tab:anita_iv_location_time}). Its direction does not match any of the four ANITA-IV event locations, further strengthening the case for either a diffuse all-sky flux or multiple rare transient sources.

IceCube is a cubic-kilometer neutrino telescope comprising 5,160 optical modules deployed along 86 strings deep within the Antarctic ice~\cite{IceCube:2016zyt}. Since beginning full-detector operations on May 13, 2011, it has accumulated $\sim 15$ years of nearly continuous exposure, providing exceptional sensitivity to rare UHE neutrino events~\cite{IceCube:2024pnx}.
The absence of any UHE neutrino event above $\sim 10$~PeV is therefore particularly striking when contrasted with the observation of KM3-230213A and the four ANITA-IV near-horizon events~\cite{ANITA:2020gmv,Li:2025tqf,KM3NeT:2025ccp,IceCube:2018fhm,IceCubeCollaborationSS:2025jbi}.
The tension between the KM3NeT observation and the IceCube non-observation has been quantified previously~\cite{KM3NeT:2025ccp,Li:2025tqf}; in this paper, we perform a joint assessment incorporating ANITA-IV, IceCube, and KM3NeT.

\section{Directional effective areas and transient exposures}
\label{sec:directional_exposure}

The effective area determines the instantaneous sensitivity of a detector to a neutrino of a given energy and arrival direction. 
Since effective areas in fine angular bins are not readily available in the literature, we use semi-analytic calculations to model their directional dependence.
The calculations for ANITA-IV, IceCube, and KM3NeT, together with the assumptions adopted in each case, are presented in Supplemental~\cref{sec:effective_area_calc}.
The derived effective areas are in good agreement with the angular-integrated and all-sky-averaged effective areas reported by the collaborations.
In what follows, we focus on the corresponding time-dependent exposure and on the comparison between the three detectors.

The exposure to a source located in the direction $\Omega$ over a time interval $\Delta t$, for a neutrino of energy $E_\nu$, is
\begin{equation}
\mathcal{E} (E_\nu,\Omega;t_0,\Delta t) = \int_{t_0}^{t_0+\Delta t} A_{\rm eff}(E_\nu,\Omega;t)\,dt \; .
\label{eq:directional_exposure}
\end{equation}
Here, $t_0$ denotes the beginning of the relevant observation window, typically set by either the onset of source activity or the start of detector data taking.
The expected number of events, for the all-sky diffuse flux scenario, is then
\begin{equation}
N_{\rm diff} = \int dE_\nu\,d\Omega\,dt\, F_\nu(E_\nu,\Omega,t) \, A_{\rm eff} (E_\nu,\Omega;t) \; ,
\label{eq:expected_events_exposure}
\end{equation}
where $F_\nu$ is the differential neutrino flux. For a point source emitting neutrinos, the expected number of events is
\begin{equation}
N_{\rm tr} = \int dE_\nu\,dt\, F_\nu(E_\nu,\Omega,t) \, A_{\rm eff} (E_\nu,\Omega;t) \; ,
\label{eq:expected_events_exposure_point_source}
\end{equation}
where the integration over the angle $\Omega$ is omitted.
Therefore, a large instantaneous effective area does not necessarily imply a large total event yield. The duration of the observation, the source duration, and the fraction of time in which the source lies within a favorable region of the detector field of view are all important factors that need to be considered.

The maximum IceCube track-like effective area is smaller by a factor of a few compared to the ANITA-IV $\nu_\tau$ effective area (see Supplemental~\cref{sec:effective_area_calc} for further details). The difference in runtime between the 2 experiments is therefore particularly important. IceCube has accumulated approximately $15$ years of data~\cite{IceCube:2016zyt}, compared with only about $27$ days for the ANITA-IV flight~\cite{ANITA:2019wyx,ANITA:2020gmv}, i.e., IceCube has been active for $\sim 200$ times longer. Thus, although ANITA-IV can have the larger instantaneous effective area in its optimal angular region, IceCube generally accumulates a much larger exposure for steady or long-duration fluxes. The ARCA-21 configuration of KM3NeT has a significantly smaller effective area than IceCube because only $21$ of the planned $230$ detection units were operating at the time of KM3-230213A~\cite{KM3Net:2016zxf,KM3NeT:2025npi,Li:2025tqf}.

\begin{figure}[t!]
    \centering
    \includegraphics[width=0.5\linewidth]
    {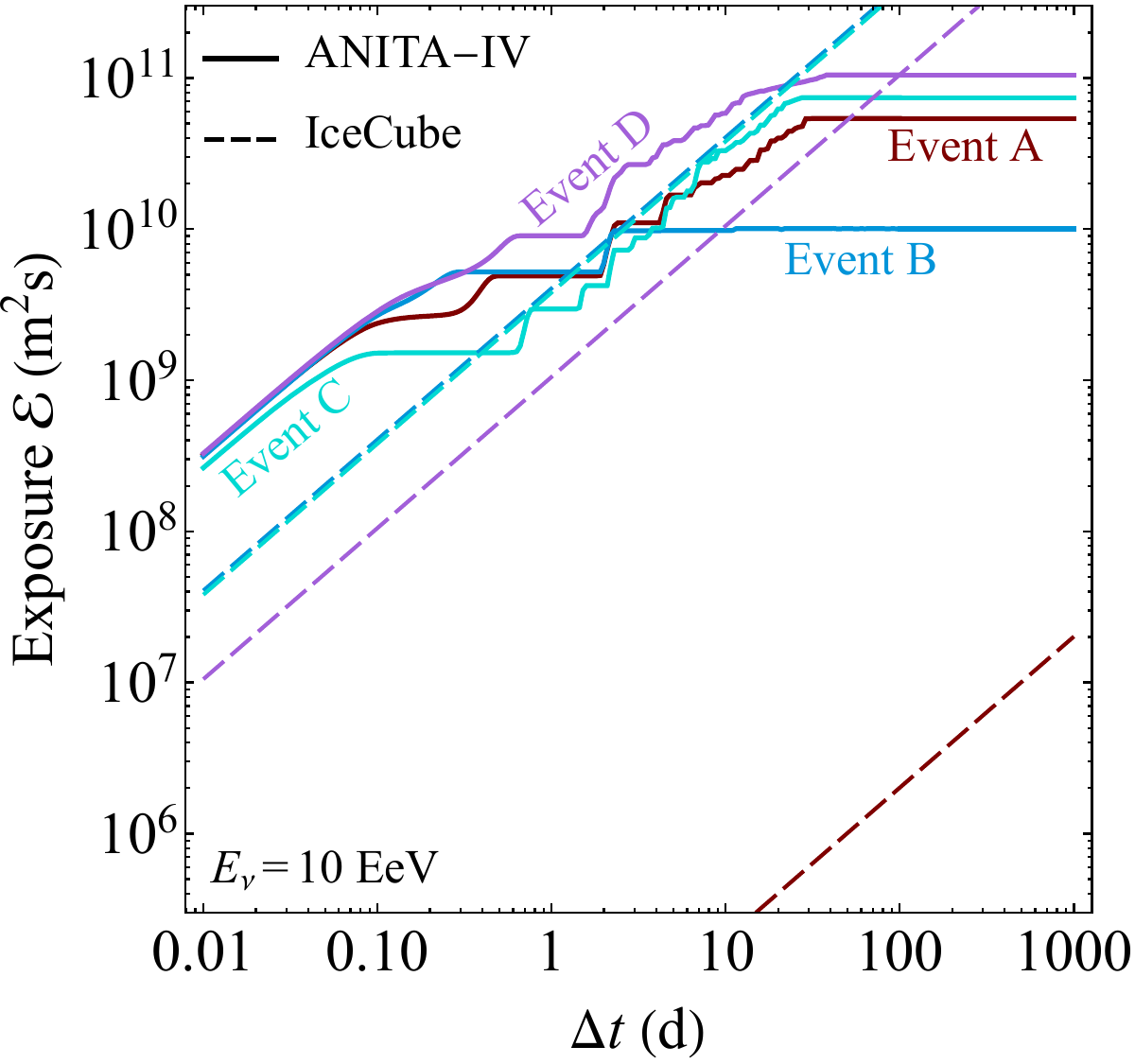}
    \hspace{0.5cm}
    \raisebox{0.067\linewidth}{
    \includegraphics[width=0.4\linewidth]
    {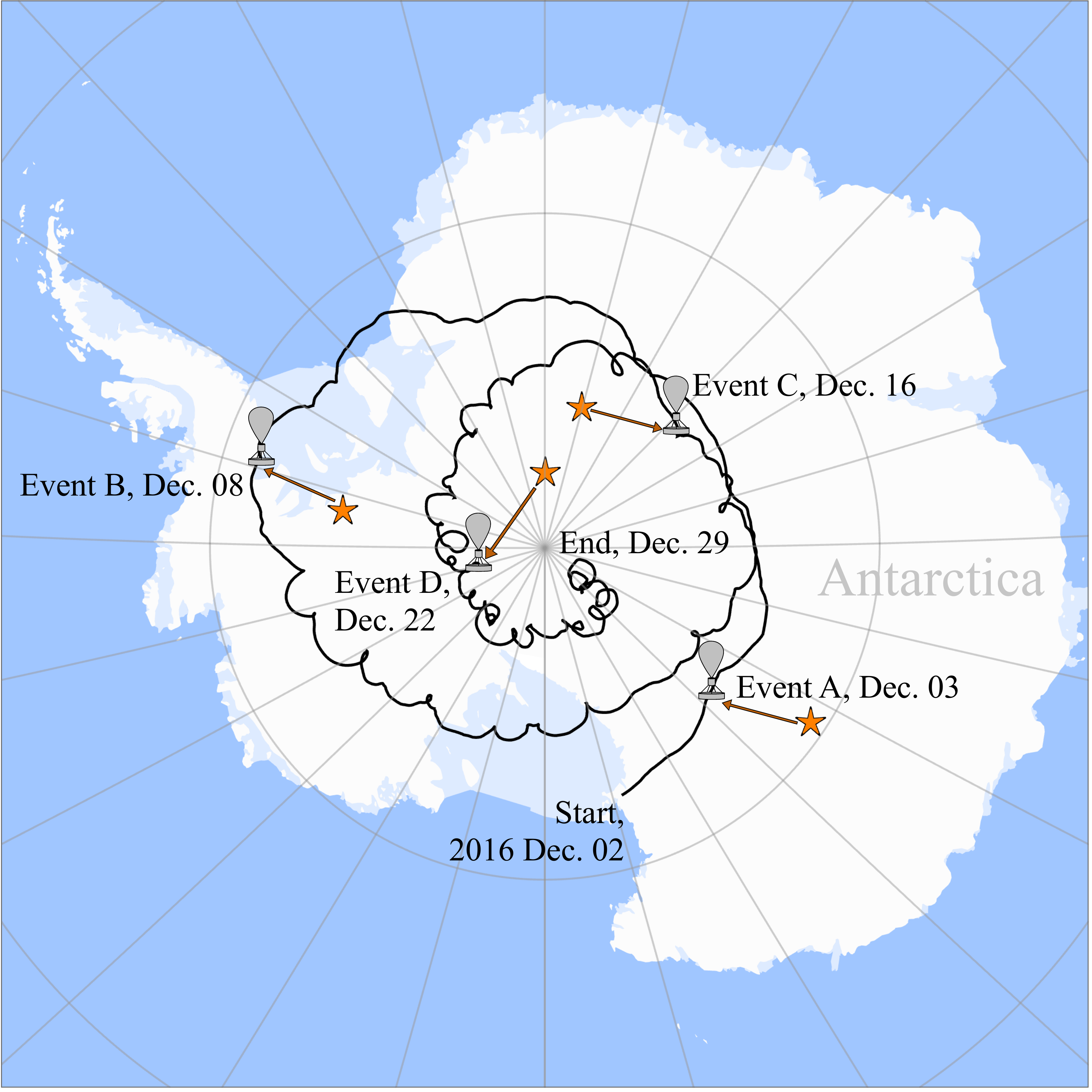}
    }
    \caption{Left panel: Exposure of ANITA-IV (solid curves) and IceCube (dashed curves) for $E_\nu=10~{\rm EeV}$ toward the directions of the four anomalous ANITA-IV events, as a function of the transient duration. Right panel: ANITA-IV flight trajectory during its approximately $27$-day flight in December 2016~\cite{ANITA:2019wyx,ANITA:2019xbi}. The payload positions at the four event times are labeled, while the orange stars indicate the corresponding apparent source directions.}
    \label{fig:anita_iv_events}
\end{figure}

In~\cref{fig:anita_iv_events} (left panel), we show the exposure for $E_\nu=10~{\rm EeV}$ toward the four ANITA-IV event directions as a function of the transient duration, for both ANITA-IV (solid lines) and IceCube (dashed lines) experiments. For simplicity, we assume that the event occurred exactly halfway through the transient duration. Note that the exposures for IceCube towards Events B and C, shown in blue and cyan dashed curves, almost overlap in the figure. This is because the two sources have almost the same elevation angle for IceCube due to their location in the sky. The balloon’s position throughout the ANITA-IV flight is essential for an accurate calculation of the exposure. The ANITA-IV balloon trajectory is shown in the right panel of ~\cref{fig:anita_iv_events}, together with the payload positions and apparent source directions at the event times. Since the four events originate from four distinct points on the sky, their exposures differ appreciably over the course of the ANITA-IV flight (see left panel).

For all four events, for sufficiently short transients, ANITA-IV can have a much larger exposure than IceCube. For longer transient durations, the ANITA-IV exposure may grow approximately linearly over some intervals, while in other cases it changes like a step function or irregularly. This is because the location of the ANITA-IV balloon and the time of observation both affect the optimal observation window in the sky.
When the transient duration becomes comparable to the flight duration, for $\Delta t \gtrsim \mathcal{O}(10)$~days, the exposure approaches a constant maximum value corresponding to continuous activity of the neutrino source throughout the full flight of ANITA-IV. In contrast, the IceCube exposure to these four points in the sky grows linearly with source duration because its location and directional response are stable over the relevant period.

Note that the comparison between ANITA-IV and IceCube exposures is strongly direction-dependent. For the directions corresponding to Events C and D, the IceCube exposure exceeds that of ANITA-IV for transient durations $\gtrsim \mathcal{O}(10)$~days. For Event B, this crossover occurs much earlier, at durations of only $\gtrsim \mathcal{O}(1)$~day. For Event A, the sky location is highly unfavorable for IceCube, and even for a transient lasting $1000$~days, the IceCube exposure remains more than $\mathcal{O}(10^3)$ times smaller than the corresponding ANITA-IV exposure.

\begin{figure}[t!]
    \centering
    \includegraphics[width=1\linewidth]
    {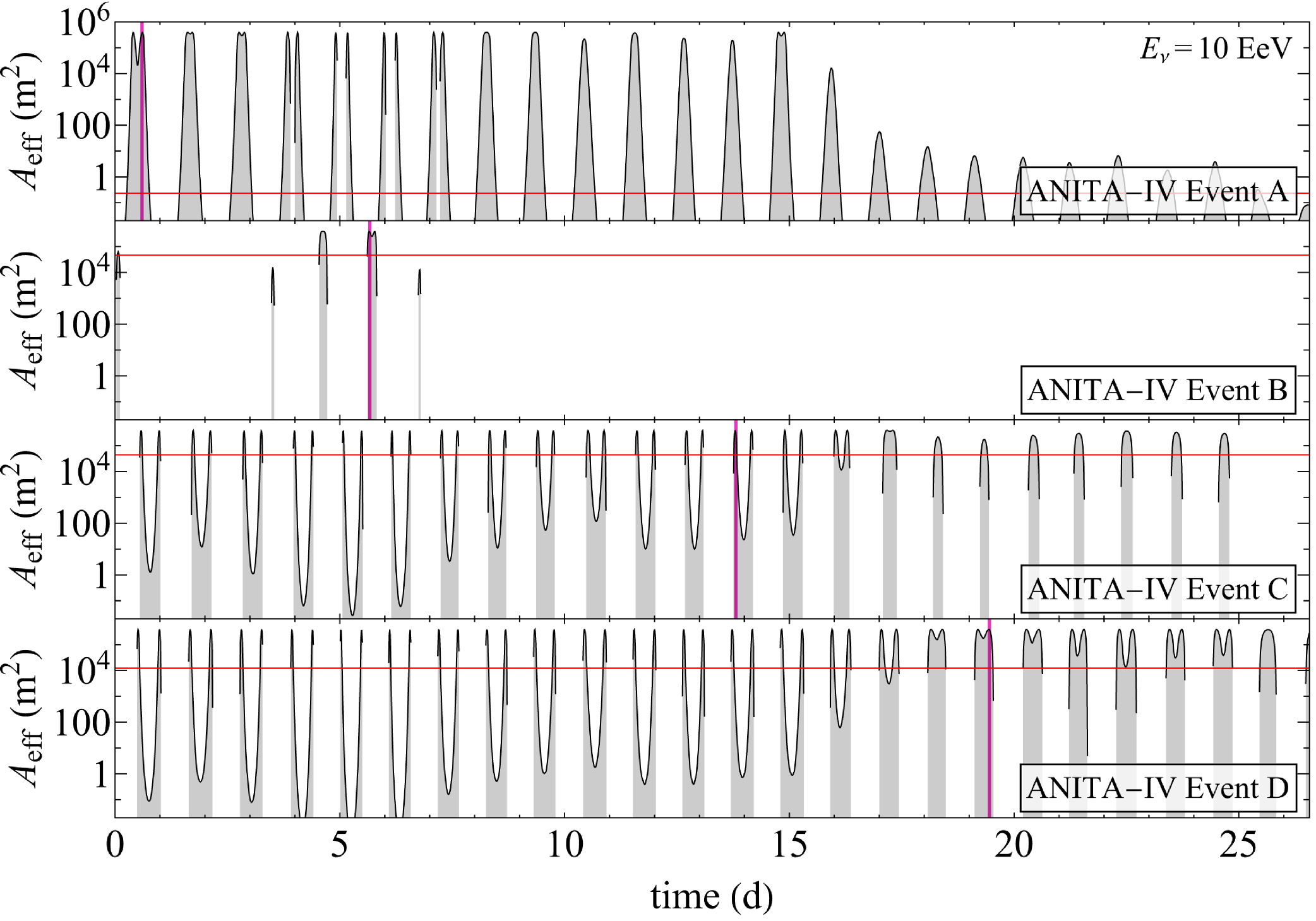}
    \caption{ANITA-IV (black curves with gray shaded regions) and IceCube (red) effective area for $E_\nu=10~{\rm EeV}$ toward the four observed event directions as a function of time during the ANITA-IV flight. The vertical magenta lines indicate the exact times at which the four ANITA-IV events were observed. The time $t=0$ corresponds to the start of the ANITA-IV experiment, 2~Dec~2016, 19:40:43 UTC.}
    \label{fig:anita_eff_area_time}
\end{figure}

In~\cref{fig:anita_eff_area_time}, we show the ANITA-IV and IceCube effective area (for a fixed energy of $E_\nu = 10$~EeV) towards each of the event directions. The ANITA-IV effective area is shown using gray shaded regions with black boundaries, while the IceCube effective area toward each fixed celestial direction is shown by the horizontal red line.
The ANITA-IV effective area consists of a non-uniform sequence of relatively narrow optimal visibility windows produced by the changing position of the balloon and by the strong near-horizon angular dependence. Between these windows, the effective area can decrease by many orders of magnitude. 

The exact event times are indicated by the vertical magenta lines in~~\cref{fig:anita_eff_area_time}, which coincide with a large ANITA-IV effective area.
The instantaneous effective areas towards the direction of the four events are not identical. In fact, Event B is in a comparatively unfavorable position in the sky for ANITA-IV. This leads to a much lower exposure for larger durations of the transients, as can be seen in~\cref{fig:anita_iv_events}.
Events A, C, and D are observed optimally by ANITA-IV; IceCube has a smaller effective area towards the direction of event A.
Since IceCube is located at the South Pole, the elevation of a source with fixed declination is essentially independent of sidereal time, i.e., Earth's rotation does not affect the optimal observation window for IceCube. Its effective area is therefore nearly constant, in contrast to the rapidly varying ANITA-IV response. While the peak ANITA-IV effective area can exceed that of IceCube, we observe that these peaks occupy only a small fraction of the flight duration.

The instantaneous and time-averaged effective areas in equatorial coordinates are shown in Supplemental~\cref{app:effective_areas_sky}, where we illustrate how the ANITA-IV flight path and the Earth’s rotation determine the detector exposure toward the observed event directions.

\begin{figure}[t!]
    \centering
    \includegraphics[width=0.5\linewidth]
    {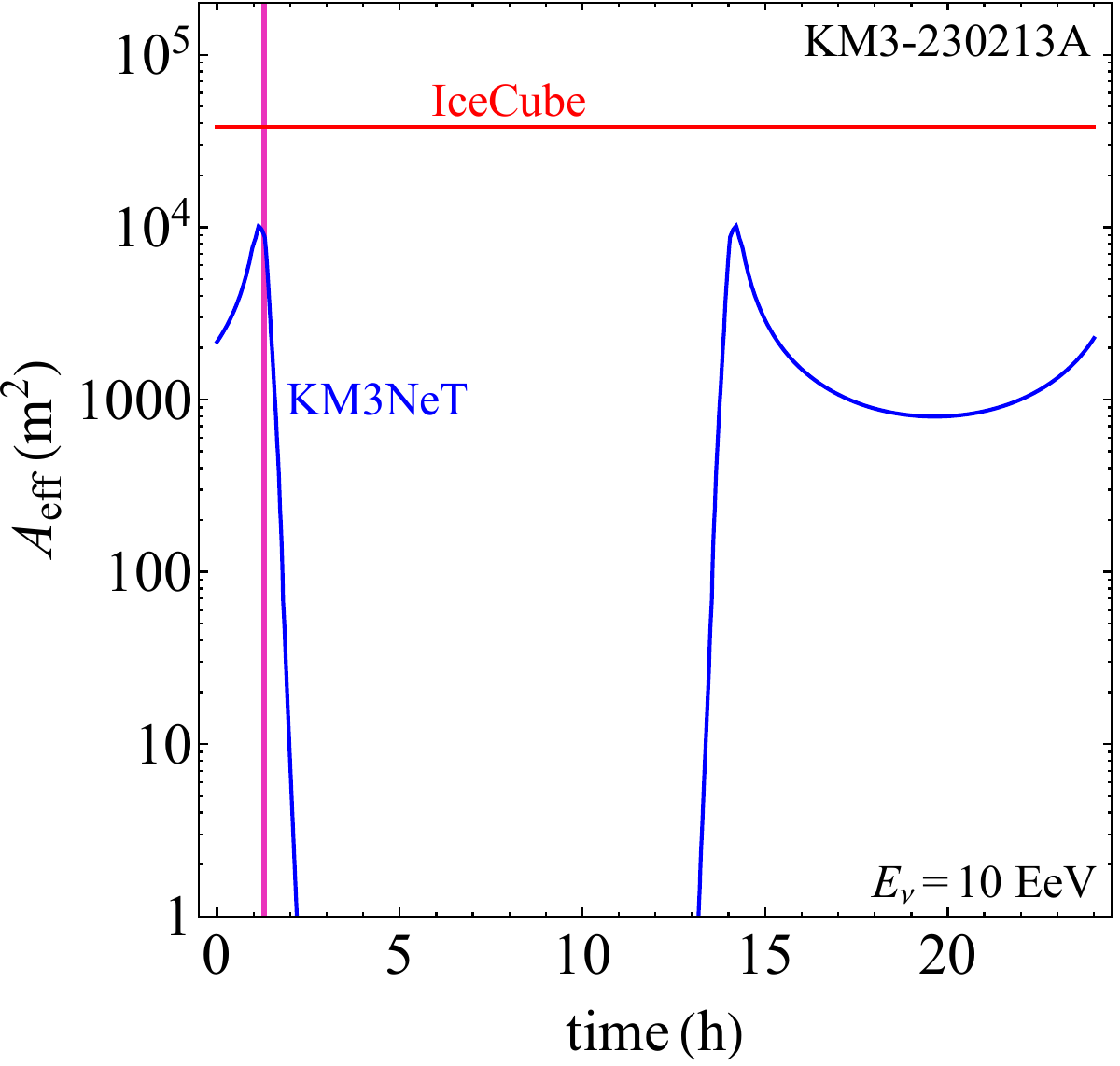}
    \caption{KM3NeT ARCA-21 (blue) and IceCube (red) effective area for $E_\nu=10~{\rm EeV}$ toward the direction of KM3-230213A as a function of time over one day. The vertical magenta line indicates the time at which KM3-230213A was observed.}
    \label{fig:km3net_eff_area_time}
\end{figure}

Figure~\ref{fig:km3net_eff_area_time} presents the comparison between the KM3NeT (ARCA-21) and IceCube effective area towards the direction of the event KM3-230213A, for $E_\nu=10~{\rm EeV}$. Due to the rotation of the Earth, the KM3NeT effective area varies strongly over a day. It features two optimal observation time windows towards the accessible directions in the sky, separated by an interval during which the neutrino trajectory through Earth crosses an extremely long path, resulting in a strong attenuation of the neutrino flux.
The KM3-230213A event occurred close to the narrow maximum of the first optimal effective area window on the $13$th of February, 2023, and therefore at a comparatively favorable time.

Even at this favorable instant, however, the KM3NeT ARCA-21 effective area remains below the corresponding IceCube effective area by a factor of a few. The observation of the UHE KM3-230213A event by the partially completed KM3NeT detector, along with the absence of corresponding IceCube events, therefore constitutes an additional source of tension for a steady or long-duration flux interpretation. This tension may be reduced only if the source is sufficiently short-lived. Additional sky-projected effective-area maps for KM3NeT, illustrating both the instantaneous and time-averaged directional effective areas, are provided in Supplemental~\cref{app:effective_areas_sky}.

Together, these results demonstrate that the anomaly in the UHE sector cannot be assessed using the peak effective areas alone. ANITA-IV benefits from a very large but highly localized and irregular effective area towards specific directions in the sky, KM3NeT observed its event during a favorable but short optimal observation window, and IceCube combines a large effective area with a much longer operating time. Any explanation of the observed data must therefore account simultaneously for the event directions, the transient duration, the detector live times, and the non-observation of corresponding events at IceCube.

\section{Diffuse flux interpretation for the ultra-high energy tension}
\label{sec:diffuse}

In this section, we present the joint likelihood analysis and expected event rates for a diffuse all-sky flux hypothesis.
Namely, we consider whether the ANITA-IV and KM3NeT events can originate from a common, isotropic diffuse neutrino flux. We parametrize the single-flavor flux as
\begin{equation}
\frac{d\Phi_\nu}{dE_\nu} = \phi_0 \left( \frac{E_\nu}{E_0} \right)^{-\gamma},
\label{eq:diffuse_flux}
\end{equation}
where $\phi_0$ is the flux normalization, $\gamma$ is the spectral index, and we take $E_0=1$~EeV. We consider a uniform flavor ratio of $\nu_e:\nu_\mu: \nu_\tau = 1:1:1$ at Earth~\cite{IceCube:2020fpi,IceCube:2024nhk}.
The expected number of events at a detector $\mathcal{D}$ (ANITA-IV, KM3NeT ARCA-21, and IceCube) is
\begin{equation}
\mu_{\mathcal{D}}(\phi_0,\gamma) = \int dt \, dE_\nu \, d\Omega \, \frac{d^2\Phi_\nu}{dE_\nu\,d\Omega} A_{\rm eff}^{\mathcal{D}}(E_\nu,\Omega,t) \; .
\label{eq:diffuse_expected_events}
\end{equation}
We consider the energy range $(0.1$--$100)$~EeV, which covers the relevant energy estimates for the four ANITA-IV events, most of the KM3NeT-inferred energy range under a $\nu_\mu$ interpretation, and the full range under a $\nu_\tau$ interpretation.
For ANITA-IV, we employ the $\nu_\tau$ effective area, while the IceCube and KM3NeT expectations include track-like contributions from both $\nu_\mu$ and $\nu_\tau$.
Under a diffuse all-sky flux, Eq.~\eqref{eq:diffuse_expected_events} shows that the expected event count also scales with detector live time. These are approximately $26.6$~days for ANITA-IV, $287$~days for KM3NeT ARCA-21, and $5400$~days for IceCube.

The joint likelihood for the observed event counts,
\begin{equation}
(n_{\rm A-IV},\, n_{\rm IC},\, n_{\rm KM3}) = (4,\, 0, \, 1)\; ,
\end{equation}
is constructed as a product of Poisson probabilities~\cite{Baker:1983tu,Cowan:2010js}
\begin{equation}
\mathcal{L}(\phi_0,\gamma) = \prod_{\mathcal{D}} \frac{ \mu_{\mathcal{D}}^{\,n_{\mathcal{D}}} \exp\left(-\mu_{\mathcal{D}}\right) }{ n_{\mathcal{D}}! }\; .
\label{eq:diffuse_likelihood}
\end{equation}
We scan over $\phi_0$ and $\gamma$ and compare the likelihood with the saturated model where the expectation at each detector is independently set equal to its observed count. Therefore, for the diffuse all-sky flux scenario
\begin{equation}
    -2\Delta \log \mathcal{L} = -2 \log \left(\frac{\mathcal L}{ \mathcal L_{\rm sat}}\right)\; , 
\end{equation}
where $\mathcal L_{\rm sat}$ corresponds to the scenario with a fixed expected number of events in each detector
\begin{equation}
(\mu_{\rm A-IV}, \, \mu_{\rm IC},\,\mu_{\rm KM3}) = (n_{\rm A-IV},\, n_{\rm IC}, \, n_{\rm KM3}) = (4,\, 0, \, 1) \; .
\end{equation}

For a spectral index $\gamma$, the expected number of events at detector $\mathcal{D}$ can be expressed as
\begin{equation}
\mu_{\mathcal{D}}(\phi_0,\gamma) = \phi_0 K_{\mathcal{D}}(\gamma) \; ,
\label{eq:expected_events_linear_normalization}
\end{equation}
where $K_{\mathcal{D}}(\gamma)$ contains the energy and direction-dependent effective area, detector live time, and all relevant exposure integrals (see~\cref{eq:diffuse_expected_events}). The joint Poisson log-likelihood is therefore
\begin{equation}
\log \mathcal{L} = \sum_{\mathcal{D}} \Big( n_{\mathcal{D}}\log \phi_0 + n_{\mathcal{D}}\log K_{\mathcal{D}} - \phi_0 K_{\mathcal{D}} - \log\left[n_{\mathcal{D}}!\right] \Big).
\end{equation}
Maximizing the log-likelihood with respect to the common flux normalization, $\phi_0$, we obtain
\begin{equation}
\frac{\partial\log\mathcal{L}}{\partial\phi_0} = \frac{N_{\rm obs}}{\phi_0} - \sum_{\mathcal{D}}K_{\mathcal{D}} = 0 \; ,
\end{equation}
where $N_{\rm obs}\equiv\sum_{\mathcal{D}}n_{\mathcal{D}}$ is the total number of observed events. Hence, for the best-fit flux, we obtain
\begin{equation}
\widehat{\phi}_0(\gamma) = \frac{N_{\rm obs}} {\sum_{\mathcal{D}}K_{\mathcal{D}}(\gamma)}\; ,
\label{eq:best_fit_total_count}
\end{equation}
which gives
\begin{equation}
    \sum_{\mathcal{D}}\widehat{\mu}_{\mathcal{D}} = N_{\rm obs} \; .
    \label{eq:ntotal_obs_expected}
\end{equation}
Here, $\widehat{\mu}_{\mathcal{D}}$ denotes the best-fit expectation value for event count at each detector.
Therefore, varying the flux normalization to determine the best-fit point yields a total expected event count equal to the total number of observed events. For the joint ANITA-IV, KM3NeT, and IceCube analysis, this gives
\begin{equation}
\widehat{\mu}_{\rm A-IV} + \widehat{\mu}_{\rm KM3} + \widehat{\mu}_{\rm IC} = 4+1+0=5.
\end{equation}

The goodness of fit is therefore determined by how the expected event counts are distributed among the detectors, which in turn depends on their relative exposure. Defining the exposure fraction
\begin{equation}
p_{\mathcal{D}}(\gamma) = \frac{K_{\mathcal{D}}(\gamma)} {\sum_{\mathcal{D}'}K_{\mathcal{D}'}(\gamma)},
\end{equation}
the best-fit expectations can be expressed as
\begin{equation}
\widehat{\mu}_{\mathcal{D}} = N_{\rm obs}\,p_{\mathcal{D}}.
\end{equation}
For the observed counts $(n_{\rm A-IV},n_{\rm KM3},n_{\rm IC})=(4,1,0)$, the profiled deviance is
\begin{equation}
-2 \Delta \log \mathcal{L} = 8\log\left[ \frac{4}{5\, p_{\rm A-IV}(\gamma)} \right] + 2\log\left[ \frac{1}{5\, p_{\rm KM3}(\gamma)} \right].
\label{eq:profiled_diffuse_deviance}
\end{equation}
Here, the fraction $p_{\rm IC}$ enters through $p_{\rm A-IV}+p_{\rm KM3}+p_{\rm IC}=1$, i.e., a large IceCube exposure reduces the fractions assigned to ANITA-IV and KM3NeT and increases the tension with the observed allocation $(4,1,0)$.
Varying the spectral index, $\gamma$, changes this ratio through the energy dependence of the detector responses. However, this cannot fully remove the tension as long as IceCube retains the dominant integrated exposure.

\begin{figure}
    \centering
    \includegraphics[width=0.48\linewidth]
    {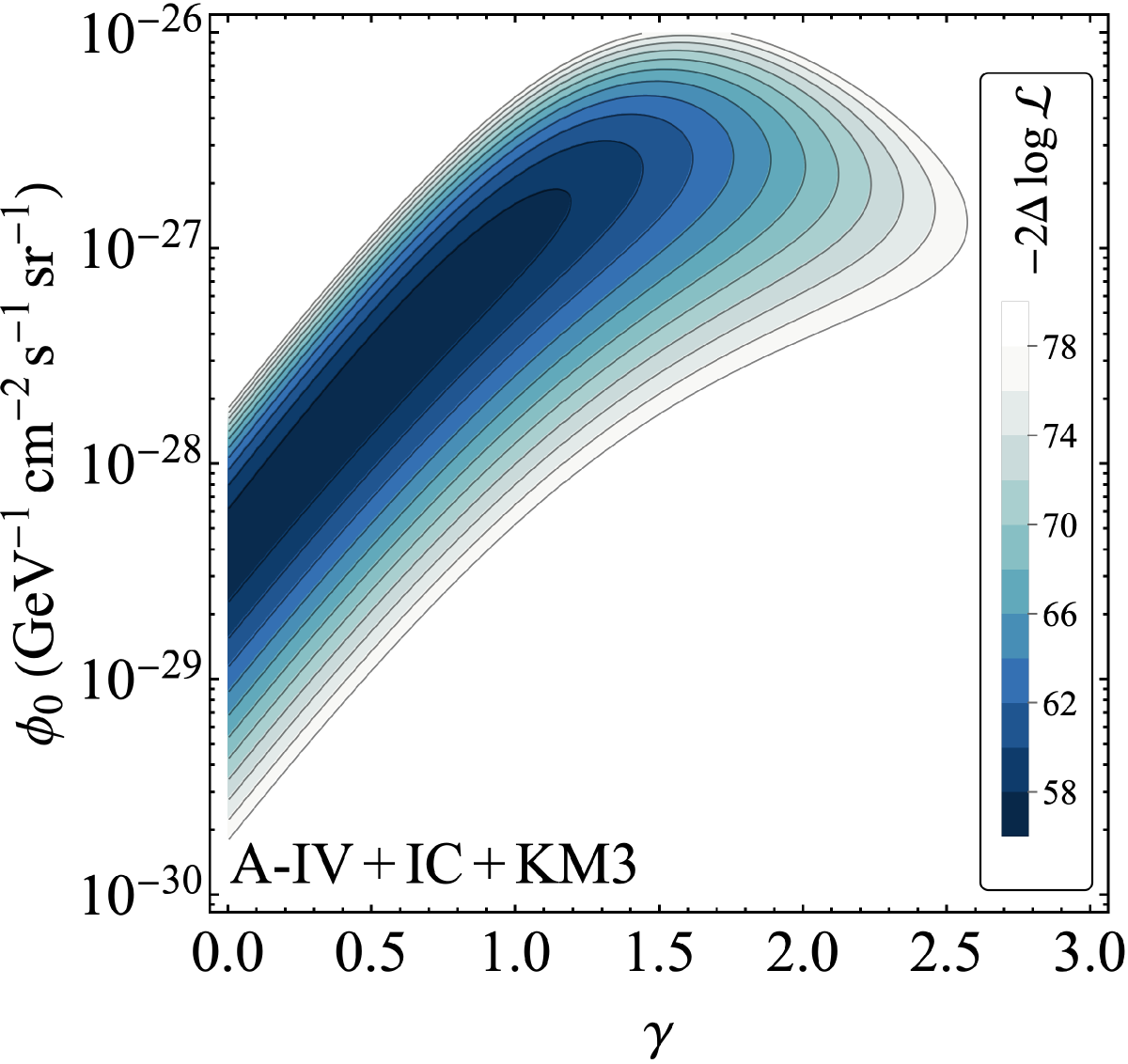}
    \hfill
    \raisebox{0.003\linewidth}{
    \includegraphics[width=0.486\linewidth]
    {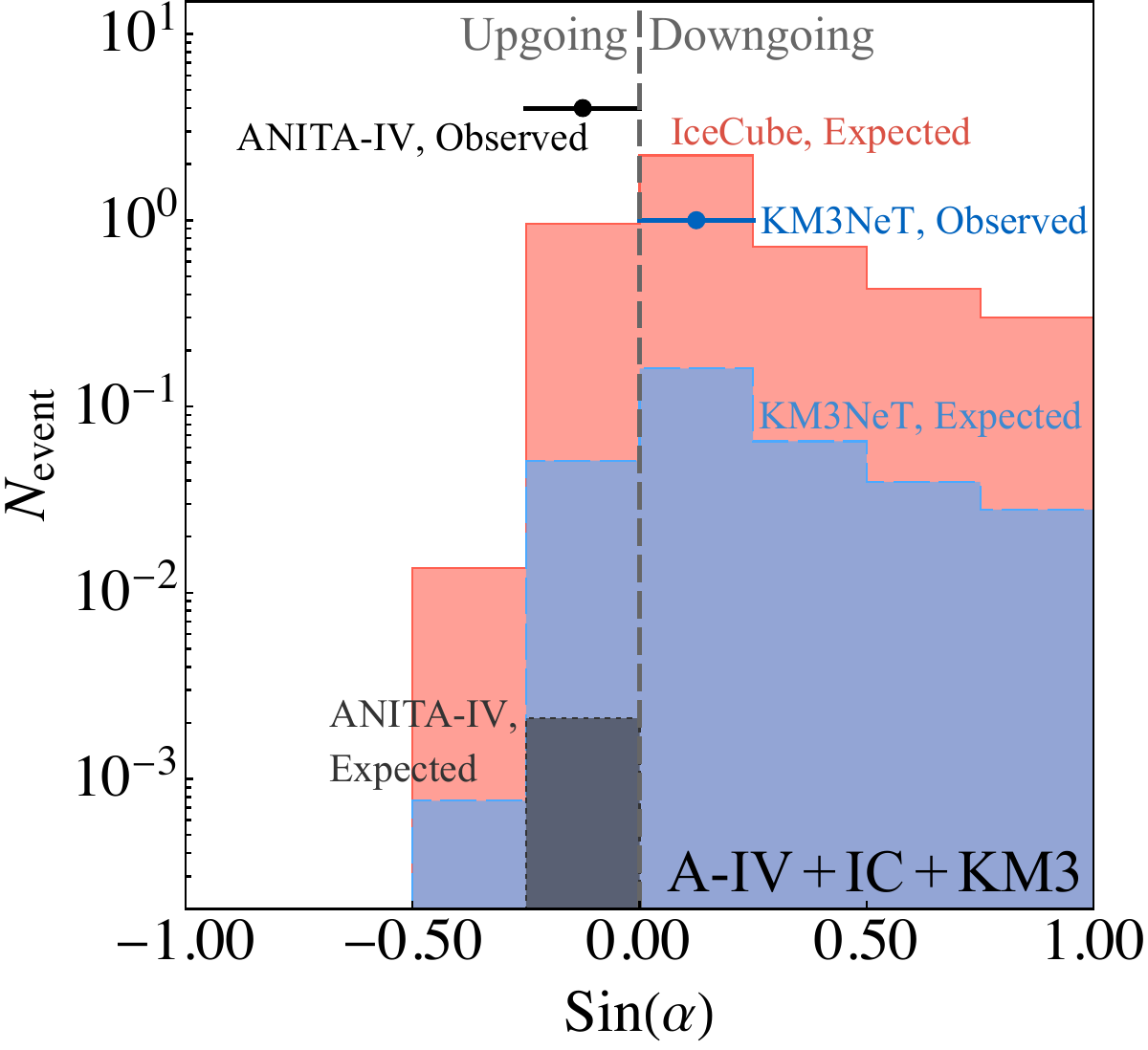}
    }
    \caption{Left panel: Joint Poisson deviance for the diffuse all-sky flux hypothesis as a function of the spectral index $\gamma$ and flux normalization $\phi_0$, comparing the expected event counts against the observed event counts $(n_{\rm A-IV},n_{\rm IC},n_{\rm KM3})=(4,0,1)$. Right panel: Angular distributions of the expected ANITA-IV, IceCube, and KM3NeT event counts at the joint best-fit point for an $E_\nu^{-2}$ neutrino energy spectrum ($\phi_0\approx 2.4\times 10^{-27}~{\rm GeV}^{-1}{\rm cm}^{-2}{\rm s}^{-1}{\rm sr}^{-1}$). The horizontal black and blue markers indicate the number of observed ANITA-IV and KM3NeT events, respectively. The vertical dashed line separates upgoing and downgoing trajectories. The best-fit diffuse flux scenario predicts $\sim5$ IceCube events and a substantially smaller number of expected events for ANITA-IV and KM3NeT (ARCA-21).}
    \label{fig:icecube_diffuse_events}
\end{figure}

The left panel of~\cref{fig:icecube_diffuse_events} shows the resulting Poisson deviance values as a function of the spectral index $\gamma$ and flux normalization $\phi_0$. Although the spectral index and normalization modify the relative detector event rates and, in turn, the likelihood, no region of parameter space reproduces four ANITA-IV events, one KM3NeT event, and the IceCube null result. 
The minimum value of $-2\Delta \log \mathcal{L} \approx 58$ is reached for $\gamma \lesssim 1$. This minimum value is reached for three independent event counts and two fitted parameters; i.e., the nominal goodness-of-fit degrees of freedom (d.o.f.) is $f=1$. For $-2\Delta \log \mathcal{L} \approx 58$, and 1 d.o.f., the $p$-value is $\approx 2.6\times 10^{-14}$, which corresponds to a tension of $7.5\sigma$.\footnote{To express the tension in terms of an equivalent Gaussian significance, we convert the corresponding $p$-value to a $\sigma$ value, assuming Wilks' theorem applies.} On the other hand, for a fixed neutrino energy spectrum that follows $E_\nu^{-2}$, i.e., for a spectral index of $\gamma =2$, we find that $-2\Delta \log \mathcal{L} \approx 68$. For 2 d.o.f, this gives a $p$-value of $1.7\times 10^{-15}$, corresponding to a tension of $7.9\sigma$.

In the right panel of~\cref{fig:icecube_diffuse_events}, the origin of this tension is illustrated.
Here, we show the angular distribution of the expected ANITA-IV, IceCube, and KM3NeT event counts at the joint best-fit flux for the $E_\nu^{-2}$ neutrino energy spectrum. We denote the local elevation angle by $\alpha$, such that $\sin\alpha<0$ and $\sin\alpha>0$ correspond to upgoing and downgoing trajectories, respectively. The ANITA-IV events occupy a narrow region immediately below the horizon, while the diffuse flux predicts a large KM3NeT event count, especially for downgoing events, and an even larger IceCube event yield over a broad angular interval. The absence of such events in IceCube therefore strongly disfavors the diffuse all-sky interpretation.

The expected event counts at ANITA-IV and KM3NeT remain significantly below their observed values, whereas a total of $\sim 5$ events are predicted at IceCube. Therefore, to match the observed values of 4, 1, and 0 events at ANITA-IV, KM3NeT (ARCA-21), and IceCube, we need simultaneous upward fluctuations at ANITA-IV and KM3NeT, and a downward fluctuation at IceCube. This behavior follows primarily from IceCube's much larger exposure (since the full detector has been running for $\sim15$ years) and broad directional acceptance, and therefore cannot be alleviated by simply varying the normalization or spectral index of a common diffuse flux. 

For the corresponding ANITA-IV--IceCube analysis of the diffuse-flux hypothesis, excluding KM3-230213A, see Supplemental~\cref{app:diffuse_anita_icecube}.
For a fixed $E_\nu^{-2}$ spectrum, we find a tension of $7.8\sigma$, with the best-fit ANITA-IV expectation remaining at $\mathcal{O}(10^{-3})$ events.

\section{Transient-source interpretation for the ultra-high energy tension}
\label{sec:transient}

Transient neutrino sources can reduce the tension found for the diffuse-flux scenario by confining the signal to the parts of the sky and short time windows during which ANITA-IV or KM3NeT (ARCA-21) has a favorable exposure.
In this section, we first investigate this possibility by placing individual transients at the observed event directions and times, thereby considering the most favorable event-by-event interpretation.
We then consider a population of rare transient sources distributed across the sky over the $5400$-day IceCube exposure and address two questions: $(i)$ how likely is ANITA-IV to detect four distinct transients, and $(ii)$ for a given transient duration and total source population over 5400 days, what is the likelihood of observing four events at ANITA-IV, one at KM3NeT, and none at IceCube?
This statistical treatment avoids artificially selecting source locations and timings that are particularly favorable to ANITA-IV or KM3NeT, allowing us to quantify how rarely a given source population and temporal configuration would arise.

We model each transient using a simple $E_\nu^{-2}$ energy spectrum,
\begin{equation}
\frac{d\Phi_s}{dE_\nu}(E_\nu,t) = \phi_0 \left( \frac{E_\nu}{E_0} \right)^{-2} W_s(t;\Delta t),
\label{eq:transient_flux}
\end{equation}
where $\phi_0$ is the single neutrino flavor flux normalization, $\Delta t$ is the transient duration, and we choose $E_0=1$~EeV. The function $W_s$ determines the time window, i.e., $W_s(t;\Delta t)=1$ between initial time $t_0$ and $t_0+\Delta t$, and is zero elsewhere. For a source at direction $\Omega_s$, the expected number of events at a detector $\mathcal{D}$ is
\begin{equation}
\mu_{\mathcal{D},s} = \int dt\,dE_\nu\, \frac{d\Phi_s}{dE_\nu}(E_\nu,t) A_{\rm eff}^{\mathcal{D}} (E_\nu,\Omega_s;t) \; ,
\label{eq:transient_expected_events}
\end{equation}
where we set $A_{\rm eff}^{\mathcal{D}}(E_\nu,\Omega_s;t)=0$ whenever $t$ falls outside the detector live-time period, which spans approximately $26.6$~days for ANITA-IV and $287$~days for KM3NeT in its ARCA-21 configuration. We consider an energy range of $(0.1-100)$~EeV.
For ANITA-IV we use the $\nu_\tau$ effective area, whereas for IceCube and KM3NeT we consider the track-like effective area contribution from both the $\nu_\mu$ and $\nu_\tau$ flavors.

\subsection{Fixed transient sources towards the observed directions}
\label{sec:fixed_transients}

\begin{figure}[t!]
    \centering
    \includegraphics[height=0.3\linewidth]
    {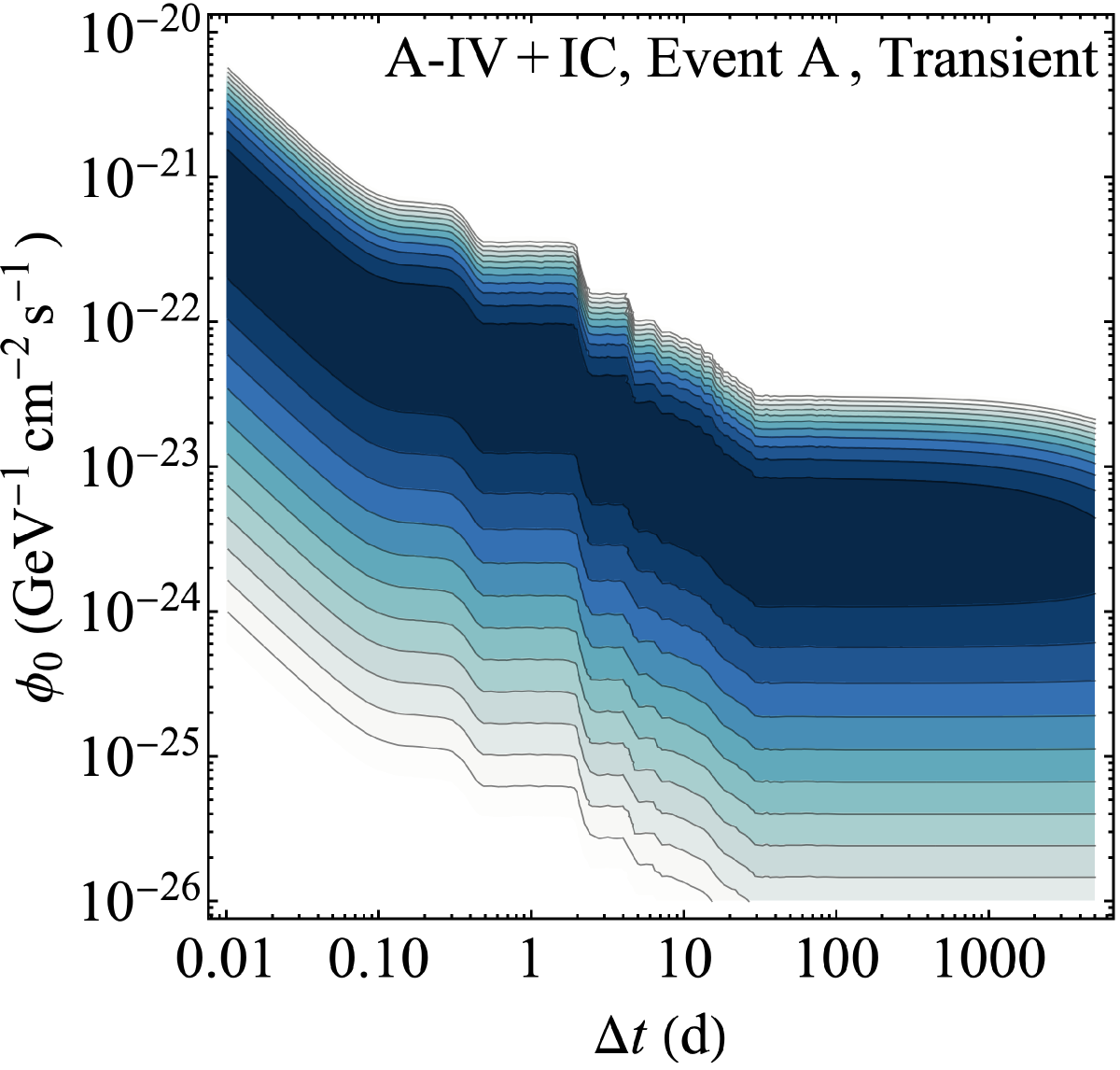}
    \hspace{0.1cm}
    \includegraphics[height=0.3\linewidth]
    {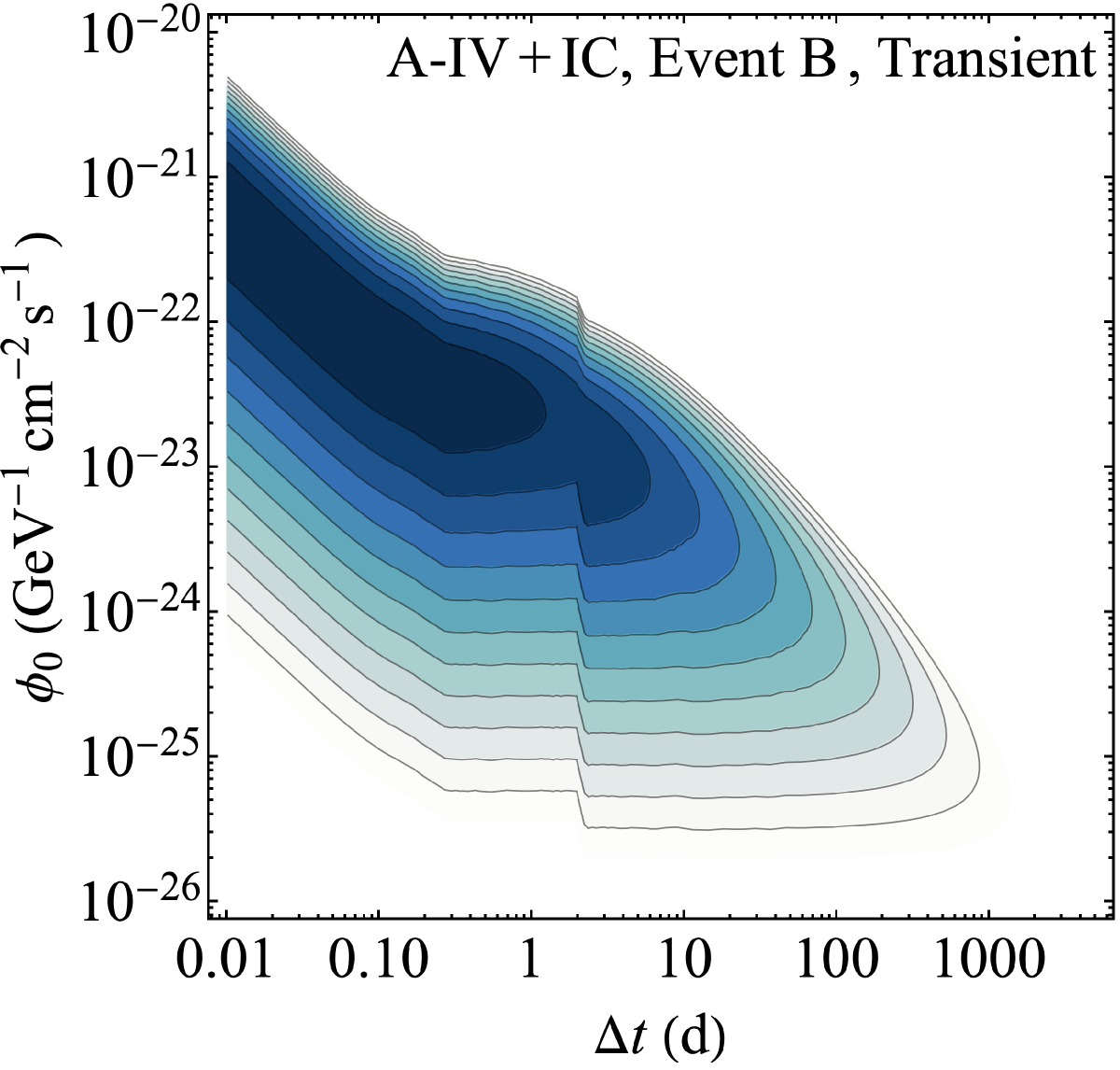}
    \hspace{0.1cm}
    \includegraphics[height=0.3\linewidth]
    {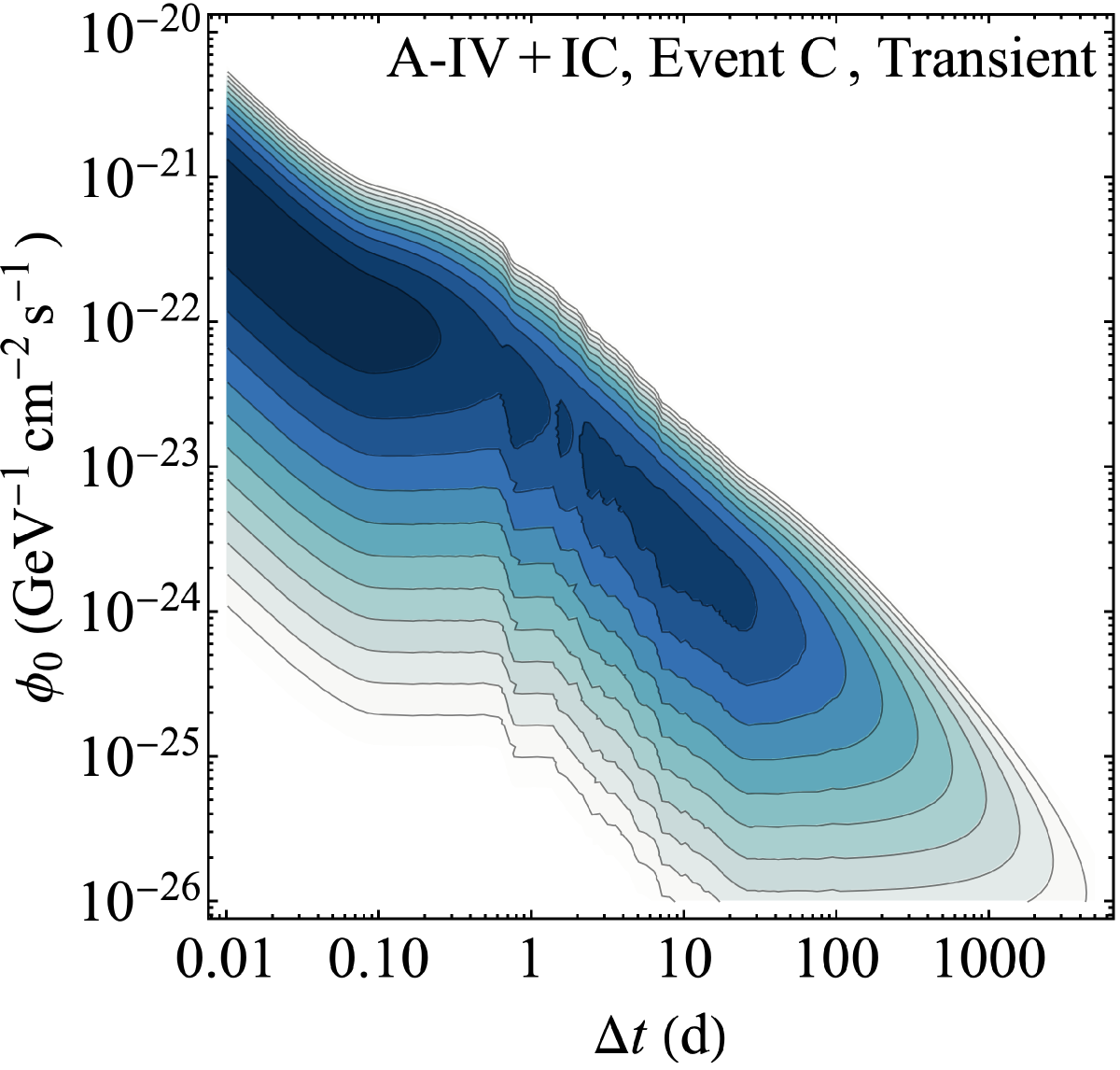}\\
    \vspace{0.2cm}
    \hspace{0.04cm}
    \includegraphics[height=0.3\linewidth]
    {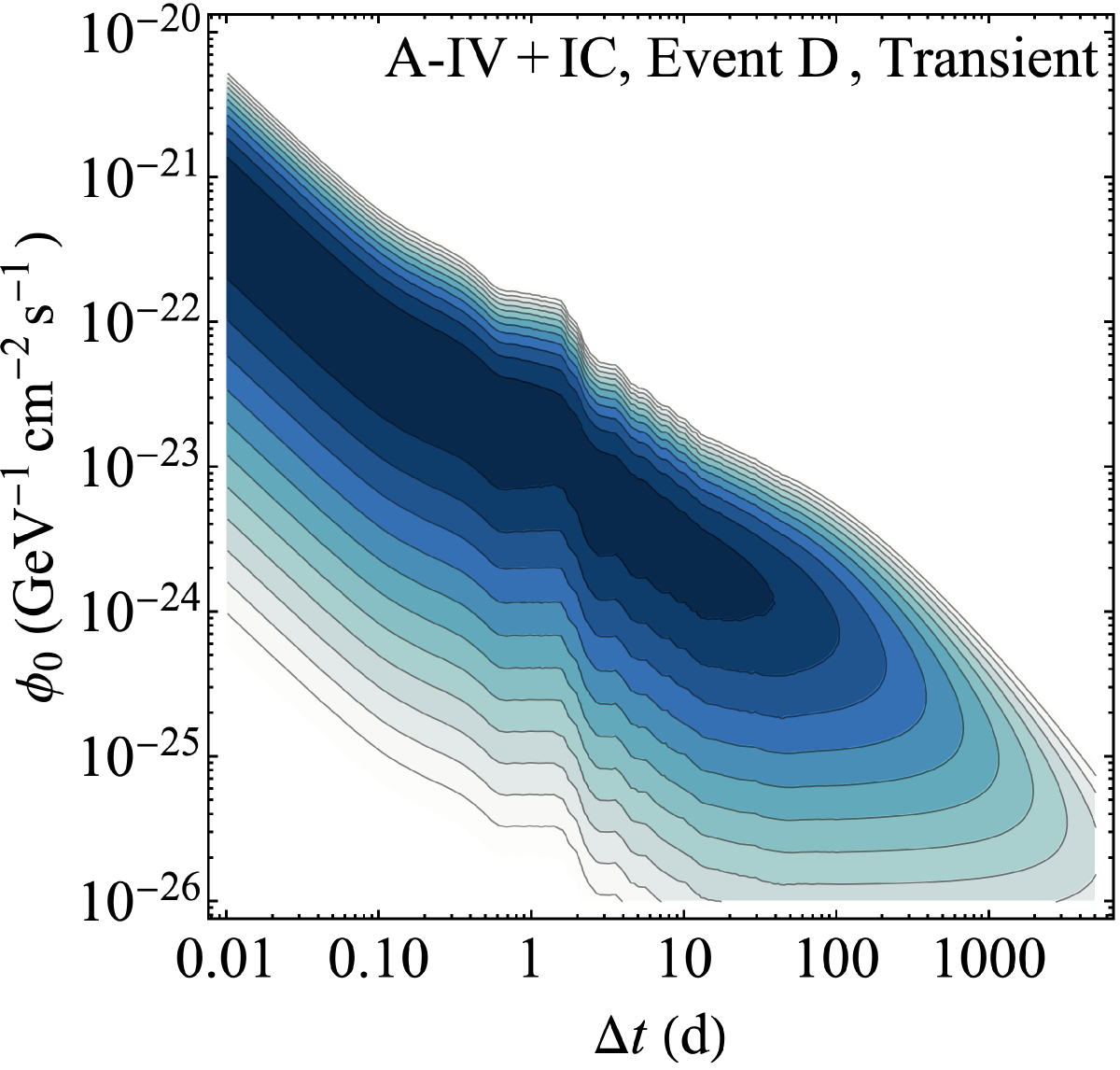}
    \hspace{0.1cm}
    \includegraphics[height=0.3\linewidth]{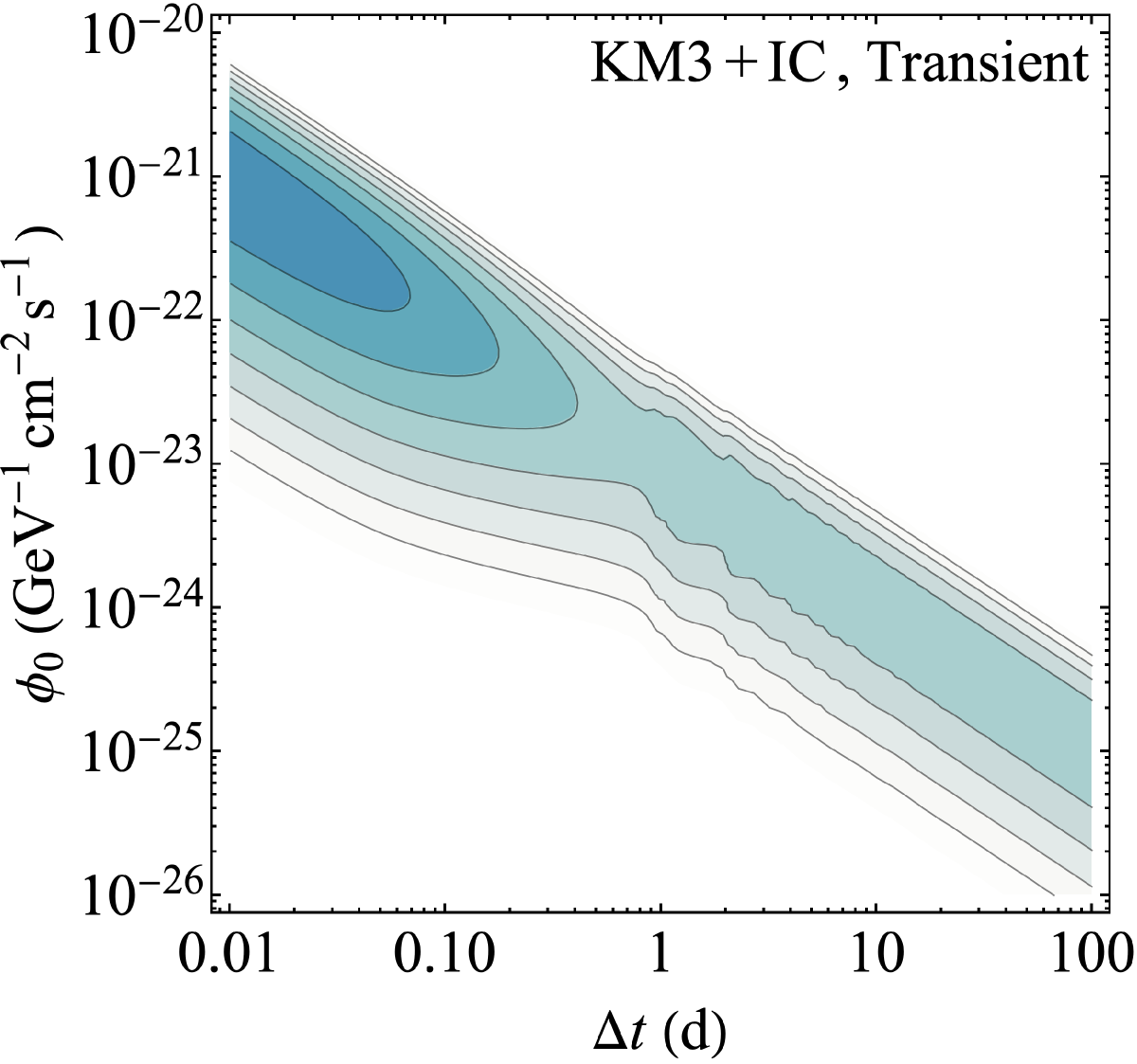}
    \hspace{0.9cm}
    \raisebox{0.035\linewidth}{
    \includegraphics[height=0.26\linewidth]{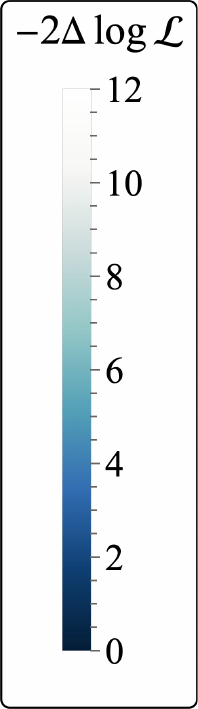}
    }
    \hspace{3.2cm}
    \caption{Poisson deviances for separate transient sources placed at the directions and times of ANITA-IV Events A, B, C, and D (top panels and the bottom left panel) and the KM3NeT event KM3-230213A (bottom middle), as a function of the transient duration $\Delta t$ and flux normalization $\phi_0$, for a fixed $E_\nu^{-2}$ spectrum. The transient duration is centered around the exact time when the events were observed at ANITA-IV and KM3NeT (ARCA-21). Each panel assumes one observed event at ANITA-IV/KM3NeT and zero events at IceCube.}
    \label{fig:log_likelihood_transient_abcd}
\end{figure}

Since the four neutrino-like events observed at ANITA-IV and the one event at KM3NeT originate from 5 distinct points in the sky, we first associate a separate transient with each of these event directions.
For each direction, we scan over flux normalization ($\phi_0$) and the transient duration using the Poisson likelihood, for one observed event at ANITA-IV/KM3NeT and 0 at IceCube per source. This simplifies to
\begin{equation}
\mathcal{L} = \mu_{\mathcal{D}} \exp \left( -\mu_{\mathcal{D}} \right) \exp \left(-\mu_{\rm IC}\right) \; ,
\label{eq:anita_transient_likelihood}
\end{equation}
where $\mathcal{D}$ denotes either the ANITA-IV or the KM3NeT experiment.
The saturated likelihood is calculated for one expected event at ANITA-IV and 0 expected events at IceCube, which gives $\mathcal{L}_{\rm sat} = 1/e$. This allows us to calculate the goodness-of-fit using $-2\Delta \log \mathcal{L} \equiv -2 \log \left( \mathcal{L}/\mathcal{L}_{\rm sat}  \right)$.

The resulting Poisson deviances for all 5 directions of the corresponding events are shown in Fig.~\ref{fig:log_likelihood_transient_abcd}. The results differ because the ANITA-IV, KM3NeT, and IceCube exposures depend strongly on the event direction. We observe that any transient source towards the direction of the ANITA-IV Event A can accommodate a very long transient duration and still be consistent with the null observation at IceCube. This is because IceCube is unfavorably located to observe that point in the sky, with the effective area significantly smaller than that of ANITA-IV (see~\cref{fig:anita_iv_events,fig:anita_eff_area_time} and the discussion in~\cref{sec:directional_exposure}).
Further, we observe that both ANITA-IV Events B and C prefer a transient duration $\lesssim \mathcal{O}(1)$~day, whereas ANITA-IV Event D prefers a transient duration of $\lesssim \mathcal{O}(10)$~days.
This is in agreement with the features observed in~\cref{fig:anita_eff_area_time,fig:anita_iv_events}.
The step-like features in the first four panels, appearing for durations of up to $\mathcal{O}(10)$ days, arise from the successive ANITA-IV optimal observation windows that become relevant as the duration of the transient source increases.

Towards the direction of KM3-230213A (\cref{fig:log_likelihood_transient_abcd}, bottom middle panel), there is a preference for a very short transient, with a duration below $\mathcal{O}(0.1)$~day. This is because the event occurred during a particularly favorable visibility window for KM3NeT, when its effective area was near its maximum (see~\cref{fig:km3net_eff_area_time}). Nevertheless, the best-fit point does not fully remove the tension, and a residual discrepancy between the KM3NeT observation and the IceCube non-observation remains. For a short-duration transient, taking $\Delta t\approx 0.01$ day, we find that the best-fit flux normalization is at $\phi_0 \approx 10^{-21}~{\rm GeV}^{-1}\,{\rm cm}^{-2}\,{\rm s}^{-1}$ (with $-2\Delta \log \mathcal{L} \approx 4.3$); this corresponds to a $p$-value of $\approx3.9\times 10^{-2}$ and a $\sim1.8\sigma$ tension. For a longer-duration transient source, the best-fit flux normalization gives $-2\Delta \log \mathcal{L}\simeq 7.3$, corresponding to a $p$-value of $6.9\times 10^{-3}$, or a tension of approximately $2.5\sigma$, in agreement with~\cite{Li:2025tqf}.

\begin{figure}[t!]
    \centering
    \includegraphics[width=0.5\linewidth]
    {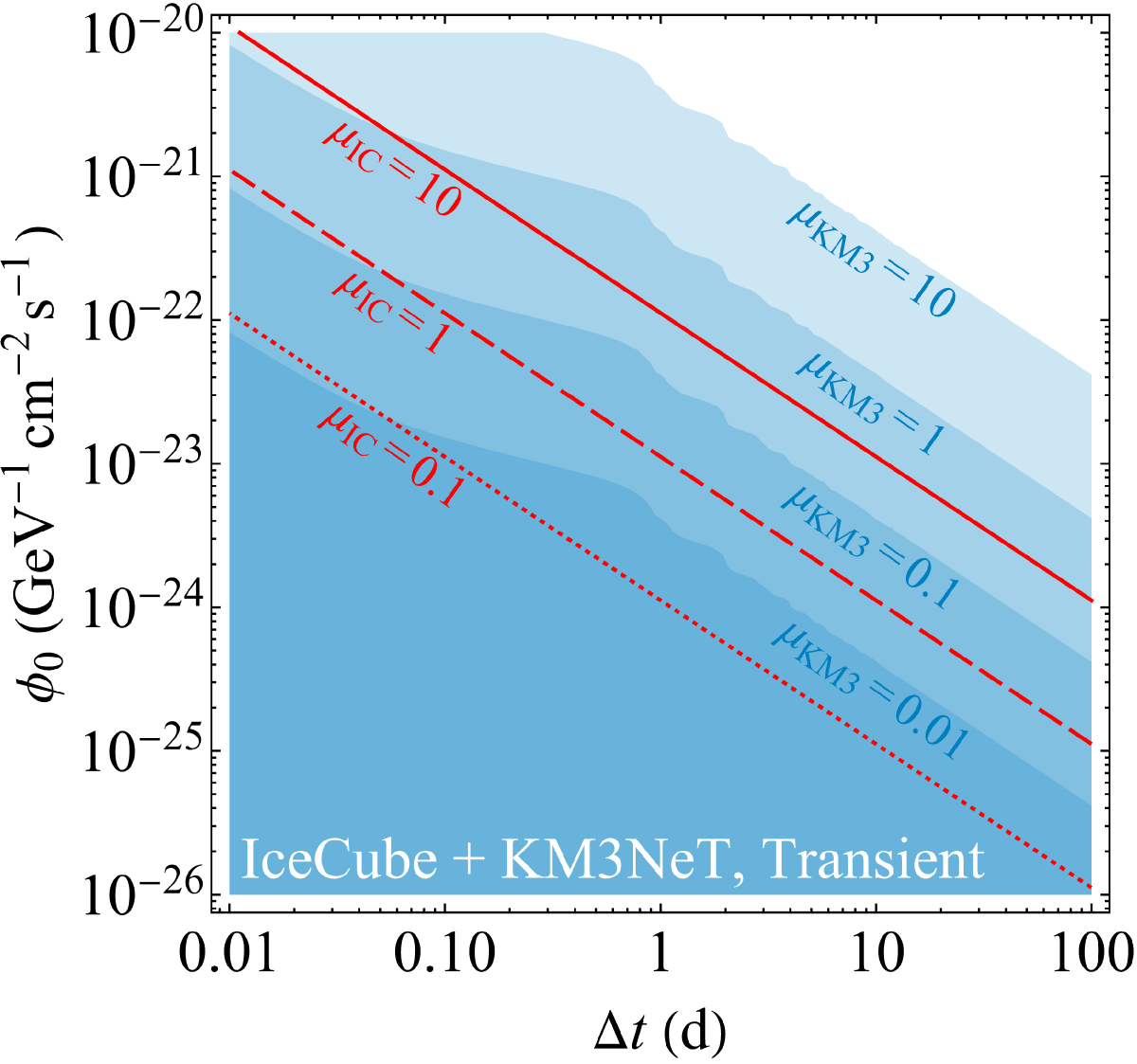}
    \caption{Expected number of KM3NeT and IceCube events as functions of the transient duration $\Delta t$ and flux normalization $\phi_0$. The statistically preferred region corresponds to a short burst of less than an hour in duration, where $\mathcal{O}(1)$ events are expected at IceCube, and $\mathcal{O}(0.1)$ events are expected at KM3NeT, in its ARCA-21 configuration.}
    \label{fig:km3net_transient_likelihood}
\end{figure}

In~\cref{fig:km3net_transient_likelihood}, we show the number of expected events at KM3NeT (ARCA-21) and IceCube, as a function of the transient duration and flux normalization $\phi_0$. The blue shaded contours indicate the expected number of events at KM3NeT, while the red solid, dashed, and dotted curves correspond to expected IceCube event counts of $10$, $1$, and $0.1$, respectively. For a transient duration of $0.01$~day, a flux normalization of $\phi_0 \simeq 10^{-21}~{\rm GeV}^{-1}\,{\rm cm}^{-2}\,{\rm s}^{-1}$ yields expectation values of $\mu_{\rm KM3}\gtrsim 0.1$ and $\mu_{\rm IC}\lesssim 1$, leading to $1.8\sigma$ tension discussed above.

For the combined fixed-transient interpretation with an $E_\nu^{-2}$ spectrum, taking a common flux normalization and transient duration, we present the results in Supplemental~\cref{app:transient_likelihood}. There, we present both the joint ANITA-IV--IceCube analysis with four fixed sources and the ANITA-IV--KM3NeT--IceCube analysis with five fixed sources, in each case placing the sources along the observed event directions.

Overall, we find that short-duration transients can reduce the IceCube exposure by concentrating the neutrino emission within the favorable effective area windows for ANITA-IV or KM3NeT. The resulting improvement, however, depends strongly on the source direction, location, and timing of the transient.

\subsection{All-sky population of rare transients}
\label{sec:all_sky_transients}

In~\Cref{sec:fixed_transients} we have shown that short-duration transients placed at the reconstructed event directions and centered on the corresponding detection times can largely remove the apparent tension with IceCube.
Such an arrangement, however, is highly fine-tuned: it does not account for the probability that the sources must occur at favorable sky locations and within the narrow time windows in which ANITA-IV or KM3NeT has optimal sensitivity.

In order to do a more realistic analysis, we therefore generate populations of $N_{\rm S}$ transient sources. First, we assign each source a random sky location and occurrence time (within the 5400~day IceCube observation period), and estimate the probability that it falls within ANITA-IV's favorable observation window, defined by the elevation range $-10^\circ<\theta<-5^\circ$.
For each transient duration, we first estimate the probability that a single source, with random sky position and occurrence time, falls within ANITA-IV's favorable observation window, using $10^4$ Monte Carlo realizations.
We then use this probability to compute, for each choice of transient duration and source abundance, the probability $P_{\rm A-IV}(N_{\rm S}\geq 4)$ that ANITA-IV has a favorable exposure to observe four or more distinct sources during its $26.57$-day flight.

\begin{figure}[t!]
    \centering
    \includegraphics[width=0.5\linewidth]
    {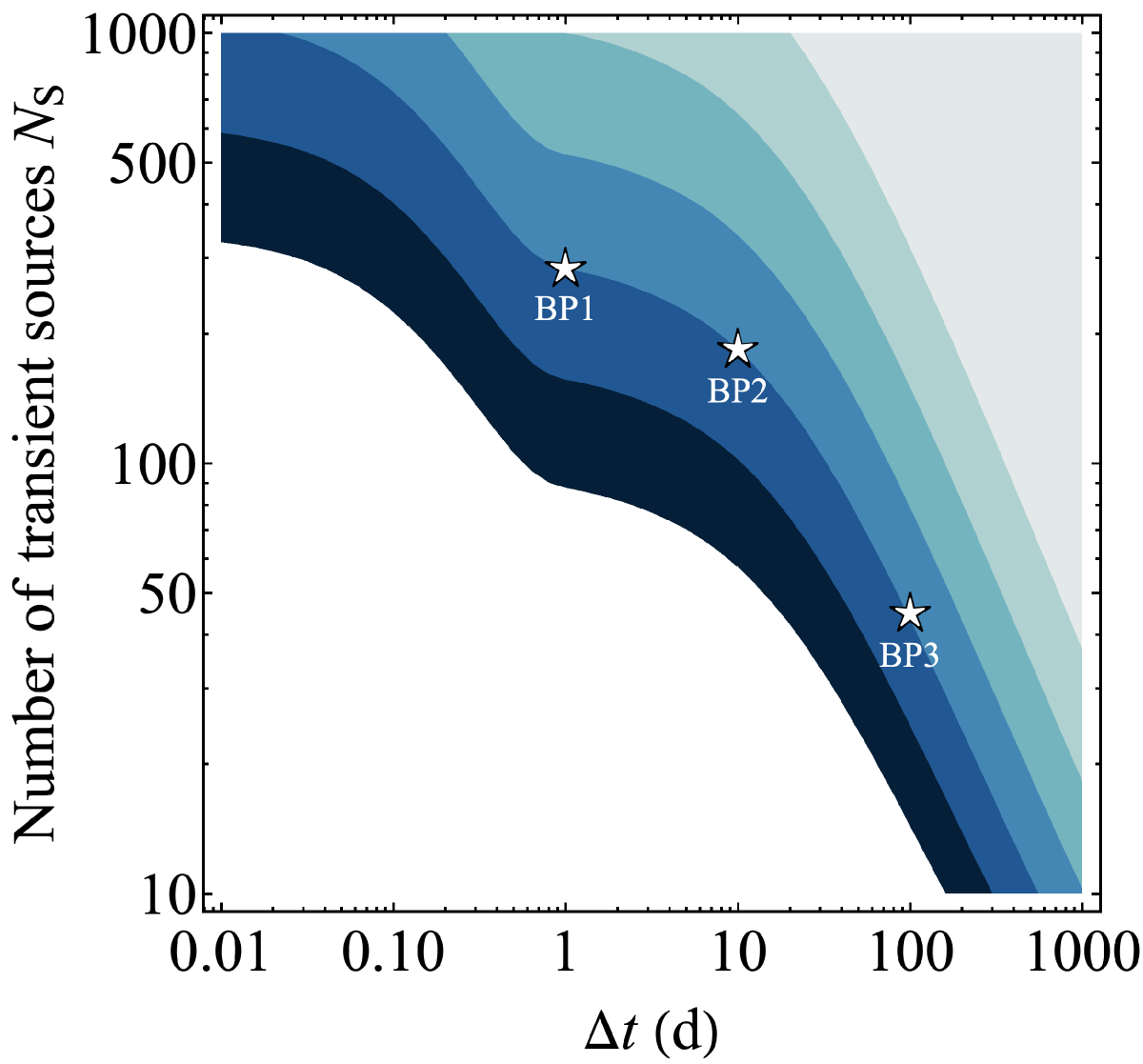}
    \raisebox{0.067\linewidth}{
    \includegraphics[width=0.072\linewidth]
    {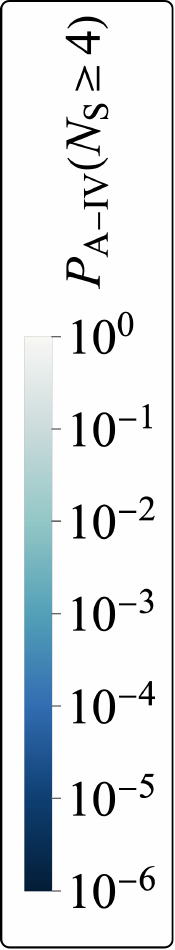}
    }
    \caption{Probability that at least four distinct transient sources overlap the optimal ANITA-IV elevation band, $-10^\circ<\theta<-5^\circ$, during the $\sim26.6$~days of ANITA-IV runtime, as a function of the transient duration and the total number of transient sources over $5400$ days. The three benchmark points, denoted by white stars, satisfy $P_{\rm A-IV}(N_{\rm S}\geq 4) \approx  10^{-4}$.}
    \label{fig:anita_four_transient_probability}
\end{figure}

In~\cref{fig:anita_four_transient_probability}, we show $P_{\rm A-IV}(N_{\rm S}\geq 4)$ as a function of the transient duration ($\Delta t$) and the number of transient sources~($N_{\rm S}$) over the $5400$ day IceCube observation period.
We select three representative benchmark points satisfying $P_{\rm A-IV}(N_{\rm S}\geq 4) \approx 10^{-4}$:
\begin{align}
{\rm BP1}:&\quad (\Delta t,N_{\rm S})=(1~{\rm d},285) \; ,\nonumber \\
{\rm BP2}:&\quad (\Delta t,N_{\rm S})=(10~{\rm d},185)\; ,\nonumber \\
{\rm BP3}:&\quad (\Delta t,N_{\rm S})=(100~{\rm d},45)\; .
\label{eq:transient_benchmark_points}
\end{align}
Longer-lived sources require fewer transients because each source has a higher probability of being active when ANITA-IV observes that region of the sky within its optimal elevation window.

For each of the benchmark points in~\cref{eq:transient_benchmark_points}, we then generate $10^5$ realizations of the corresponding transient population, with source locations sampled uniformly over the full sky and occurrence times sampled uniformly over the 5400-day IceCube observation period. We then select the realizations relevant for the ANITA-IV observations by requiring at least four distinct transient sources to fall within ANITA-IV's optimal elevation window, $-10^\circ<\theta<-5^\circ$. 
For each benchmark point, we find $\mathcal{O}(10)$ MC realizations, out of $10^5$, that contain four or more sources within ANITA-IV's optimal observation window. By construction, this is consistent since the benchmark points were chosen to satisfy $P_{\rm A-IV}(N_{\rm S}\geq 4)\simeq 10^{-4}$.
Once a particular realization is generated, every transient can in principle contribute to any of the detectors according to its time-dependent directional exposure. We further restrict the sample to realizations for which KM3NeT, in its ARCA-21 configuration, has nonzero exposure to at least one of the selected transient sources.
Our procedure ensures that a transient, even if optimally located, will only be visible to ANITA-IV/KM3NeT if it emits neutrinos when the experiment is running.
The total expected count is therefore
\begin{equation}
\mu_{\mathcal{D}}^{(r)} = \sum_{i=1}^{N_{\rm S}} \mu_{\mathcal{D},i}^{(r)},
\label{eq:population_expected_events}
\end{equation}
where $r$ denotes the MC realization, $i$ denotes the source, and $\mathcal{D}$ denotes the detector. The quantity $\mu_{\mathcal{D}, i}^{(r)}$ depends on the source position, occurrence time, and transient duration due to the time- and direction-dependent effective area of each detector.
In our statistical analysis, we profile the likelihood over a common flux normalization $\phi_0$ and find the best-fit point for an $E_\nu^{-2}$ neutrino energy spectrum. This analysis closely follows the diffuse flux one, described in~\cref{sec:diffuse}.

In~\cref{tab:transient_benchmark_all}, we list the most favorable realizations for each of the three benchmark points considered, for the joint transient analysis with four observed events at ANITA-IV, one at KM3NeT, and none at IceCube. Even after restricting the analysis to realizations with four or more ANITA-IV favorable transients, the best-fit expectation value for the event count remains dominated by IceCube.
The smallest tension is obtained for BP2, whose best-performing realization yields the expected event counts of
\begin{equation}
\mu_{\rm A-IV}=0.067 \; , \qquad \mu_{\rm IC}=4.911 \, , \qquad \mu_{\rm KM3}=0.022 \; ,
\end{equation}
at each detector.
The four ANITA-IV events and the one KM3NeT event must therefore be attributed to large upward fluctuations, accompanied by a simultaneous downward fluctuation at IceCube to account for its null observation.
Scanning over the flux normalization sets the total expected number of events to the total observed count. However, the relative detector exposures assign nearly all of this expected event count to IceCube.
Even for the best-fit source configuration chosen amongst $10^5$ MC realizations, for BP2, a value of $-2\Delta \log \mathcal{L}\approx 40.4$ gives a $p$-value of $1.70 \times 10^{-9}$, which corresponds to a tension of $5.9\sigma$.
With a larger number of MC realizations, one may find more favorable spatial and temporal arrangements of rare transient sources, further reducing the inferred tension.
By construction, however, such arrangements correspond to increasingly rare realizations of the transient population.

\begin{table}[t!]
    \centering
    {\renewcommand{\arraystretch}{1.2}
    \begin{tabular}{|c|c|c|c|c|c|}
    \hline
    \; Scenario \; & \; $\phi_0~({\rm GeV}^{-1}~{\rm cm}^{-2}~{\rm s}^{-1})$ \; & \qquad $\mu_{\rm A-IV}$ \quad \; & \qquad $\mu_{\rm IC}$ \quad \; & \qquad $\mu_{\rm KM3}$ \quad \;  & \; $-2\Delta \log \mathcal{L}$  \; \\
    \hline
    \hline
    \multirow{3}{*}{BP1} & $5.48 \times 10^{-25}$ & $0.035$ & $4.945$ & $0.020$ & $45.80$\\
         \cline{2-6}
         & $4.55 \times 10^{-25}$ & $0.032$ & $4.955$ & $0.013$ & $47.37$\\
         \cline{2-6}
         & $5.48 \times 10^{-25}$ & $0.025$ & $4.948$ & $0.027$ & $47.90$\\
    \hline
    \hline
    \multirow{3}{*}{BP2} & $8.39 \times 10^{-26}$ & $0.067$ & $4.911$ & $0.022$ & $40.38$\\
         \cline{2-6}
         & $7.16 \times 10^{-26}$ & $0.054$ & $4.928$ & $0.018$ & $42.42$\\
         \cline{2-6}
         & $8.57 \times 10^{-26}$ & $0.041$ & $4.917$ & $0.042$ & $42.96$\\
    \hline
    \hline
    \multirow{3}{*}{BP3} & $3.02 \times 10^{-26}$ & $0.048$ & $4.914$ & $0.038$ & $41.93$\\
         \cline{2-6}
         & $4.49 \times 10^{-26}$ & $0.046$ & $4.943$ & $0.011$ & $44.78$\\
         \cline{2-6}
         & $4.33 \times 10^{-26}$ & $0.032$ & $4.931$ & $0.038$ & $45.26$\\
         \hline
    \end{tabular}
    }
    \caption{The best-fit flux normalization $(\phi_0)$, and the number of expected event counts at ANITA-IV, IceCube, and KM3NeT for the 3 most favorable realizations, out of $10^5$ MC samples, for the 3 benchmark points considered. Each of the samples is chosen such that ANITA-IV sees 4 different transients within its optimal observation band.}
    \label{tab:transient_benchmark_all}
\end{table}

The transient-source scenario therefore alleviates the diffuse-flux tension only modestly, reducing the significance from $7.9\sigma$ under the diffuse all-sky flux hypothesis to $5.9\sigma$ for the transient benchmark, for an $E_\nu^{-2}$ spectrum.
The corresponding best-fit event yields, for both the diffuse and transient scenarios, are summarized in Fig.~\ref{fig:best_fit_event_all_diffuse_transient} (see~\cref{sec:introduction}). For the diffuse-flux scenario, we show the expected event counts obtained for the best-fit normalization assuming an $E_\nu^{-2}$ spectrum. The joint ANITA-IV, KM3NeT, and IceCube fit predicts approximately five events at IceCube, but only $\mathcal{O}(10^{-3})$ events at ANITA-IV and $\mathcal{O}(10^{-1})$ events at KM3NeT in its ARCA-21 configuration. These expectations are in severe tension with the observed counts of four, one, and zero events, respectively.

In~\cref{fig:best_fit_event_all_diffuse_transient}, for the transient-source scenario, we show the results for the benchmark choice $\Delta t=10$~days and $N_{\rm S}=185$ sources (BP2) distributed over the 5400-day IceCube observation period.
This scenario utilizes the time- and direction-dependent detector exposures, increasing the expected ANITA-IV yield to $\mathcal{O}(10^{-1})$ compared to the diffuse flux hypothesis. However, it also simultaneously reduces the expected KM3NeT yield by approximately one order of magnitude. In conclusion, in both the diffuse and transient scenarios, the observed event counts require large simultaneous upward fluctuations at ANITA-IV and KM3NeT, together with a downward fluctuation at IceCube.

The presented all-sky transient-population treatment can be repeated for ANITA-IV and IceCube alone, excluding KM3-230213A. We show these results in Supplemental~\cref{app:transient_likelihood}. There, assuming an $E_\nu^{-2}$ spectrum, we report the best-fit flux normalization, expected event counts, and Poisson deviance for the three benchmark points (see~\cref{tab:transient_benchmark_anita_ic}). We also compare the best-fit event yields in the combined ANITA-IV and IceCube analysis in~\cref{fig:transient_best_fit_counts}. For the full three-detector, five-event analysis, the best-fit transient benchmark gives a tension of $5.9\sigma$; for the two-detector (ANITA-IV and IceCube) analysis, we find a very similar tension of $5.8\sigma$.

In summary, short, favorably timed transients can reduce the tension associated with the observed events. However, once the probability of realizing the required source directions and timings, together with the all-sky transient population, is taken into account, the IceCube non-observation remains in strong disagreement with the event counts reported by ANITA-IV and KM3NeT.

\section{Summary and conclusions}
\label{sec:conclusion}

In this work, we investigated whether the four near-horizon ANITA-IV events and KM3-230213A can be consistently interpreted as ultra-high energy (UHE) neutrino events, given the absence of corresponding UHE neutrino observations at IceCube.
Treating the four anomalous ANITA-IV events as Earth-skimming $\nu_\tau$-induced showers and KM3-230213A as a track-like $\nu_\mu$ or $\nu_\tau$ event, we calculated the energy-, direction-, and time-dependent effective areas of ANITA-IV, IceCube, and KM3NeT (ARCA-21) using a semi-analytic framework, presented in Supplemental~\cref{sec:effective_area_calc}.

The directional exposures make the origin of this tension particularly clear. ANITA-IV can achieve the largest instantaneous effective area, but only within a narrow near-horizon band, and its sensitivity to a given sky direction changes substantially as the balloon moves. IceCube has a somewhat smaller peak effective area, but remains sensitive over a much broader angular range and has operated for roughly $200$ times longer than ANITA-IV. The KM3NeT detector, in the partially completed ARCA-21 configuration, meanwhile, has a smaller effective area and run time when compared to IceCube.

For a common diffuse power-law flux, no choice of normalization and spectral index accounts for the observed event counts of
\begin{equation}
n_{\rm A-IV}=4 \; , \qquad n_{\rm IC}=0 \, , \qquad n_{\rm KM3}=1 \; .
\end{equation}
The joint best-fit predicts approximately five events at IceCube, while the expected ANITA-IV and KM3NeT counts remain far below their observed values. This corresponds to a tension of approximately $7.5\sigma$. Meanwhile, for an $E_\nu^{-2}$ spectrum, we find a tension of approximately $7.9\sigma$, with the best-fit expected event rates of
\begin{equation}
\mu_{\rm A-IV}=0.002 \; , \qquad \mu_{\rm IC}=4.654 \, , \qquad \mu_{\rm KM3}=0.344 \; ,
\end{equation}
at the three detectors.
The diffuse scenario therefore requires simultaneous and large upward fluctuations at ANITA-IV and KM3NeT and a downward fluctuation at IceCube.

Short transients placed directly at the observed event locations in the sky and centered around the detection times can substantially reduce the tension for individual events. The required durations are, however, strongly direction dependent, and the KM3NeT event favors a particularly short burst. Indeed, KM3-230213A occurred during a particularly favorable KM3NeT visibility window, when the detector’s effective area for the observed event direction was close to its maximum.
Such a configuration of source location, timing, and duration is highly fine-tuned, as it does not account for the probability of obtaining the required source locations and timings.

We, therefore, also considered populations of rare transients distributed over the full sky and throughout the IceCube observation period of $5400$ days. We selected benchmark populations for which the probability of obtaining at least four separate favorable transients at ANITA-IV is approximately $10^{-4}$.
Even after selecting realizations containing four or more separate transient sources observed favorably at ANITA-IV (out of $10^5$ MC realizations), the best-fit expectation remains dominated by IceCube.
For the least anomalous benchmark realization, corresponding to a choice of 10-day transient duration and 185 sources over a period of $5400$ days, we obtain
\begin{equation}
\mu_{\rm A-IV}=0.067 \; , \qquad \mu_{\rm IC}=4.911 \, , \qquad \mu_{\rm KM3}=0.022 \; .
\end{equation}
This leads to a tension of approximately $5.9\sigma$.
Therefore, accounting for transient timing and directional exposure weakens the diffuse-flux tension only modestly.

In summary, for an $E_\nu^{-2}$ spectrum, the combined ANITA-IV--KM3NeT--IceCube analysis yields a tension in the range of $(5.9-7.9)\sigma$, depending on whether the sources are modeled as rare transients or as a diffuse all-sky flux. This tension is driven primarily by the mismatch between the four ANITA-IV events and the IceCube null result. In fact, even when KM3-230213A is excluded, the ANITA-IV--IceCube-only analysis gives a comparable tension of $(5.8-7.8)\sigma$.
In contrast, KM3-230213A alone corresponds to a mild tension of $1.8\sigma$ for a short-duration transient and $2.5\sigma$ for a longer-duration transient. 
We only find strong evidence for tension in the joint fit when including the ANITA-IV observations.

We conclude that neither an all-sky diffuse flux nor an all-sky population of rare identical transients provides a satisfactory explanation of the combined event counts.
Resolving the anomaly may require a more structured source population, unaccounted-for detector or event-classification effects, or new physics effects that modify the relative propagation or detection probabilities among the three experiments. 
Future UHE neutrino observations will be essential for clarifying the origin of the experimental observations.


\acknowledgments
We would like to thank K.S. Babu, Toni Bertólez-Martínez, P. S. Bhupal Dev, Bhaskar Dutta, Raj Gandhi, André de Gouvêa, Nick Kamp, Kevin Kelly, and Yago Porto for the stimulating discussions.
We would especially like to thank Stephanie Wissel and Andrew Zeolla for their help in the ANITA-IV effective area calculations.
The work of VB is supported by the United States Department of Energy Grant No. DE-SC0016013.
CAA is supported by the Faculty of Arts and Sciences of Harvard University, the National Science Foundation, the Research Corporation for Science Advancement, and the David \& Lucile Packard Foundation.
DSC and VB would also like to thank CETUP* and the Institute for Underground Science at SURF for their warm hospitality during the 2025 and 2026 Summer Workshop, where part of this work was completed.

\bibliographystyle{JHEP}
\bibliography{refs}

\clearpage

\setcounter{figure}{0}
\setcounter{table}{0}
\setcounter{equation}{0}
\setcounter{page}{1}
\renewcommand{\thepage}{Supplemental Methods and Tables -- S\arabic{page}}

\renewcommand{\figurename}{SUPPL. FIG.}
\renewcommand{\tablename}{SUPPL. TABLE}
\renewcommand{\theequation}{A\arabic{equation}}

\newcounter{SIfig}
\setcounter{SIfig}{1}
\renewcommand{\theSIfig}{SUPPL. FIG. \arabic{SIfig}}

\begin{center}
    \textbf{\large Supplemental Material}
\end{center}


\setcounter{equation}{0}
\setcounter{figure}{0}
\setcounter{table}{0}
\setcounter{section}{0}

\makeatletter
\renewcommand{\theequation}{S\arabic{equation}}
\renewcommand{\thefigure}{S\arabic{figure}}
\renewcommand{\thetable}{S\arabic{table}}
\renewcommand{\thesection}{S\arabic{section}}

\makeatother


\section{Semi-analytic effective area estimates}
\label{sec:effective_area_calc}

In what follows, we develop semi-analytic calculations of the energy- and direction-dependent effective areas for ANITA-IV, IceCube, and KM3NeT. The effective area calculations allow us to compare the relative exposures of the three experiments.

\subsection{ANITA-IV}

The ANITA effective area for an incident tau neutrino with energy $E_\nu$ and payload elevation angle $\theta$ can be expressed as~\cite{Romero-Wolf:2018zxt,ANITA:2021xxh}
\begin{equation}
\graymathbox{
A_{\rm eff}(E_\nu,\theta) = A_{\rm geo} \otimes P_{\rm exit} \otimes P_{\rm dec} \otimes P_{\rm trig} \; .
}
\label{eq:anita_effective_area}
\end{equation}
Here, $A_{\rm geo} \otimes P_{\rm exit}$ is the geometrical projected area weighted by the probability that the incident neutrino produces a tau lepton that emerges from the Earth, $P_{\rm dec}$ is the probability that the emerging tau decays in the atmosphere before reaching the payload, and $P_{\rm trig}$ denotes the detector trigger efficiency.

The quantity  $A_{\rm geo} \otimes P_{\rm exit}$ is evaluated jointly because the emergence angle, Earth-crossing distance, and projected surface area vary across the region viewed by the payload, particularly close to the horizon, where the effective area of the ANITA experiment is largest.
The projected area $A_{\rm geo}$ is specified by the Earth radius $R_{\oplus}=6371~{\rm km}$, the Antarctic ice thickness, and the ANITA payload altitude; we take the latter two quantities to be $d_{\rm ice}=2.5~{\rm km}$, and $h_{\rm A}=38~{\rm km}$. We further define 
\begin{equation}
R_{\rm s}=R_{\oplus}+d_{\rm ice}\; , \qquad R_{\rm A}=R_{\oplus}+h_{\rm A}\; .
\end{equation}
For a trajectory observed at payload elevation angle $\theta<0$, the distance from the emergence point to the payload is
\begin{equation}
L_{\rm atm}(\theta) = -R_{\rm A}\sin\theta - \sqrt{ R_{\rm s}^{2} - R_{\rm A}^{2}\cos^{2}\theta} \; .
\label{eq:anita_atmospheric_distance}
\end{equation}
The corresponding tau-emergence angle above the local horizontal is
\begin{equation}
\alpha(\theta) = \cos^{-1} \left( \frac{R_{\rm A}}{R_{\rm s}} \cos\theta \right) \; .
\label{eq:anita_emergence_angle}
\end{equation}
The total chord length through the Earth is therefore
\begin{equation}
L_{\rm chord}(\alpha) = 2R_{\rm s}\sin\alpha \; .
\label{eq:anita_chord_length}
\end{equation}
A neutrino arriving at an angle $\theta$ propagates along this chord. We take the Preliminary Reference Earth Model (PREM)~\cite{Dziewonski:1981xy}, supplemented by a $2.5$~km thick layer of ice on the surface, to model the density of matter along the path. At such UHEs, where the effects of oscillation can be ignored, the survival probability of a neutrino up to a position $s$ along the trajectory is given by
\begin{equation}
P_{\nu}^{\rm surv}(E_{\nu},s) = \exp \left[ -N_{\rm A}\sigma_{\nu N}^{\rm CC}(E_{\nu}) \int_{0}^{s} \rho(s')\,ds' \right]\; .
\label{eq:anita_neutrino_survival}
\end{equation}
Here, $N_{\rm A}$ is Avogadro's number, $\rho(s')$ is the local matter density at a position $s'$, and $\sigma_{\nu N}^{\rm CC}$ is the charged-current deep-inelastic neutrino--nucleon cross section~\cite{Connolly:2011vc,Valera:2022ylt,Xie:2023suk,Weigel:2024gzh}.

Let $x$ denote the distance of the neutrino interaction point from the Earth-exit point, measured backward along the trajectory. The differential probability that the neutrino survives to this point and undergoes a charged-current interaction is
\begin{equation}
\frac{dP_{\rm int}}{dx} = N_{\rm A}\rho(x)\sigma_{\nu N}^{\rm CC}(E_{\nu}) P_{\nu}^{\rm surv}(E_{\nu},x) \; .
\label{eq:anita_interaction_probability}
\end{equation}
Here,
\begin{equation}
P_{\nu}^{\rm surv}(E_{\nu},x) = \exp \left[ -N_{\rm A}\sigma_{\nu N}^{\rm CC}(E_{\nu}) \int_{x}^{L_{\rm chord}} \rho(\ell)\,d\ell \right] \; ,
\label{eq:anita_survival_to_interaction}
\end{equation}
where $x=0$ corresponds to the exit point and $x=L_{\rm chord}$ to the entry point.

To simplify our calculations we assume that the $\tau$-lepton takes $80\%$ of the incident $\nu_\tau$ energy
\begin{equation}
E_{\tau,0}=0.8E_{\nu}\; .
\label{eq:anita_initial_tau_energy}
\end{equation}
The $\tau$-lepton's energy loss in matter is described using the continuous-loss approximation~\cite{Dutta:2000hh,Bigas:2008ff,Dutta:2005yt}
\begin{equation}
-\frac{dE_{\tau}}{dX} = a_{\tau}+b_{\tau}E_{\tau} \; ,
\label{eq:anita_tau_energy_loss}
\end{equation}
where $X$ is the column depth and
\begin{equation}
a_{\tau} = 3.5\times10^{-3} ~{\rm GeV\,cm^{2}\,g^{-1}}, \qquad b_{\tau} = 1.0\times10^{-6} ~{\rm cm^{2}\,g^{-1}} \;.
\label{eq:anita_tau_loss_parameters}
\end{equation}
For a constant-density layer of density $\rho$, the tau energy after propagating a distance $L$ is
\begin{equation}
E_{\tau}(L) = \left( E_{\tau,0}+\frac{a_{\tau}}{b_{\tau}} \right) \exp \left( -b_{\tau}\rho L \right) - \frac{a_{\tau}}{b_{\tau}} \;.
\label{eq:anita_tau_energy_solution}
\end{equation}
The maximum distance over which a tau lepton can propagate while remaining above the threshold detection energy $E_{\tau}^{\rm th}$, for a constant-density medium, is
\begin{equation}
L_{\rm attn} = \frac{1} {b_{\tau}\rho} \log \left[ \frac{a_{\tau}+b_{\tau}E_{\tau,0}}{a_{\tau}+b_{\tau}E_{\tau}^{\rm th}} \right]\; ,
\label{eq:anita_tau_loss_length}
\end{equation}
where we take $E_{\tau}^{\rm th}=0.1~{\rm EeV}$. While an exact estimate of the threshold energy for the ANITA-IV experiment is not readily available in the literature, this is consistent with~\cite{Wissel:2019ot,ANITA:2021xxh,ANITA:2020gmv}.
Furthermore, we also treat the ice and rock segments separately to account for the difference in densities between the two media~\cite {2025CoPhC.31609799S}. The $\tau$-lepton decay length can be approximated to be~\cite{ParticleDataGroup:2024cfk}
\begin{equation}
L_{\tau}^{\rm dec}(E_{\tau}) \simeq 49~{\rm km} \left( \frac{E_{\tau}}{{\rm EeV}} \right).
\label{eq:anita_tau_decay_length}
\end{equation}
The maximum interaction depth capable of producing an emerging $\tau$-lepton is taken to be
\begin{equation}
L_{\rm max} = \min \left[ L_{\rm attn}, \, L_{\tau}^{\rm dec}(E_{\tau,0}), \, L_{\rm chord} \right] \; .
\label{eq:anita_maximum_interaction_depth}
\end{equation}
The $\tau$-lepton exit probability is therefore
\begin{equation}
\graymathbox{
P_{\rm exit}(E_{\nu},\alpha) = \int_{0}^{L_{\rm max}} dx\, N_{\rm A}\rho(x)\sigma_{\nu N}^{\rm CC}(E_{\nu}) P_{\nu}^{\rm surv}(E_{\nu},x) \; .
}
\label{eq:anita_exit_probability}
\end{equation}

The projected geometric area is calculated by integrating over the portion of the Antarctic surface visible ($\mathcal{A}_{\rm vis}$) from the payload. For each surface element $dA$, the contribution is weighted by the projected-area factor, the radio-viewing condition, and the local $\tau$-lepton exit probability, which yields
\begin{equation}
\graymathbox{
A_{\rm geo} \otimes P_{\rm exit} = \int_{\mathcal{A}_{\rm vis}} dA\, \max \left[ \hat{\mathbf{n}}\cdot\hat{\mathbf{u}}, \, 0 \right] \Theta \left( \hat{\mathbf{v}}\cdot\hat{\mathbf{u}} - \cos\theta_{\rm view} \right) P_{\rm exit}(E_{\nu},\alpha) \; .
}
\label{eq:anita_geometrical_acceptance}
\end{equation}
Here, $\hat{\mathbf{n}}$ is the normal to the surface, $\hat{\mathbf{u}}$ is the direction of the emerging shower, and $\hat{\mathbf{v}}$ points from the surface element toward the payload. We fix the radio-viewing half-angle to $\theta_{\rm view}=1.5^{\circ}$.
Near the horizon, Eq.~\eqref{eq:anita_geometrical_acceptance} is evaluated using an MC integration over \(2\times10^{5}\) points uniformly distributed on the visible spherical cap. For trajectories further below the payload horizon, the viewed surface region is sufficiently small such that a naive projected-disk approximation can be used, leading to
\begin{equation}
\graymathbox{
A_{\rm geo}(\theta) \simeq \pi \left( L_{\rm atm}(\theta) \tan\theta_{\rm view} \right)^{2} \; .
}
\label{eq:anita_projected_area}
\end{equation}

For an interaction occurring at depth $x$, the exiting $\tau$-lepton energy can be approximated by
\begin{equation}
E_{\tau}^{\rm exit}(x) \simeq 0.8E_{\nu} \exp \left[ -b_{\tau} X(x) \right] \; ,
\label{eq:anita_tau_exit_energy}
\end{equation}
where
\begin{equation}
X(x) = \int_{0}^{x} \rho(\ell)\,d\ell \; ,
\label{eq:anita_column_depth}
\end{equation}
is the column depth between the interaction point and the surface. This expression corresponds to the radiative loss dominated limit of Eq.~\eqref{eq:anita_tau_energy_loss}, appropriate at UHEs.

The probability that the emerging $\tau$-lepton decays before reaching the payload, and therefore leads to an observable EAS in the detector, is
\begin{equation}
\graymathbox{
P_{\rm dec}(E_{\tau}^{\rm exit},\theta) = 1- \exp \left[ -\frac{ L_{\rm atm}(\theta) }{ L_{\tau}^{\rm dec}(E_{\tau}^{\rm exit}) } \right] \;.
}
\label{eq:anita_atmospheric_decay_probability}
\end{equation}
For the decay probability, we use the interaction-weighted average, ensuring that $P_{\rm dec}$ is a conditional probability and that the interaction probability is not counted twice.

\begin{figure}
    \centering
    \includegraphics[width=0.4\linewidth]{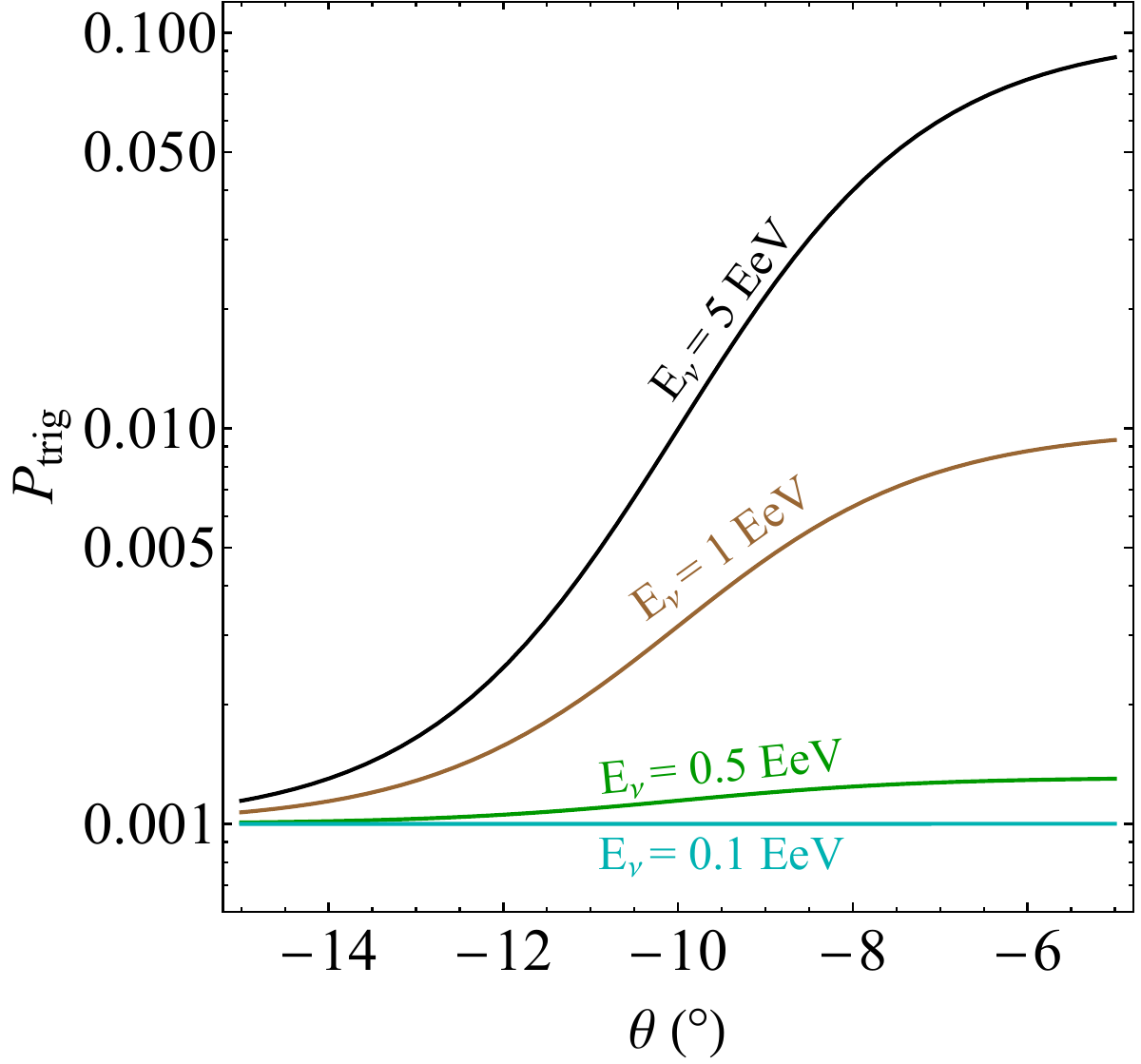}
    \caption{Energy and angular dependence of the ANITA-IV trigger efficiency, $P_{\rm trig}$, used in our work.}
    \label{fig:ptrig}
\end{figure}

We parametrize the trigger efficiency as
\begin{equation}
\graymathbox{
P_{\rm trig}(E_\nu,\theta) = 10^{-3} \, 10^{\, S_{\theta}(\theta) \, S_{E}(E_\nu)} \; ,
}
\label{eq:anita_trigger_probability}
\end{equation}
where the angular behavior is regulated by
\begin{equation}
    S_{\theta} (\theta)  = 1+\tanh \left[0.35 \left( \frac{\theta+10^\circ}{1^{\circ}} \right) \right] \; ,
\label{eq:anita_trigger_angle}
\end{equation}
and the energy dependence is
\begin{equation}
S_E(E_{\nu}) = \frac{1}{2} \left( 1+ \tanh \left[ \frac{ \ln(E_{\nu}/E_{0}) }{ \Delta_{E}} \right] \right) \;,
\label{eq:anita_trigger_energy}
\end{equation}
with
\begin{equation}
E_{0}=1~{\rm EeV} \;, \qquad \Delta_E=0.5 \; .
\end{equation}
At UHEs ($E_\nu > 1$~EeV), $S_E(E_{\nu})\rightarrow1$. In this regime, the trigger efficiency approaches $\sim 10^{-3}$ deep below the horizon and $\sim10^{-1}$ near the horizon. This reproduces the expected qualitative behavior: near-horizon events have the largest acceptance, while steeply emerging events are strongly suppressed.
The energy-dependent factor $S_E(E_\nu)$ encodes the reduced trigger efficiency at $\lesssim 1$~EeV shower energies. At $E_\nu=1~{\rm EeV}$, the near-horizon efficiency is reduced from $\sim10^{-1}$ to $\sim10^{-2}$, i.e. by roughly one order of magnitude.

We show the energy and angular dependence of the trigger efficiency, $P_{\rm trig}$, defined in~\cref{eq:anita_trigger_probability}, in~\cref{fig:ptrig}. The trigger efficiency at $\mathcal{O}(10)$~EeV closely follows the inferred behavior from Fig.~10 in~\cite{ANITA:2021xxh}. In contrast, for $\mathcal{O}(1)$~EeV energies, it follows the expected drop from a $P_{\rm trig}\approx 10^{-2}$ near the horizon, to $P_{\rm trig}\approx 10^{-3}$ for more steeply incoming signals.\footnote{We thank Stephanie Wissel and Andrew Zeolla for helpful correspondence regarding ANITA trigger efficiencies at $\sim$EeV energies.}
At lower energies, our trigger-efficiency model approaches the conservative floor of $10^{-3}$.

Our calculation includes the dominant geometrical, neutrino-attenuation, charged-current interaction, $\tau$-lepton propagation, atmospheric-decay, and trigger effects.
Note, however, that we do not include neutral-current energy degradation, tau-neutrino regeneration, stochastic tau-energy losses, local surface features, and a detailed simulation of the radio signal and detector response. Nevertheless, our semi-analytic calculation reproduces the effective-area curves reported in the literature within a factor of $\sim 2$ near the horizon, where the effective area is largest and most relevant for our analysis. At energies above a few EeV, the elevation-angle dependence is reproduced particularly well, enabling a denser scan of the effective area with respect to the neutrino energy and elevation angles. This further allows the calculation to be extended to steeper trajectories.

\subsection{IceCube and KM3NeT}
\label{sec:icecube_km3net_effective_area}

For IceCube and KM3NeT, the dominant track-like events are produced by charged-current interactions of $\nu_\mu$ and $\nu_\tau$. A $\nu_\mu$ charged-current interaction produces a muon that may enter or cross the instrumented volume, while a $\nu_\tau$ interaction can produce a sufficiently energetic tau lepton whose path through the detector is also track-like at such UHEs. For lepton flavor $\ell=\{\mu,\tau\}$, we calculate the corresponding effective area as~\cite{Li:2025tqf,Gaisser:2016uoy}
\begin{equation}
\graymathbox{
A_{\rm eff}^{({\rm track},\,\ell)}(E_\nu,\theta) = A_{\rm geo} N_{\rm A} \sigma_{\nu N}^{\rm CC}(E_\nu) \int_{0}^{L_{\ell}(E_\nu,\theta)} \rho(\theta,x) P_{\nu}^{\rm surv}(E_\nu,\theta,x)\,dx \; .
}
\label{eq:track_effective_area}
\end{equation}
Here, $x$ is the distance from the detector measured backward along the incoming neutrino trajectory, $\rho(\theta,x)$ is the density at the corresponding interaction point, and $L_{\ell}$ is the maximum distance from which the produced charged lepton can reach the detector above the threshold energy.

We approximate the instrumented volume of the IceCube detector as a sphere of volume $V=1~{\rm km}^{3}$. For the KM3NeT detector, we perform the calculations with the full volume of $V=1~{\rm km}^{3}$, and then normalize to match the existing all-sky effective area for the ARCA-21 configuration.\footnote{The spherical approximation is not exact, since the detector geometry introduces additional angular dependence in the effective area. Such corrections are particularly relevant for ARCA-21 (KM3NeT), whose sparse 21-string configuration is strongly anisotropic and far from spherical. By comparison, IceCube has a much denser, more compact, and approximately azimuthally symmetric instrumented volume, making the spherical approximation more accurate.
}
For the ARCA-21 configuration of the KM3NeT detector, we find that a normalization factor of $1/16$ allows us to match the existing all-sky effective area estimates within $\sim(20-30)\%$ at UHEs~\cite{Muller:2023xO,KM3NeT:2025npi}. Such a factor arises from a combination of a smaller detector size and the reduced efficiency caused by the smaller size and the different (elongated) shape.
The corresponding radius and projected geometrical area are
\begin{equation}
R = \left( \frac{3V}{4\pi} \right)^{1/3}, \qquad A_{\rm geo} = \pi R^{2} \; .
\label{eq:detector_geometrical_area}
\end{equation}
The spherical approximation makes the projected area independent of the arrival direction. The angular dependence of the effective area, instead, arises from the neutrino attenuation and the density profile along the trajectory.

Here, $\theta$ denotes the elevation angle relative to the local horizon, with $\theta<0$ corresponding to an upgoing neutrino. For a detector at radial position
\begin{equation}
r = R_{\oplus}+h+z \; ,
\end{equation}
where $h$ is the thickness of the surface ice or water layer and $z<0$ is the detector depth below the surface, the distance from the detector to the entry point of the neutrino trajectory is
\begin{equation}
L_{\rm los}(\theta) = \sqrt{ \left(R_{\oplus}+h\right)^{2} - r^{2}\cos^{2}\theta } - r\sin\theta \; .
\label{eq:detector_line_of_sight}
\end{equation}
For IceCube, we take $h_{\rm IC}=2.835~{\rm km}$ and $z_{\rm IC}=-1.95~{\rm km}$, while for KM3NeT we use $h_{\rm KM3}=3.45~{\rm km}$ and $z_{\rm KM3}=-3.10~{\rm km}$. The Earth density is described by the PREM profile, supplemented by an outer ice layer for IceCube and a seawater layer for KM3NeT.

The survival probability of a neutrino up to an interaction point located at distance $x$ from the detector is
\begin{equation}
\graymathbox{
P_{\nu}^{\rm surv}(E_\nu,\theta,x) = \exp \left[ -N_{\rm A}\sigma_{\nu N}^{\rm CC}(E_\nu) \int_{x}^{L_{\rm los}(\theta)} \rho(\theta,s)\,ds \right] \; .
}
\label{eq:detector_neutrino_survival}
\end{equation}
The integrand in Eq.~\eqref{eq:track_effective_area} therefore gives the probability that the neutrino survives to the interaction point and undergoes a charged-current interaction within the region from which the resulting lepton can reach the detector.

For both muons and tau leptons, we approximate the initial charged lepton energy as
\begin{equation}
E_{\ell,0}=0.8E_\nu \; ,
\label{eq:detector_initial_lepton_energy}
\end{equation}
and describe propagation using the continuous-energy-loss relation
\begin{equation}
-\frac{dE_\ell}{dX} = a_\ell+b_\ell E_\ell \; .
\label{eq:detector_lepton_energy_loss}
\end{equation}
For a constant-density layer, the distance over which the lepton energy decreases from $E_{\ell,0}$ to the threshold energy $E_\ell^{\rm th}$ is
\begin{equation}
L_{\ell}^{\rm attn} = \frac{1}{b_\ell\rho} \log \left[ \frac{ a_\ell+b_\ell E_{\ell,0} }{ a_\ell+b_\ell E_\ell^{\rm th} }
\right] \; .
\label{eq:detector_lepton_range}
\end{equation}
For trajectories where the incoming neutrino crosses both rock and ice/seawater, we evaluate the energy loss sequentially in the two media. We take $E_\mu^{\rm th}=1~{\rm TeV}$ and $E_\tau^{\rm th}=5~{\rm PeV}$. The threshold energy for the $\tau$ lepton is taken to be much higher to ensure that the tau decay length is at least comparable to the size of the detectors, i.e., the $\tau$-lepton can produce a track-like signature before decaying.

The maximum contributing distance for a muon track is
\begin{equation}
L_{\mu}(E_\nu,\theta) = \min \left[ L_{\mu}^{\rm attn} (E_\nu,\theta), L_{\rm los}(\theta) \right] \; .
\label{eq:detector_muon_range}
\end{equation}
For a $\tau$-lepton track, its finite lifetime must also be included (see~\cref{eq:anita_tau_decay_length}), which leads to
\begin{equation}
L_{\tau}(E_\nu,\theta) = \min \left[ L_{\tau}^{\rm attn}(E_\nu,\theta), L_{\tau}^{\rm dec}(0.8E_\nu), L_{\rm los}(\theta) \right] \; .
\label{eq:detector_tau_range}
\end{equation}

The expected number of track-like events, in the presence of both $\nu_\mu$ and $\nu_\tau$ flux, is
\begin{equation}
N_{\rm track} = \int dt\, dE_\nu \, d\Omega \Big( F_{\nu_\mu}(E_\nu,\Omega) A_{{\rm eff}}^{({\rm track},\,\mu)}(E_\nu,\theta) + F_{\nu_\tau}(E_\nu,\Omega) A_{{\rm eff}}^{({\rm track},\,\tau)} (E_\nu,\theta) \Big) \; .
\label{eq:detector_track_event_number}
\end{equation}
For equal $\nu_\mu$ and $\nu_\tau$ fluxes at Earth~\cite{IceCube:2020fpi,IceCube:2024nhk}, it is convenient to define the total track-like effective area as
\begin{equation}
\graymathbox{
A_{\rm eff}^{({\rm track})} = A_{{\rm eff}}^{({\rm track},\,\mu)} + A_{{\rm eff}}^{({\rm track},\,\tau)} \; .
}
\label{eq:detector_total_track_area}
\end{equation}

Cascade events arise from charged-current interactions of electron neutrinos and neutral-current interactions of all three flavors. Since the interaction vertex must lie inside the instrumented volume, the effective area does not contain the extended charged-lepton range appearing in Eq.~\eqref{eq:track_effective_area}. The flavor-dependent cascade effective areas are approximated by~\cite{Kistler:2013my}
\begin{align}
A_{{\rm eff}}^{({\rm cascade},e)} & = V \rho^{\rm det} N_{\rm A} P_{\nu}^{\rm surv}(E_\nu,\theta,0) \left( \sigma_{\nu N}^{\rm CC}(E_\nu) + \sigma_{\nu N}^{\rm NC}(E_\nu) \right), \nonumber\\
A_{{\rm eff}}^{({\rm cascade},\mu)} = A_{{\rm eff}}^{({\rm cascade},\tau)} & =
V \rho^{\rm det} N_{\rm A} P_{\nu}^{\rm surv}(E_\nu,\theta,0) \, \sigma_{\nu N}^{\rm NC}(E_\nu) \; .
\label{eq:detector_cascade_flavors}
\end{align}
Thus, for an equal flux in all three flavors,
\begin{equation}
A_{\rm eff}^{({\rm cascade})} = V \rho^{\rm det} N_{\rm A} P_{\nu}^{\rm surv}(E_\nu,\theta,0) \left( \sigma_{\nu N}^{\rm CC}(E_\nu) + 3\sigma_{\nu N}^{\rm NC}(E_\nu) \right) \; .
\label{eq:detector_total_cascade_area}
\end{equation}
The neutral current (NC) cross-section at UHE can be approximated as~\cite{Gandhi:1998ri,2011JHEP...08..042C}
\begin{equation}
\sigma_{\nu N}^{\rm NC}(E_\nu) \simeq 0.42\,\sigma_{\nu N}^{\rm CC}(E_\nu) \; .
\end{equation}
As expected, we find that the dominant contribution to the total effective area, at UHEs, comes from the $\mu$ and $\tau$ track-like events.

In this work, the effective areas are evaluated for the energy range $0.1~{\rm EeV}\leq E_\nu\leq100~{\rm EeV}$. At such high energies, we can safely assume the detection efficiency to be 1.
Furthermore, to simplify our calculations, we also neglect stochastic charged-lepton energy losses, neutral-current energy degradation during propagation, neutrino regeneration, and the detailed trigger and reconstruction efficiencies of the two detectors.
Further, we neglect local topographic effects near the KM3NeT and IceCube detectors. For KM3NeT, such effects can enhance the signal from the direction of KM3-230213A by a factor of $\sim 3$ in certain new-physics scenarios~\cite{Arguelles:2025ewg}, especially in the presence of an intermediate long-lived particle that can propagate from the nearby underwater mountain.
For the Standard Model interpretation considered in this work, this effect is expected to be relevant only if charged leptons produced near the underwater cliff, located roughly $34$~km from KM3NeT, can survive propagation over this distance and reach the detector. We conservatively ignore this contribution.
Our approach reproduces the all-sky-averaged and angular-bin-averaged effective areas reported for IceCube and KM3NeT to within $\sim(20-30)\%$~\cite{2014ApJ...796..109A,2024EPJC...84..885K,KM3NeT:2025npi,IceCubeCollaborationSS:2025jbi}.

\subsection{Comparing the effective areas of all 3 detectors}

\begin{figure}[t!]
    \centering
    \includegraphics[height=0.3\linewidth]{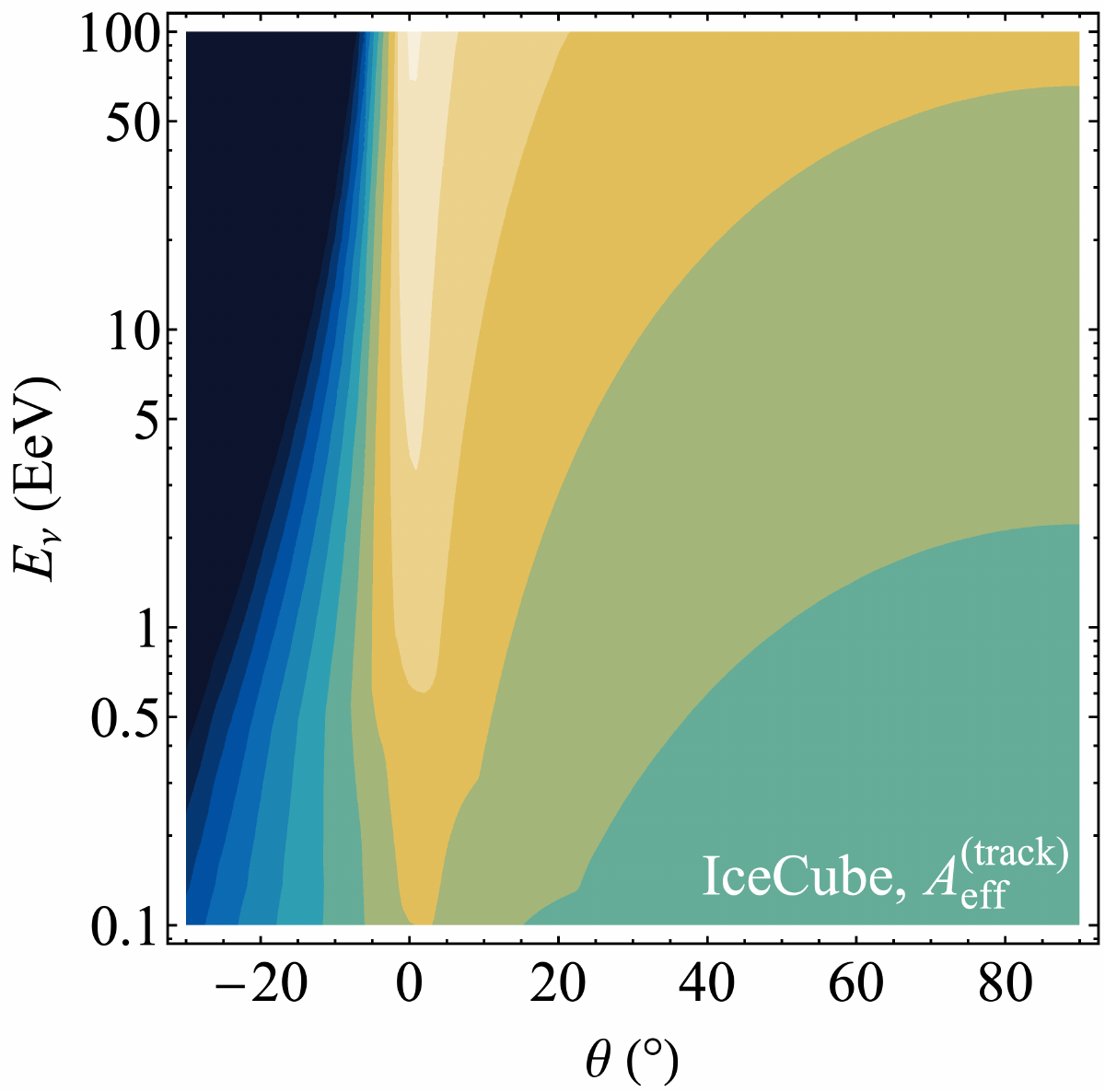}
    \hfill
    \includegraphics[height=0.3\linewidth]{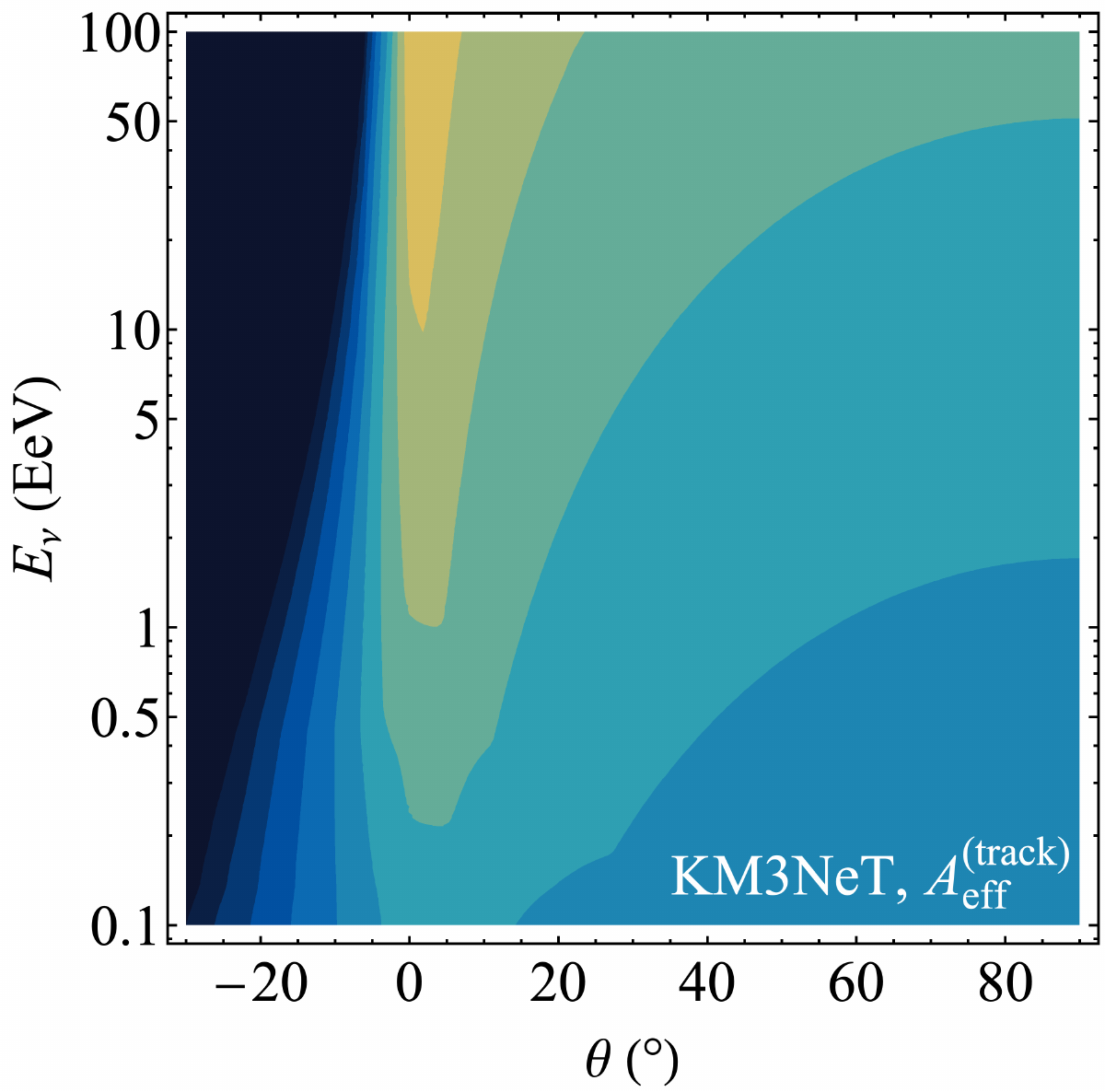}
    \hfill
    \includegraphics[height=0.3\linewidth]{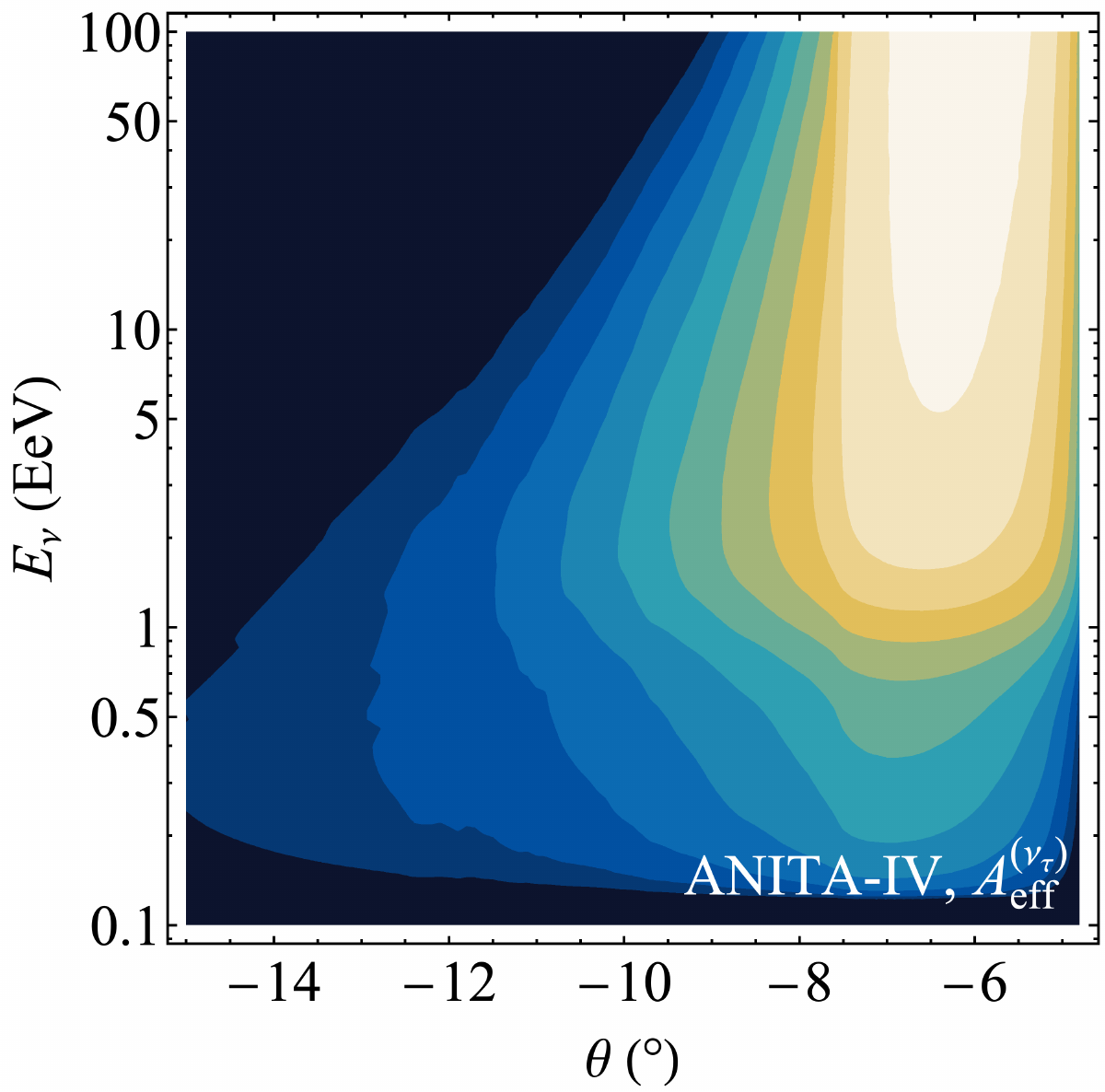}
    \hfill
    \raisebox{0.038\linewidth}{
    \includegraphics[height=0.26\linewidth]{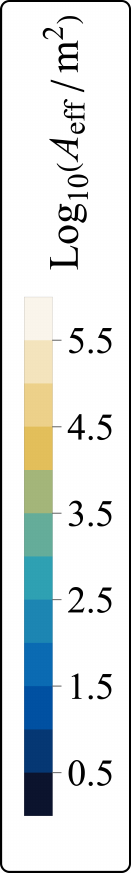}
    }
    \caption{Effective area for track-like $\nu_\mu$ and $\nu_\tau$ events at IceCube (left) and KM3NeT in its ARCA-21 configuration (middle), and $\nu_\tau$ effective area for ANITA-IV (right), as a function of elevation angle $\theta$ and energy of the incoming neutrino $E_\nu$.}
    \label{fig:eff_area_icecube}
\end{figure}

Figure~\ref{fig:eff_area_icecube} shows the effective areas of the three detectors, as a function of the neutrino energy and elevation angle. We show the $\nu_\tau$ effective area for ANITA-IV. For IceCube and KM3NeT, we include track-like contributions from both $\nu_\mu$ and $\nu_\tau$ flavors. ANITA-IV reaches the largest peak effective area, concentrated within a narrow range of near-horizon directions where the Earth-skimming geometry is optimal. The IceCube effective area is almost comparable over its most favorable angular region and remains sizeable over a much broader range of elevation angles.

IceCube has accumulated $\sim 15$ years of data, compared with only about $27$ days for the ANITA-IV flight, corresponding to a difference in run time of $\sim 200$. Therefore, even though the maximum instantaneous effective area of ANITA-IV is larger, the time-integrated IceCube exposure is generally expected to dominate. The ARCA-21 configuration of KM3NeT has a significantly smaller effective area than IceCube, primarily because only a fraction of the final detector volume was installed during the observation of  KM3-230213A.

\section{Effective areas in equatorial coordinates}
\label{app:effective_areas_sky}

\begin{figure}[t!]
    \centering
    \begin{minipage}[c]{0.045\linewidth}
        \centering
        \includegraphics[width=\linewidth]
        {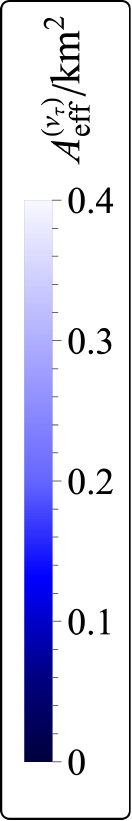}
    \end{minipage}
    \hfill
    \begin{minipage}[c]{0.29\linewidth}
        \centering
        \includegraphics[width=\linewidth]
        {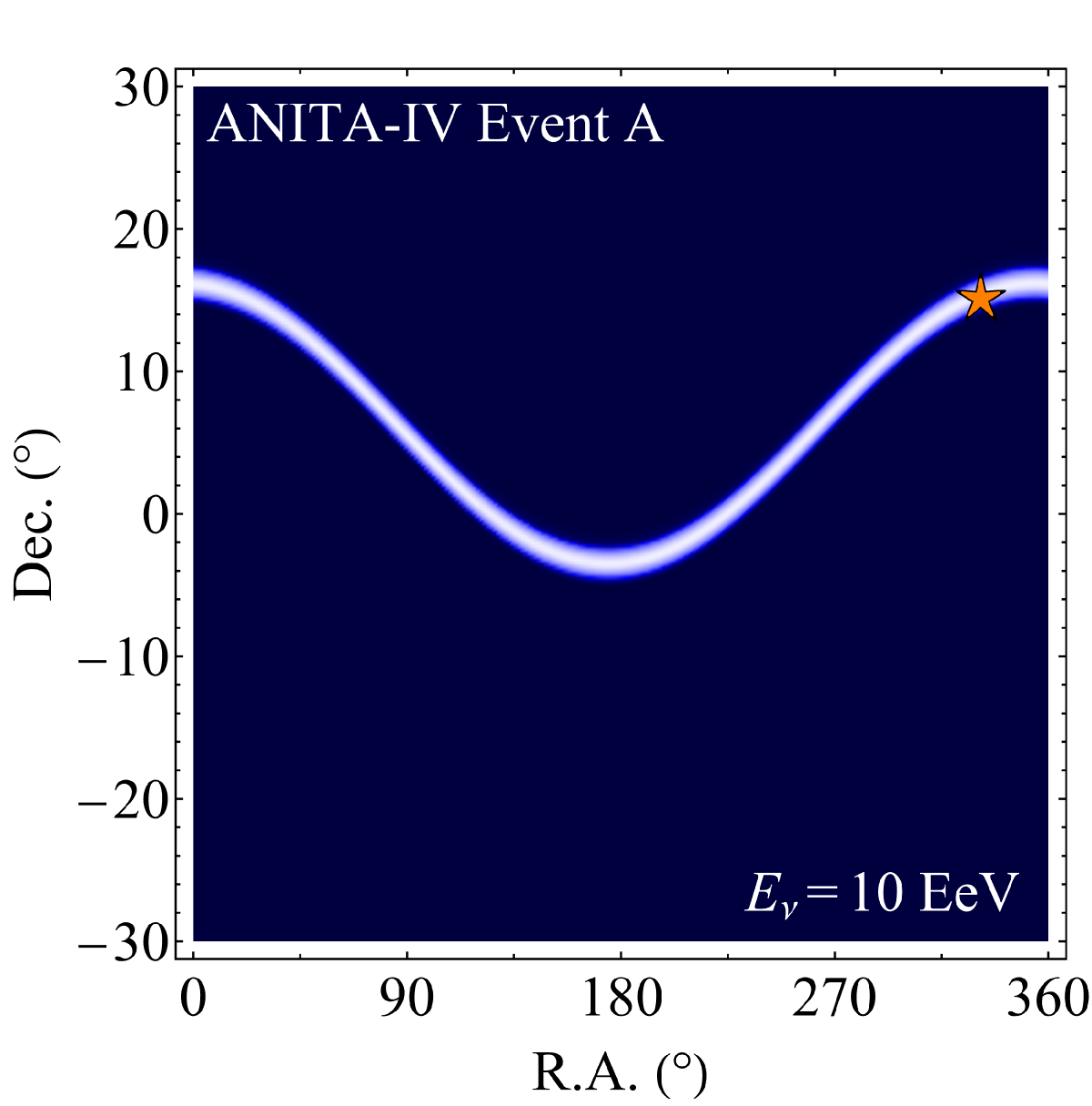}

        \vspace{0.1cm}

        \includegraphics[width=\linewidth]
        {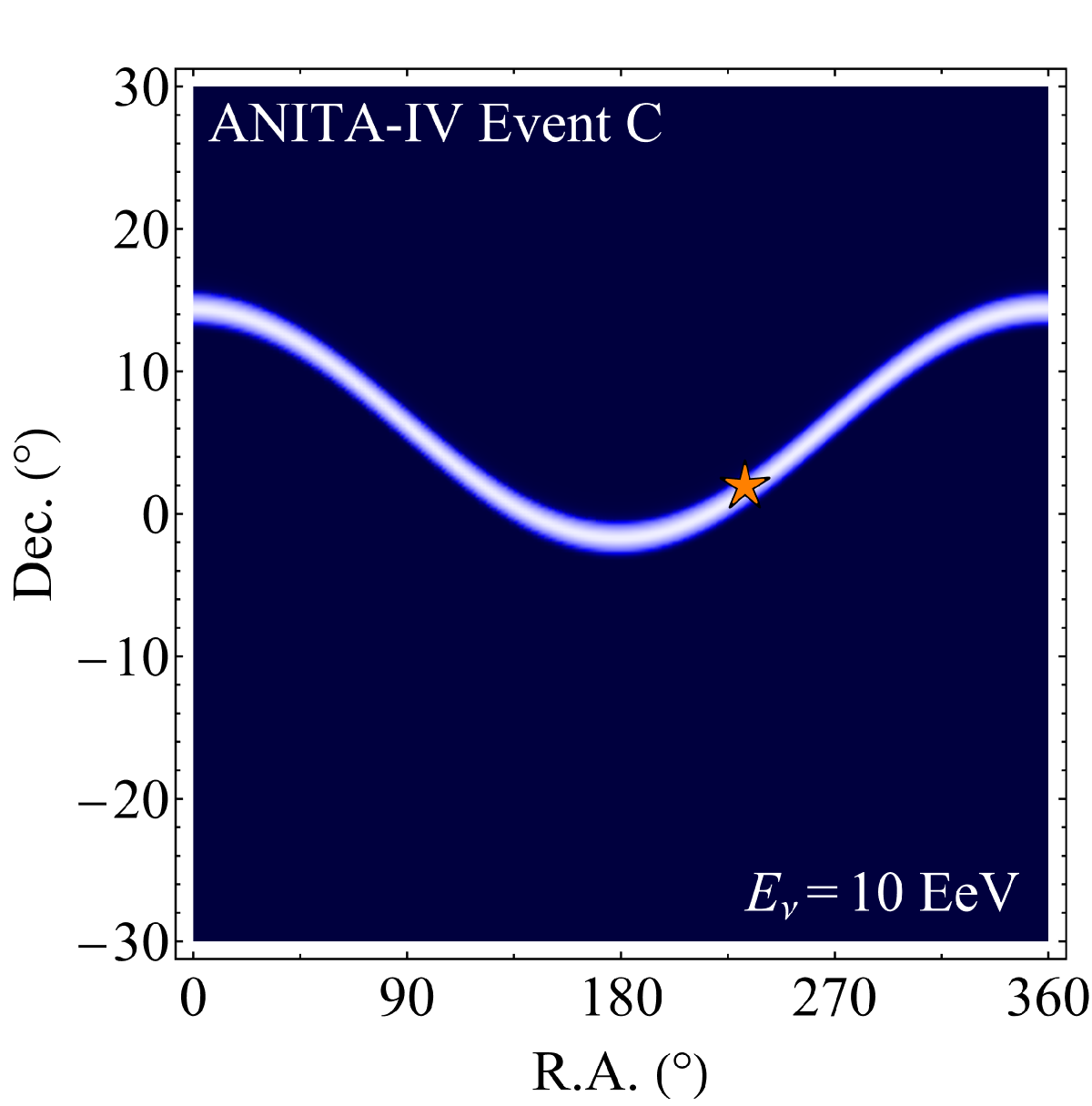}
    \end{minipage}
    \hfill
    \begin{minipage}[c]{0.29\linewidth}
        \centering
        \includegraphics[width=\linewidth]
        {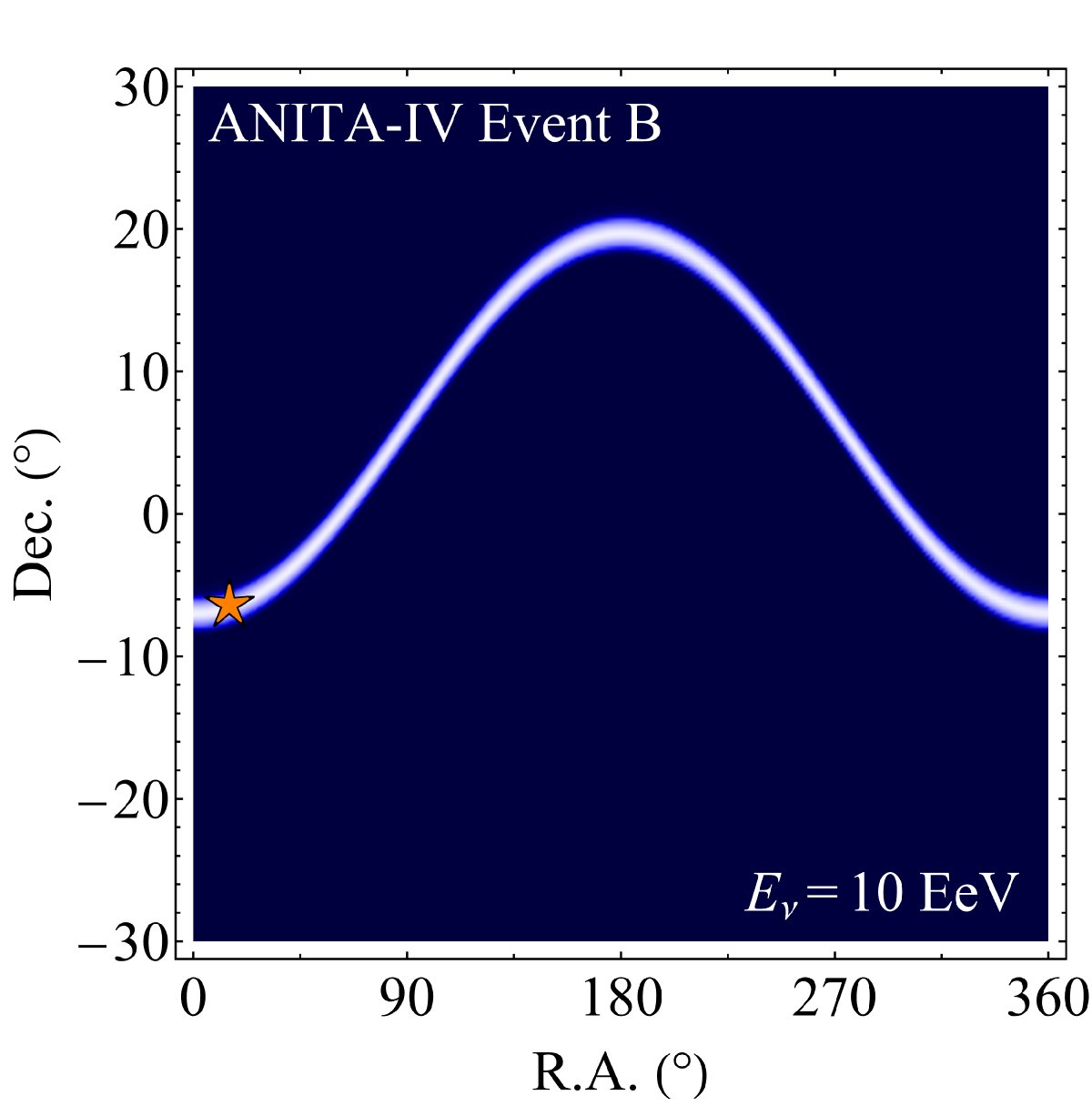}

        \vspace{0.1cm}

        \includegraphics[width=\linewidth]
        {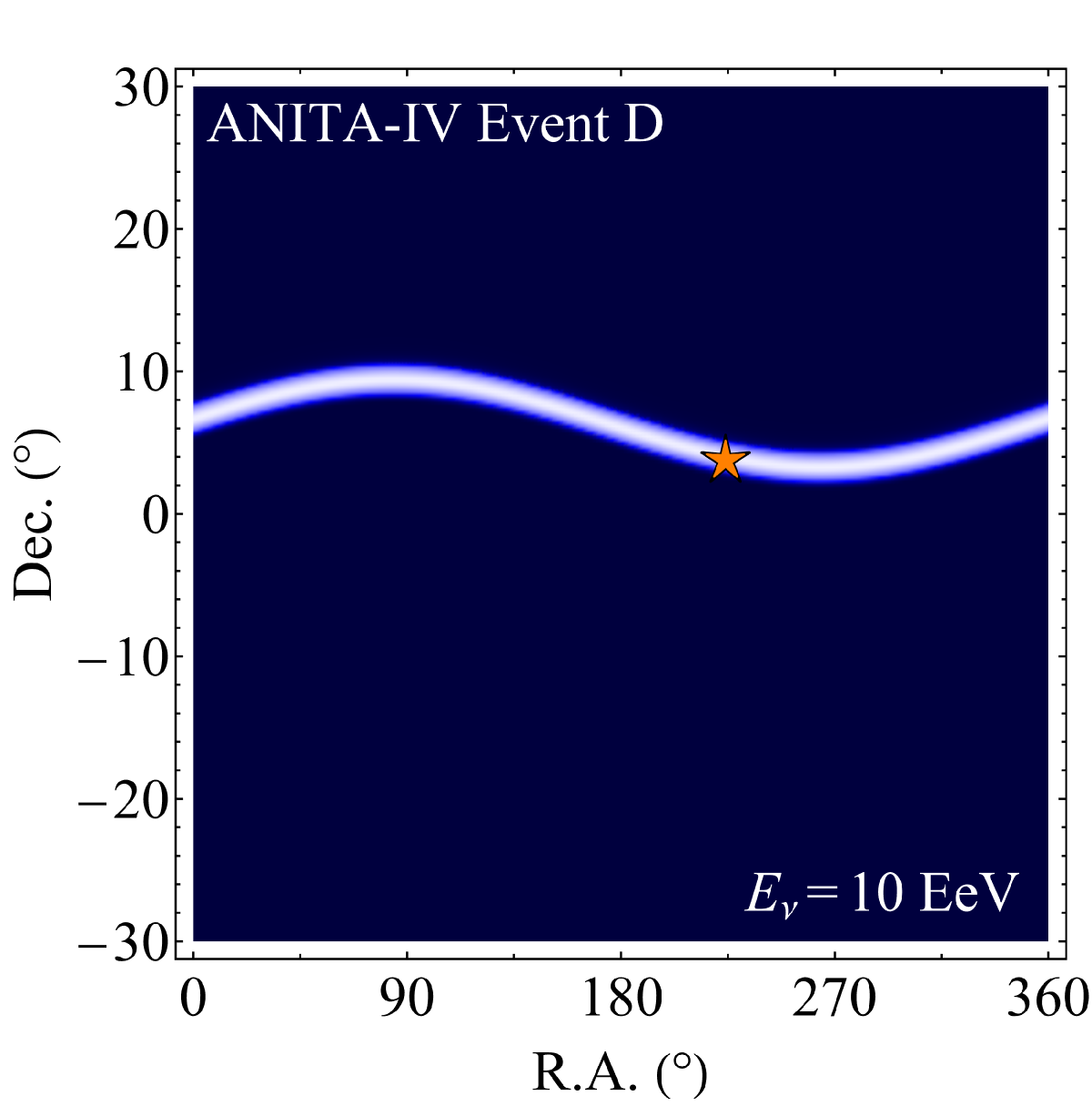}
    \end{minipage}
    \hfill
    \begin{minipage}[c]{0.005\linewidth}
    \centering
    {\color{gray!30}\rule{1pt}{10.05cm}}
    \end{minipage}
    \hfill
    \begin{minipage}[c]{0.29\linewidth}
        \centering
        \includegraphics[width=\linewidth]
        {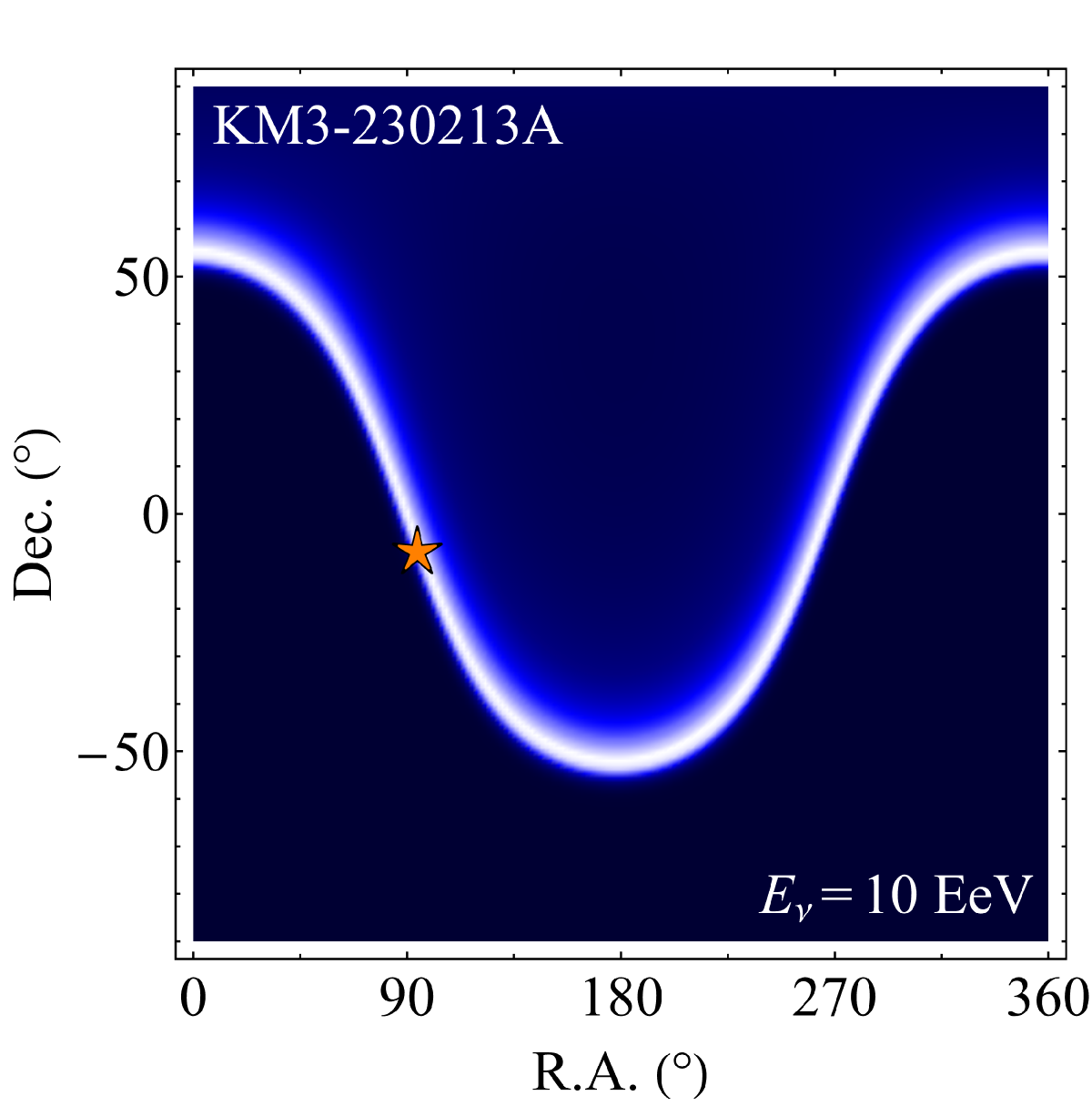}
    \end{minipage}
    \hfill
    \begin{minipage}[c]{0.05\linewidth}
        \centering
        \raisebox{0.2cm}{
        \includegraphics[width=\linewidth]
        {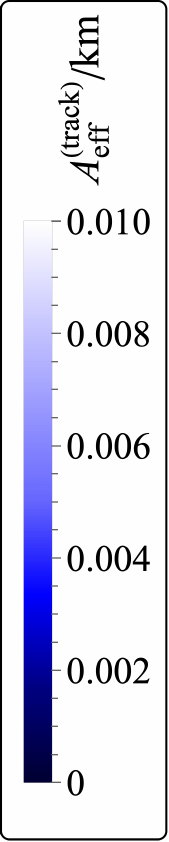}
        }
    \end{minipage}
    \caption{Instantaneous $\nu_\tau$ effective areas for the ANITA-IV (left and middle panels) and track-like effective area for KM3NeT (ARCA-21) (right panel), in equatorial coordinates, at the corresponding event times, evaluated for a fixed neutrino energy of $E_\nu=10~{\rm EeV}$. The orange stars indicate the reconstructed arrival directions of the events.}
    \label{fig:anita_instant_skymap}
\end{figure}

In this section, we present the effective areas in equatorial coordinates, complementing the exposure calculations discussed in the main text.

To compare the parts of the sky seen most effectively by ANITA-IV and KM3NeT with the reconstructed directions of the observed events, we convert the local detector coordinates into equatorial coordinates (right ascension $\alpha$ and declination $\delta$). The resulting effective area can be expressed as
\begin{equation}
A_{\rm eff}(E_\nu,\alpha,\delta;t) = A_{\rm eff} \left[ E_\nu, \theta(\alpha,\delta;t) \right] \; ,
\label{eq:sky_projected_effective_area}
\end{equation}
where the relation between $(\alpha,\delta)$ and the local elevation angle depends on the exact position of the detector as well as the observation time. Note that, for ANITA-IV, the detector position also changes continuously along the balloon trajectory, whereas the KM3NeT detector position is fixed.

In~\cref{fig:anita_instant_skymap}, we show the instantaneous ANITA-IV effective area in equatorial coordinates at the detection times of the four neutrino-like events. The optimal effective area window is concentrated in a restricted $(\alpha,\delta)$ region determined by the location of the detector during the event.
Neutrinos arriving too far below the local horizon are strongly attenuated while crossing the Earth, whereas directions above the horizon do not produce Earth-emerging $\tau$-leptons. This leads to a small band where ANITA-IV would be sensitive at a given instant.
The shape and the range of the optimal band in equatorial coordinates change across the four events because both the payload location and the local orientation of the horizon evolve during the flight.
The reconstructed direction of each event is indicated by an orange star. All four ANITA-IV events lie within the optimal effective area band. This is expected, since they were observed close to the horizon, where the experiment is most sensitive.

The instantaneous KM3NeT sky map at the time of KM3-230213A is shown in the rightmost panel of~\cref{fig:anita_instant_skymap}. At UHEs, the detection probability of steeply upgoing neutrinos is strongly suppressed due to Earth's absorption. In contrast, nearly horizontal trajectories can have a substantial survival probability and still provide a sufficiently long interaction region for the production of a detectable charged lepton. For the KM3NeT detector, rotation of the Earth determines the orientation of the local directional sensitivity in equatorial coordinates. As the Earth rotates, this local sensitivity pattern sweeps across the sky.

\begin{figure}[t!]
    \centering
    \raisebox{0.067\linewidth}{
    \includegraphics[height=0.32\linewidth]{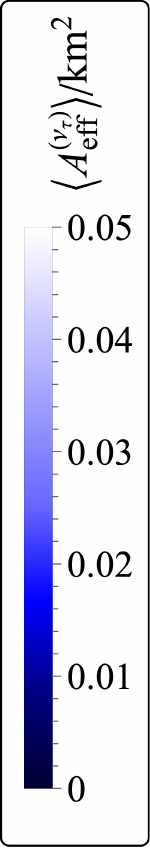}
    }
    \hfill
    \includegraphics[height=0.42\linewidth]{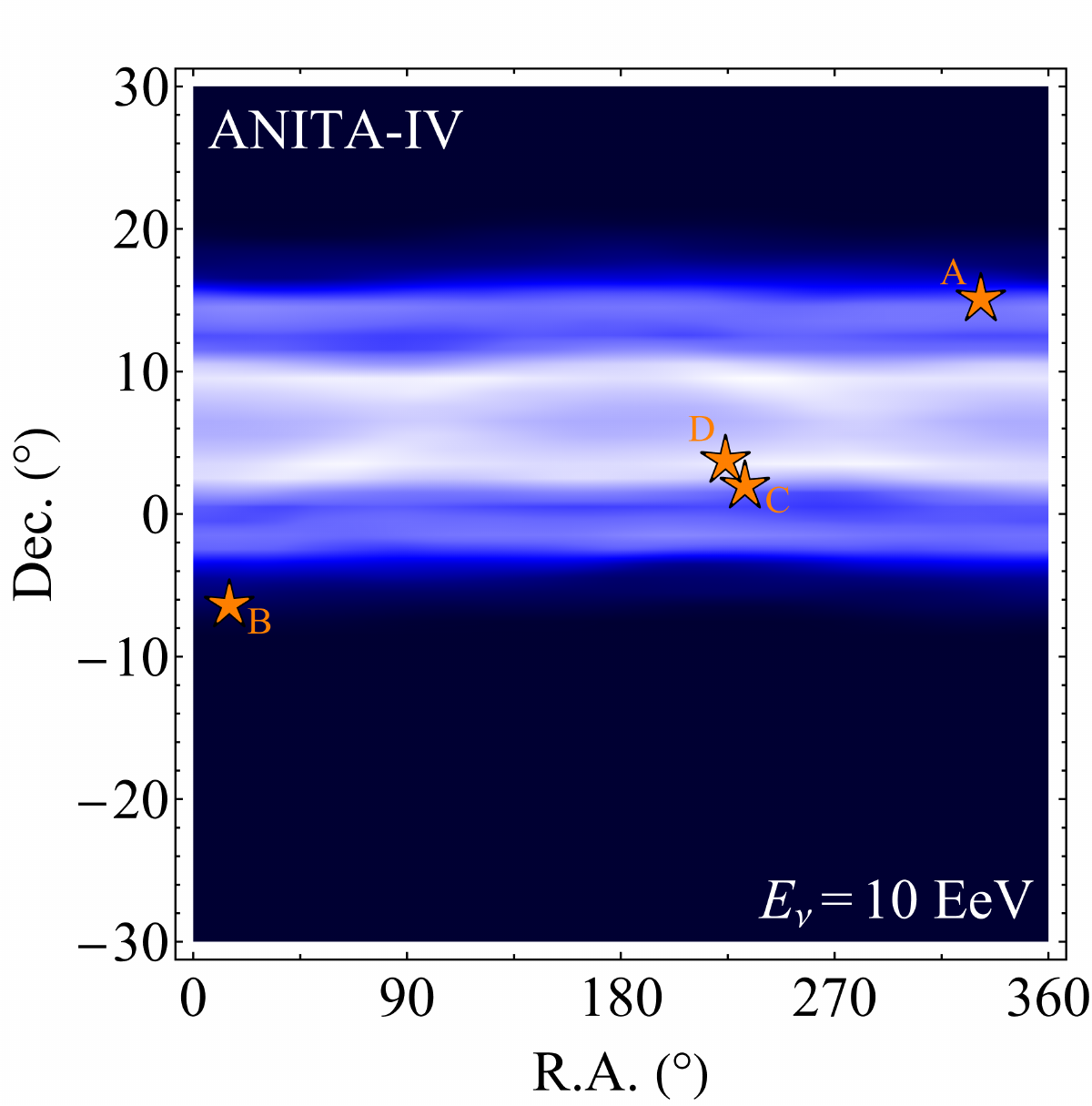}
    \hfill
    \includegraphics[height=0.42\linewidth]{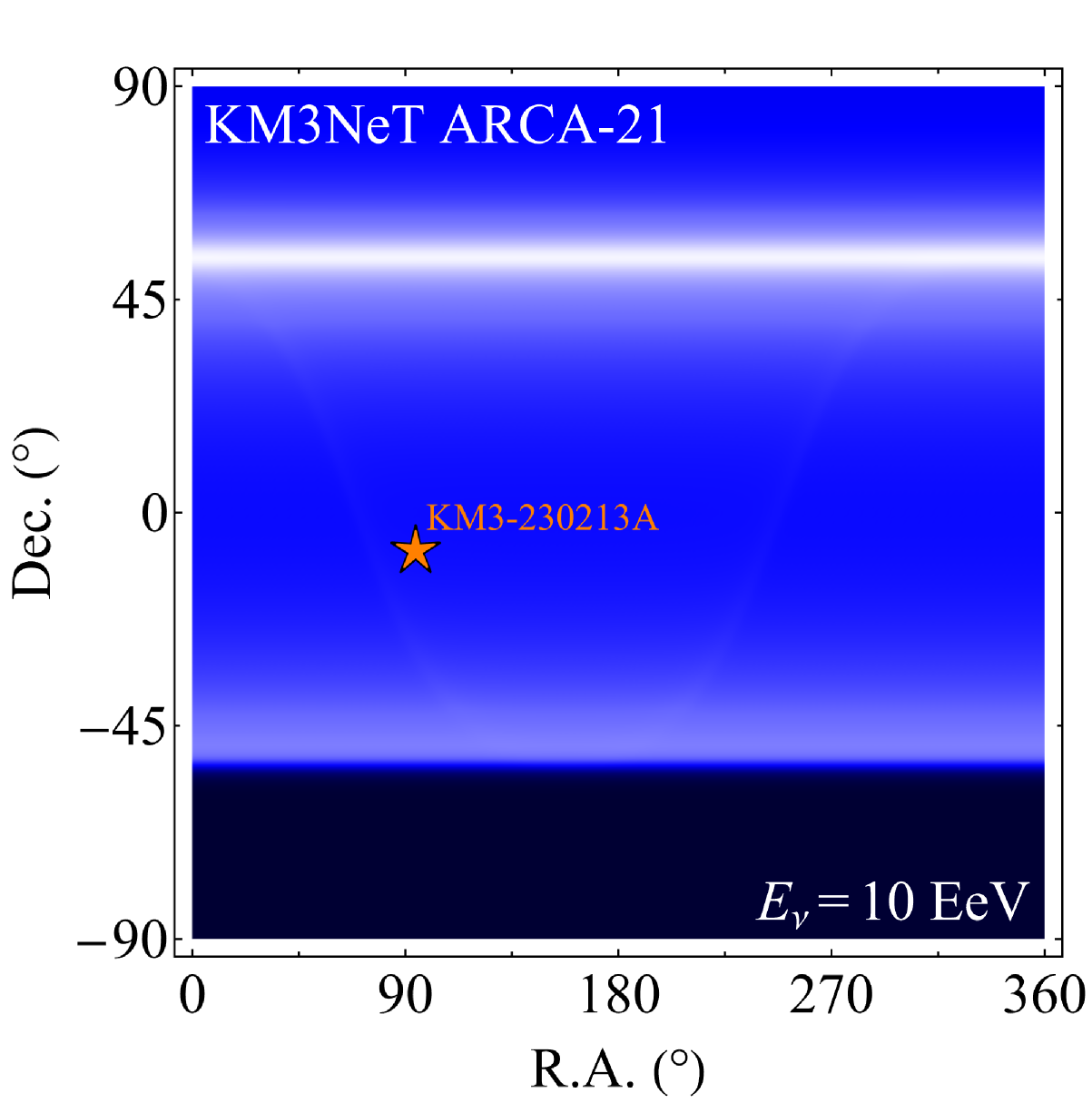}
    \hfill
    \raisebox{0.067\linewidth}{
    \includegraphics[height=0.32\linewidth]{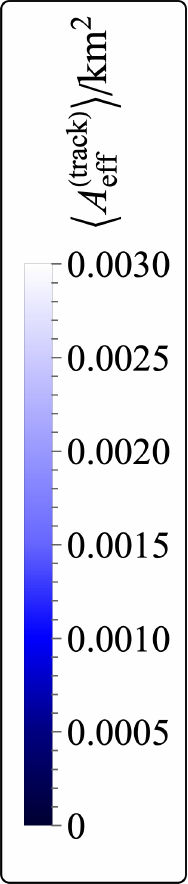}
    }
    \caption{Time-averaged effective area in equatorial coordinates for ANITA-IV and KM3NeT (ARCA-21).}
    \label{fig:anita_km3net_avg_eff_area}
\end{figure}

For a steady or a sufficiently long-lived source, or for a diffuse all-sky flux, the relevant quantity is the effective area averaged over the detector observation period,
\begin{equation}
\langle {A}_{\rm eff}(E_\nu,\alpha,\delta) \rangle = \frac{1}{T} \int_{t_0}^{t_0+T} dt \, A_{\rm eff}(E_\nu,\alpha,\delta;t)\;,
\label{eq:time_averaged_effective_area}
\end{equation}
where $T$ is the detector run time.
For ANITA-IV, this average would also include the time dependence of the payload latitude, longitude, and altitude along the flight trajectory.
For KM3NeT, the detector position is fixed, and the time dependence is generated by the rotation of the Earth.

We show the resulting time-averaged effective areas in~\cref{fig:anita_km3net_avg_eff_area}. Here, we average over the full ANITA-IV flight duration of approximately $26.6$~days. For KM3NeT, we instead show the effective area averaged over February 13, 2023, the day on which KM3-230213A was detected. Time averaging spreads the localized instantaneous sensitivity over the larger regions of the sky that pass through the detector's optimal field of view during the exposure.  For a fixed ground-based detector, such as KM3NeT, with approximately uniform coverage in sidereal time, the dependence on right ascension fades away, leaving the time-averaged exposure primarily as a function of declination. For ANITA-IV, the finite flight duration and changing payload trajectory produce a unique sky pattern. It is worth noting that Events C and D lie close to the region of maximum time-averaged ANITA-IV effective area, while Event A is farther from this region, and Event B falls in an unfavorable part of the sky.

\section{Diffuse-flux likelihood for ANITA-IV and IceCube}
\label{app:diffuse_anita_icecube}

In what follows, we present how anomalous the four ANITA-IV events are in the absence of any such UHE events at IceCube~\cite{ANITA:2021xxh,ANITA:2020gmv,Bertolez-Martinez:2023scp} under the diffuse-flux hypothesis. 
Specifically, the analysis performed here is analogous to the one in~\cref{sec:diffuse}, but considering only the four ANITA-IV events and the IceCube null observation, while not taking into account KM3-230213A.
For a power-law flux, the expected event counts can be expressed as
\begin{equation}
\mu_{\mathcal D}(\phi_0,\gamma) = \phi_0 K_{\mathcal D}(\gamma) \; ,
\end{equation}
where $\mathcal{D}$ denotes ANITA-IV and IceCube detectors, and $K_{\mathcal D}$ contains the energy- and time-integrated effective area of detector $\mathcal D$. The joint likelihood is~\cite{Cowan:2010js}
\begin{equation}
\mathcal L(\phi_0,\gamma) = \frac{\mu_{\rm A-IV}^{4} \exp\left(-\mu_{\rm A-IV}\right)}{4!} \exp\left(-\mu_{\rm IC}\right) \; .
\end{equation}
For any value of $\gamma$, profiling over the common flux normalization gives (see~\cref{eq:ntotal_obs_expected} and the corresponding discussions in~\cref{sec:diffuse})
\begin{equation}
\widehat{\mu}_{\rm A-IV} + \widehat{\mu}_{\rm IC} = 4 \;.
\end{equation}
The fit therefore reproduces the total observed count, while the goodness of fit is determined entirely by the fraction of the expected events assigned to ANITA-IV compared to that of IceCube. Defining
\begin{equation}
p_{\rm A-IV}(\gamma) = \frac{K_{\rm A-IV}(\gamma)} {K_{\rm A-IV}(\gamma)+K_{\rm IC}(\gamma)} \; ,
\end{equation}
the best-fit value takes the form
\begin{equation}
-2\Delta \log \mathcal{L} = 8\log\left[\frac{1}{p_{\rm A-IV}(\gamma)}\right] = 8\log\left[ 1+\frac{K_{\rm IC}(\gamma)}{K_{\rm A-IV}(\gamma)} \right] \; .
\label{eq:anita_icecube_profiled_deviance}
\end{equation}
Thus, the ANITA-IV--IceCube anomaly is directly regulated by the IceCube-to-ANITA-IV exposure ratio.

\begin{figure}[t!]
    \centering
    \includegraphics[width=0.48\linewidth]
    {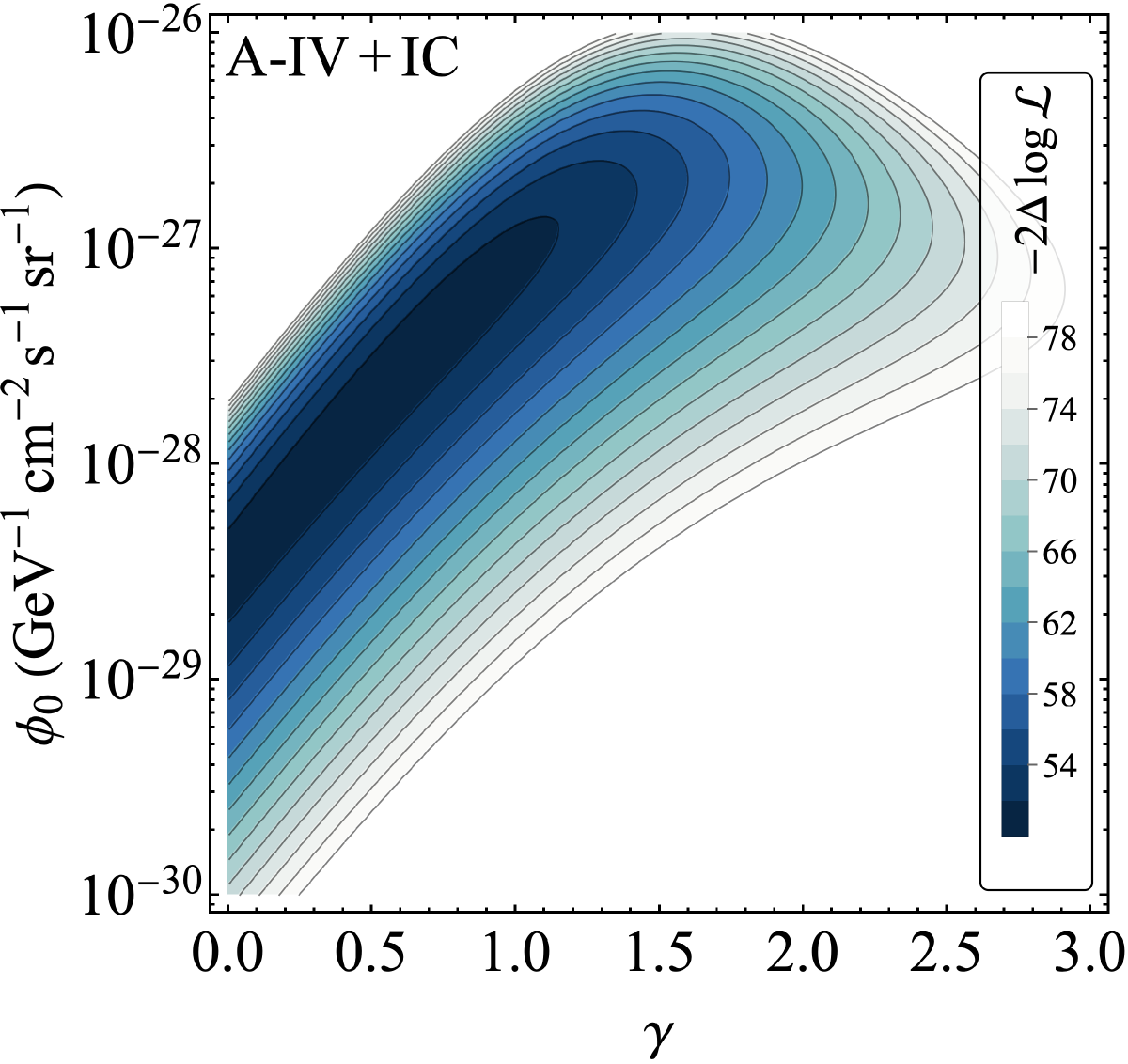}
    \hfill
    \raisebox{0.003\linewidth}{
    \includegraphics[width=0.486\linewidth]
    {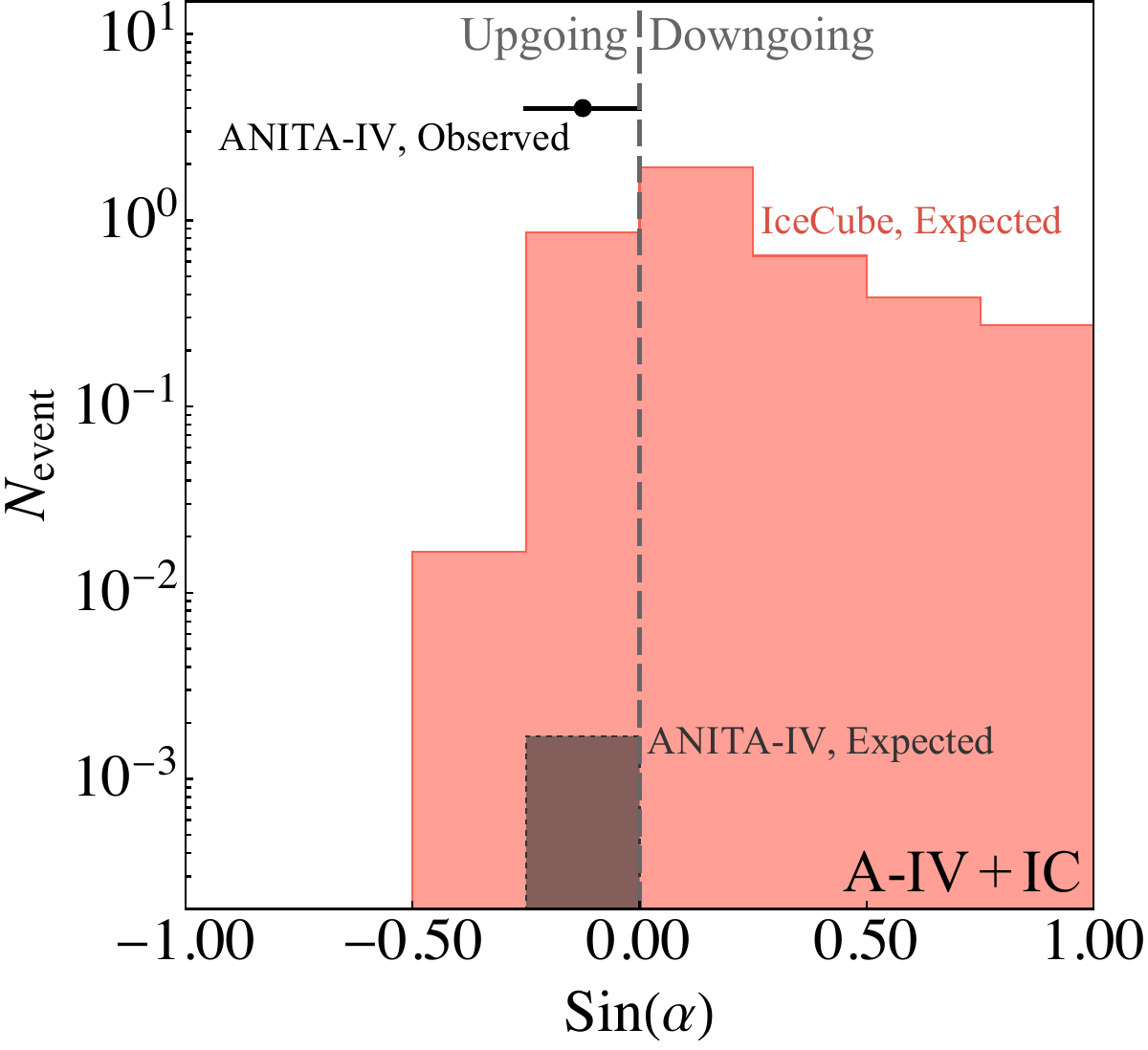}
    }
    \caption{Left panel: Joint Poisson deviance for the diffuse all-sky flux hypothesis, as a function of the spectral index $\gamma$ and the flux normalization $\phi_0$, using the observed counts $(n_{\rm A-IV},n_{\rm IC})=(4,0)$. Right panel: Angular distributions of the expected ANITA-IV and IceCube event counts at the best-fit flux normalization, for an $E_\nu^{-2}$ spectrum. The horizontal red marker denotes the four observed ANITA-IV events, while the vertical dashed line separates upgoing and downgoing trajectories. The best-fit normalization assigns nearly all of the total expected event counts to IceCube and strongly suppresses the ANITA-IV observations.}
    \label{fig:icecube_diffuse_events_ic_anita}
\end{figure}

In the left panel of~\cref{fig:icecube_diffuse_events_ic_anita}, we present the Poisson deviance as a function of the spectral index $\gamma$ and flux normalization $\phi_0$.
Varying $\gamma$ changes the ratio between the two detector exposures, but no region of the scanned parameter space yields a sufficiently large fraction of the expected events at ANITA-IV while remaining consistent with the IceCube null observation.
The right panel of~\cref{fig:icecube_diffuse_events_ic_anita} shows the angular distributions of the expected events at IceCube and ANITA-IV for the best-fit normalization for a fixed $E_\nu^{-2}$ spectrum.
While the best-fit value predicts a total of four events, almost all of this expectation is assigned to IceCube, while the expected ANITA-IV event count is at $\mathcal{O}(10^{-3})$.
Reproducing the observations, therefore, requires a very large upward fluctuation at ANITA-IV together with a downward fluctuation at IceCube.
For a fixed $E_\nu^{-2}$ spectrum, the minimum value of $-2\Delta \log \mathcal{L} = 62$ leads to a $p$-value of $3.4\times 10^{-15}$, which corresponds to $\approx 7.8\sigma$ tension.
For comparison, in the combined ANITA-IV--KM3NeT--IceCube analysis with a diffuse $E_\nu^{-2}$ flux, we found a tension of approximately $7.9\sigma$. The fact that a comparable tension persists when KM3-230213A is not considered indicates that the overall UHE tension is driven primarily by the four ANITA-IV events.

\section{Likelihood analysis for rare-transient scenarios}
\label{app:transient_likelihood}

In what follows, we present the supplemental results for the transient interpretation. First, we summarize the combined likelihood analysis for five fixed transient sources: four placed in the directions of the ANITA-IV events~\cite{ANITA:2021xxh,ANITA:2020gmv} and one placed in the direction of KM3-230213A~\cite{KM3NeT:2025npi}.

\begin{figure}
    \centering
    \includegraphics[height=0.42\linewidth]{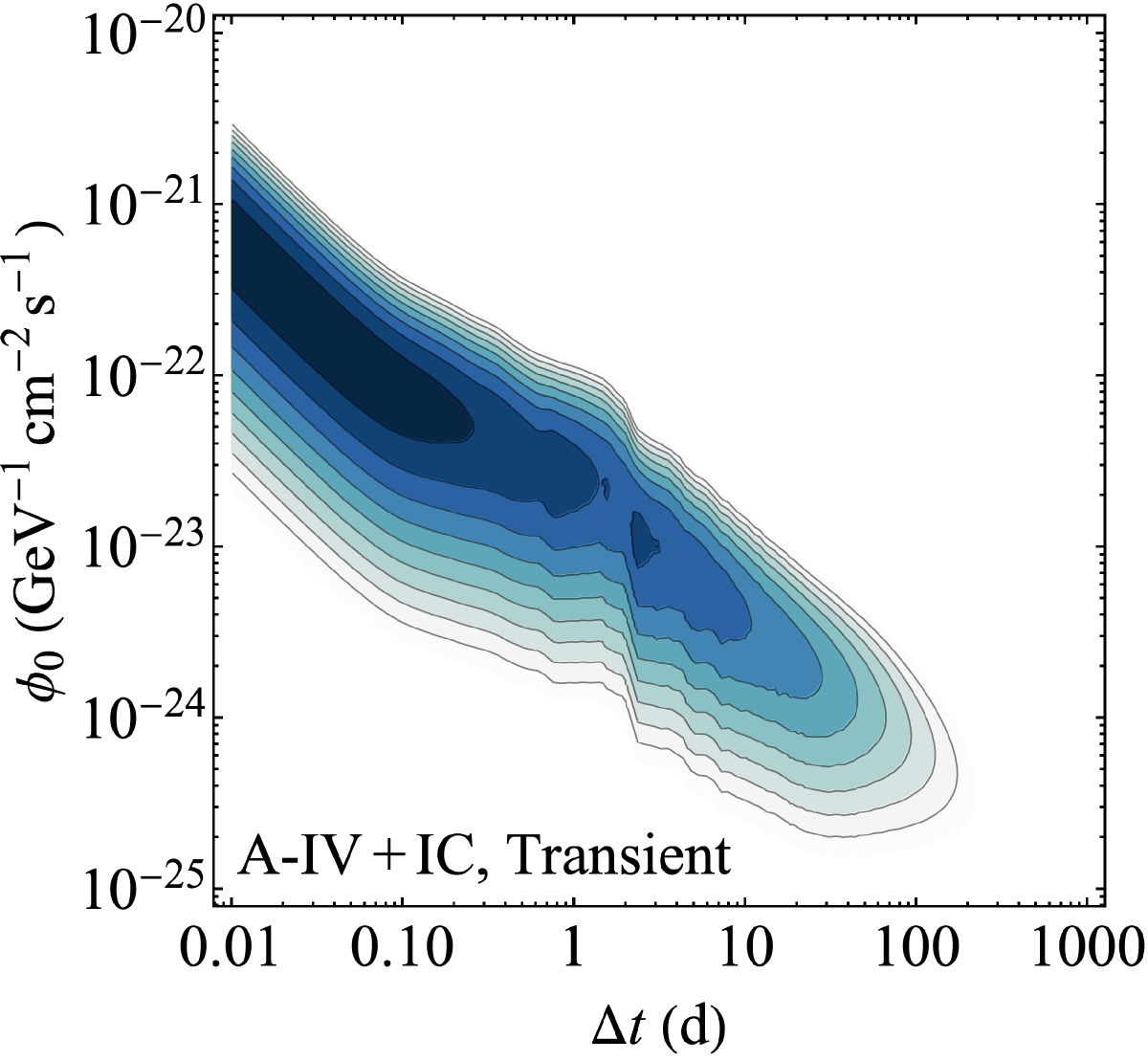}
    \hfill
    \includegraphics[height=0.42\linewidth]{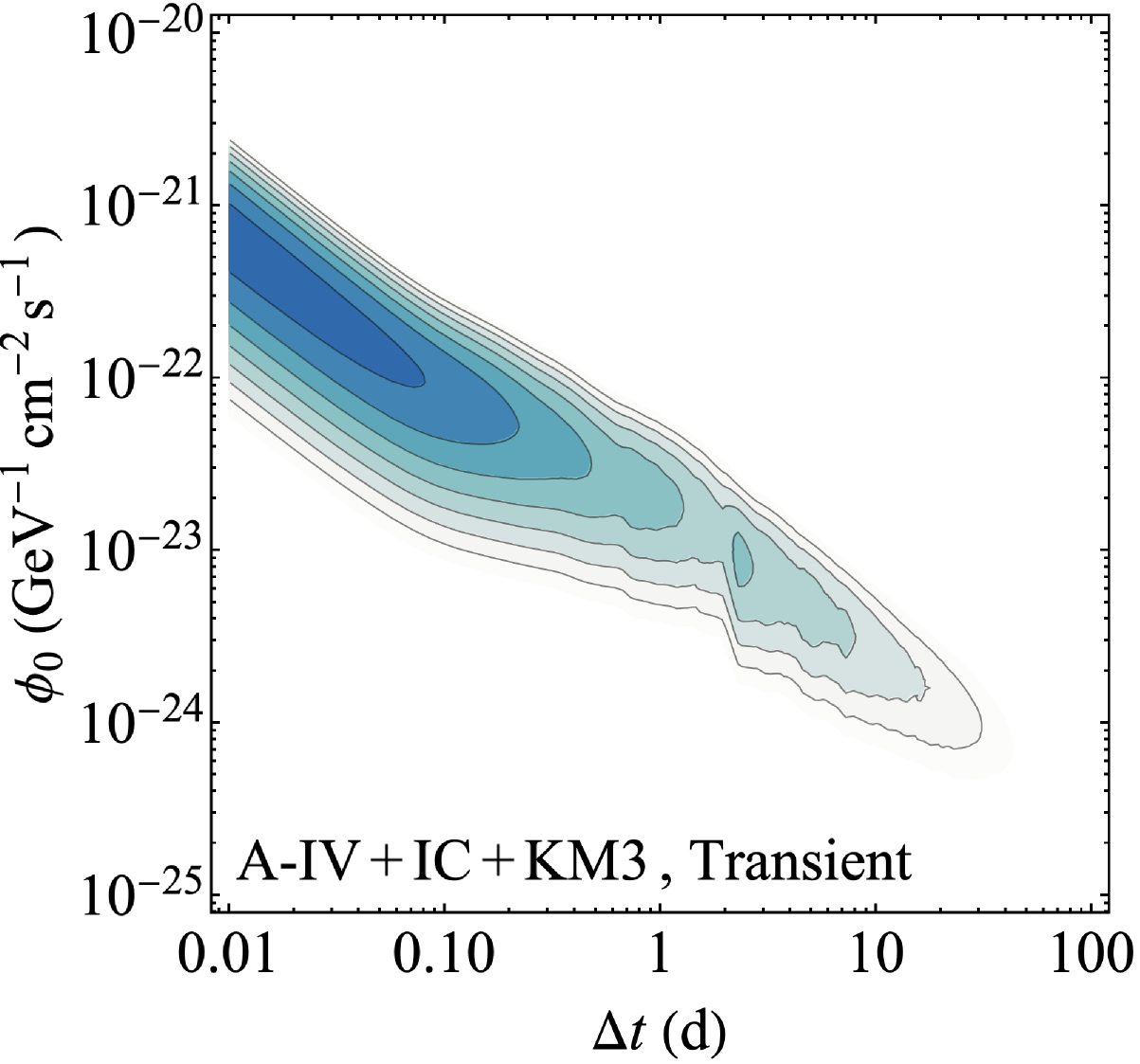}
    \hfill
    \raisebox{0.06\linewidth}{
    \includegraphics[height=0.355\linewidth]{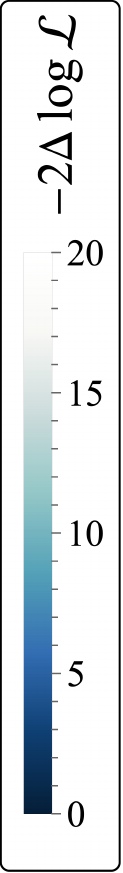}
    }
    \caption{Poisson deviance for a common flux normalization and a common transient duration considered across the fixed transient sources. Left panel: ANITA-IV and IceCube only, assuming four transient sources placed at the four ANITA-IV event directions and times, and using the observed counts $(n_{\rm A-IV},n_{\rm IC})=(4,0)$. Right panel: Result of the joint ANITA-IV, IceCube, and KM3NeT analysis, with an additional transient placed at the direction and time of KM3-230213A, using $(n_{\rm A-IV},n_{\rm IC},n_{\rm KM3})=(4,0,1)$. In both cases, we assume a fixed $E_\nu^{-2}$ spectrum.}
    \label{fig:log_likelihood_transient}
\end{figure}

Figure~\ref{fig:log_likelihood_transient} shows the combined Poisson deviance as a function of transient duration and flux normalization. The left panel corresponds to the ANITA-IV and IceCube comparison, where four transients are placed at the four observed ANITA-IV event directions and times.
The right panel shows the analogous result for the joint ANITA-IV, IceCube, and KM3NeT analysis, where a fifth source is added in the direction and time of KM3-230213A.
In both cases, shorter transients provide the most favorable fit, since they can overlap the ANITA-IV or KM3NeT optimal effective area windows while limiting the integrated IceCube exposure.
In the right panel, however, the tension persists even in the preferred region. The required normalization still predicts an appreciable IceCube event yield while underpredicting the KM3NeT count. This residual discrepancy is driven by the $\sim 1.8\sigma$ tension associated with KM3-230213A for short-duration transients, as discussed in~\cref{sec:fixed_transients}.

We now turn to the all-sky transient-population analysis described in~\cref{sec:all_sky_transients}. Here, we present the joint ANITA-IV and IceCube analysis, following the approach used for the combined ANITA-IV, IceCube, and KM3NeT analysis presented in the main text.
For each benchmark point, BP1, BP2, and BP3 (see~\cref{eq:transient_benchmark_points}), we generate $10^5$ realizations over the $5400$-day IceCube observation period and retain those in which ANITA-IV observes at least four distinct transients within its optimal elevation band. For each retained realization, we profile the likelihood over $\phi_0$ and rank the realizations according to the resulting test statistic.

We consider a population of rare transients with a fixed $E_\nu^{-2}$ spectrum, and vary only the common flux normalization $\phi_0$. For a given transient duration $\Delta t$ and a fixed realization of source positions and times, the expected number of events at detector $\mathcal D$ can be written as
\begin{equation}
\mu_{\mathcal D}(\phi_0,\Delta t) = \phi_0 K_{\mathcal D}(\Delta t) \; ,
\end{equation}
where $K_{\mathcal D}$ contains the corresponding time-dependent directional exposure. The joint likelihood is then constructed as a product of Poisson probabilities over the relevant detectors.

As in the diffuse case, profiling over the common normalization forces the total expected event count to equal the total observed count. Therefore, for the ANITA-IV and IceCube joint fit, we obtain
\begin{equation}
\widehat{\mu}_{\rm A-IV} + \widehat{\mu}_{\rm IC} = 4 \; .
\end{equation}
The deviance, and hence the inferred tension, is therefore determined by how the expected events are distributed among the detectors, rather than by the total event count alone.

\begin{table}[t!]
    \centering
    {\renewcommand{\arraystretch}{1.2}
    \begin{tabular}{|c|c|c|c|c|}
    \hline
    \; Scenario \; & \; $\phi_0~({\rm GeV}^{-1}~{\rm cm}^{-2}~{\rm s}^{-1})$ \; & \qquad $\mu_{\rm A-IV}$ \quad \; & \qquad $\mu_{\rm IC}$ \quad \; & \; $-2\Delta \log \mathcal{L}$ \; \\
    \hline
    \hline
    \multirow{3}{*}{BP1} & $4.40 \times 10^{-25}$ & $0.028$ & $3.972$  & $39.72$\\
         \cline{2-5}
         & $ 3.65 \times 10^{-25}$ & $0.025$ & $3.975$  & $40.50$\\
         \cline{2-5}
         & $4.50 \times 10^{-25}$ & $0.023$ & $3.977$  & $41.17$\\
    \hline
    \hline
    \multirow{3}{*}{BP2} & $6.74 \times 10^{-26}$ & $0.054$ & $3.946$  & $34.51$\\
         \cline{2-5}
         & $5.75 \times 10^{-26}$ & $0.043$ & $3.957$ & $36.19$\\
         \cline{2-5}
         & $7.87 \times 10^{-26}$ & $0.038$ & $3.962$ & $37.32$\\
    \hline
    \hline
    \multirow{3}{*}{BP3} & $2.44 \times 10^{-26}$ & $0.039$ & $3.961$ & $37.14$\\
         \cline{2-5}
         & $3.60 \times 10^{-26}$ & $0.037$ & $3.963$ & $37.46$\\
         \cline{2-5}
         & $2.70 \times 10^{-26}$ & $0.030$ & $3.970$ & $39.15$\\
         \hline
    \end{tabular}
    }
    \caption{Best three realizations, out of $10^5$ MC samples, for the three transient-population benchmark points (see~\cref{eq:transient_benchmark_points}) in the joint ANITA-IV and IceCube analysis.
    We consider only those realizations in which ANITA-IV observes at least four distinct transients within its favorable observation window.
    For each realization, the best-fit flux normalization, the corresponding expected event counts, and the Poisson deviance are shown.}
    \label{tab:transient_benchmark_anita_ic}
\end{table}

\begin{figure}[b!]
    \centering
    \includegraphics[width=0.5\linewidth]
    {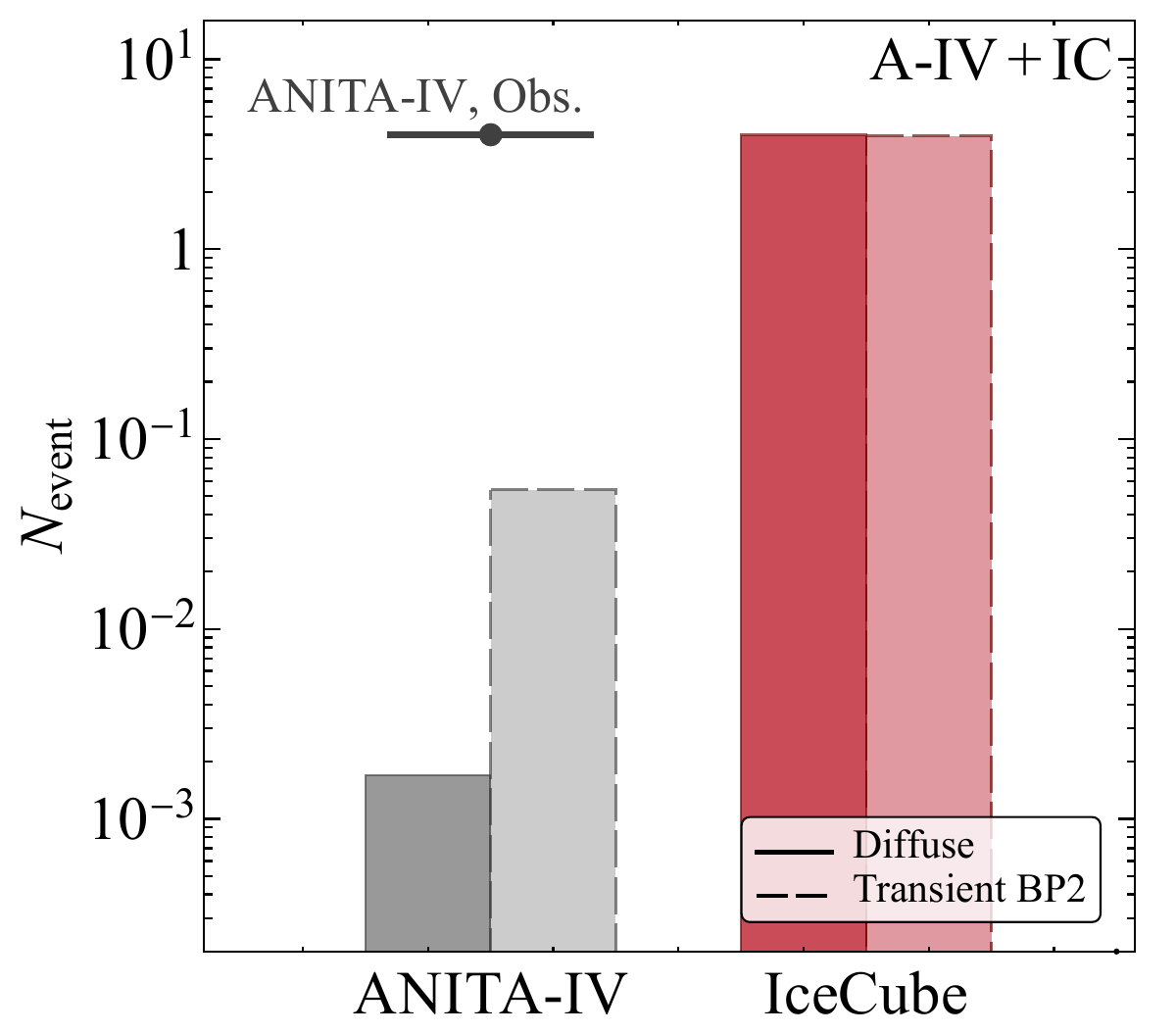}
    \caption{Best-fit expected event counts for the ANITA-IV--IceCube analysis, comparing the diffuse-flux hypothesis with the transient benchmark BP2. The horizontal red marker denotes the four observed ANITA-IV events. Although the transient benchmark improves the relative ANITA-IV exposure, the best-fit expectation remains dominated by IceCube.}
    \label{fig:transient_best_fit_counts}
\end{figure}

The best three realizations for the joint ANITA-IV and IceCube fit are listed in Table~\ref{tab:transient_benchmark_anita_ic}. Among the three benchmarks, BP2 provides the closest agreement with the observed event counts. However, even in the best-fit case, the ANITA-IV expectation value for the event counts remains far below the observed four events, while the IceCube expectation is close to four events. Thus, although the transient source hypothesis improves the ANITA-IV-to-IceCube exposure ratio relative to the diffuse scenario, it only alleviates the tension modestly. This is in agreement with our findings presented in~\cref{sec:all_sky_transients}, where KM3-230213A was also included.

In Fig.~\ref{fig:transient_best_fit_counts}, best-fit event counts for the transient benchmark BP2 (dashed boundary, lighter shades) are compared with the diffuse case (solid boundary, darker shades). This figure is analogous to~\cref{fig:best_fit_event_all_diffuse_transient}, presented in~\cref{sec:introduction}, and discussed further in ~\cref{sec:all_sky_transients}, where KM3-230213A is also included.
The transient-source scenario raises the ANITA-IV expectation from $\mathcal{O}(10^{-3})$ to $\mathcal{O}(10^{-1})$, but the best-fit event yield remains overwhelmingly dominated by IceCube.
The ANITA-IV observation must therefore still be interpreted as a large upward fluctuation relative to the best-fit expectation value.

Overall, for the joint ANITA-IV and IceCube case, and for a fixed $E_\nu^{-2}$ spectrum, the transient interpretation reduces the tension from $7.8\sigma$ to $5.8\sigma$.
The all-sky MC analysis for the benchmark points shows that, even after selecting realizations with four favorable transient sources at ANITA-IV, the best-fit event count remains concentrated towards IceCube.
In summary, both the full ANITA-IV--KM3NeT--IceCube analysis and the ANITA-IV--IceCube analysis yield a substantial residual tension under the all-sky transient-population hypothesis.

\end{document}